\documentclass[aps, prd, twocolumn, preprintnumbers, superscriptaddress, nofootinbib, floatfix]{revtex4-1}

\usepackage{amsmath}
\usepackage{bm}
\usepackage{amsfonts}
\usepackage{graphicx}
\usepackage{epstopdf}
\usepackage{hyperref}
\usepackage{array}

\usepackage{soul}
\usepackage{color}

\enlargethispage{\baselineskip}

\hypersetup{
     colorlinks   = true,
     citecolor    = blue
}

\renewcommand\({\left(}
\renewcommand\){\right)}
\renewcommand\[{\left[}
\renewcommand\]{\right]}
\newcommand{\bx}{{\bf x}}
\newcommand{\Te}{T_{\rm eff}}

\newcommand{\ns}{n_1^{\rm mis}}

\newcommand{\mA}{\mathcal{A}}

\newcommand{\mF}{\mathcal{F}}

\newcommand{\OCDM}{\Omega_{\rm CDM}}
\newcommand{\TRH}{T_{\rm RH}}
\newcommand{\MPl}{M_{\rm Pl}}

\begin{document}
\newcommand{\FIRSTAFF}{\affiliation{The Oskar Klein Centre for Cosmoparticle Physics,
	Department of Physics,
	Stockholm University,
	AlbaNova,
	10691 Stockholm,
	Sweden}}
\newcommand{\SECONDAFF}{\affiliation{Nordita,
	KTH Royal Institute of Technology and Stockholm University,
	Roslagstullsbacken 23,
	10691 Stockholm,
	Sweden}}

\title{Light axion-like dark matter must be present during inflation}

\author{Luca Visinelli}
\email[Electronic address: ]{luca.visinelli@fysik.su.se}
\FIRSTAFF
\SECONDAFF
\preprint{NORDITA-2017-026}
\date{\today}

\begin{abstract}
Axion-like particles (ALPs) might constitute the totality of the cold dark matter (CDM) observed. The parameter space of ALPs depends on the mass of the particle $m$ and on the energy scale of inflation $H_I$, the latter being bound by the non-detection of primordial gravitational waves. We show that the bound on $H_I$ implies the existence of a mass scale $\bar{m}_\chi = 10\,n{\rm eV} \div 0.5{\rm \,peV}$, depending on the ALP susceptibility $\chi$, such that the energy density of ALPs of mass smaller than $\bar{m}_\chi$ is too low to explain the present CDM budget, if the ALP field has originated after the end of inflation. This bound affects Ultra-Light Axions (ULAs), which have recently regained popularity as CDM candidates. Light ($m < m_\chi$) ALPs can then be CDM candidates only if the ALP field has already originated during the inflationary period, in which case the parameter space is constrained by the non-detection of axion isocurvature fluctuations. We comment on the effects on these bounds from additional physics beyond the Standard Model, besides ALPs.
\end{abstract}

\maketitle

\section{Introduction} \label{introduction}

In the era of precision cosmology, the cold dark matter (CDM) budget in our Universe has been established at about 84$\%$ of the total matter in the Universe, yet its composition remains unknown. Among the proposed hypothetical particles which could address this fundamental question is the QCD axion~\cite{Weinberg:1977ma,Wilczek:1977pj}, the quantum of the axion field arising from the spontaneous breaking of a U(1) symmetry conjectured by Peccei and Quinn (PQ~\cite{Peccei:1977hh, Peccei:1977ur}) to solve the strong-CP problem in quantum chromodynamics (QCD). The symmetry breaking occurs at a yet unknown energy scale $f_a$, which is constrained by measurements to be much larger than the electroweak energy scale~\cite{Raffelt2008}. The mass of the QCD axion at zero temperature $m_0$ is related to the axion energy scale $f_a$ by $m_0\,f_a = \Lambda_a^2$, where the energy scale $\Lambda_a$ is related to the QCD parameter $\Lambda_{\rm QCD}$. Realistic ``invisible'' axion models introduce new particles that further extend the Standard Model: examples include the coupling of the axion to heavy quarks~\cite{Kim:1979if, Shifman1980493} or to a Higgs doublet~\cite{Dine:1981rt, Zhitnitsky:1980tq}. 

The history and the properties of axions produced in the early Universe depend on the relative magnitude of the energy scale $f_a$ compared to the inflation energy scale $H_I$~\cite{Linde:1987bx, Linde:1991km, Turner:1991, Wilczek:2004cr, Tegmark:2005dy, Hertzberg:2008wr, Freivogel:2008qc, Mack:2009hv, Visinelli:2009zm, Acharya:2010zx}. In facts, if $f_a > H_I/2\pi$, the breaking of the U(1)$_{\rm PQ}$ symmetry occurs before reheating begins and axions must be present during inflation, while, if $f_a < H_I/2\pi$, the axion field originates after the end of inflation. Measurements of the CMB properties constrain the parameter space of the axion, including the scale of inflation $H_I$ and axion isocurvature fluctuations. Dense structures like axion miniclusters~\cite{Hogan:1988mp, Kolb:1993hw, Kolb:1993zz, Fairbairn:2017sil, Visinelli:2018wza, Vaquero:2018tib} or axion stars~\cite{Helfer:2016ljl, Braaten:2015eeu, Visinelli:2017ooc} could be used as laboratories for axion searches in the near future. Laboratory searches have developed strategies that involve axion electrodynamics~\cite{Wilczek:1987mv, Krasnikov:1996bm, Li:2009tca, visinelli:2013fia, Tercas:2018gxv, Visinelli:2018zif} for promising detection methods~\cite{Stern:2016bbw, Raggi:2014zpa, Majorovits:2016yvk, Kahn:2016aff, Alesini:2017ifp}. See Refs.~\cite{Raffelt:1995ym, Raffelt:2006rj, Sikivie:2006ni, Kim:2008hd, Wantz:2009it, Kawasaki:2013ae, Marsh:2015xka,Kim:2017yqo} for reviews of the QCD axion.

Besides the QCD axion, other Axion-Like Particles (ALPs) arise from various ultra-violet completion models, in which additional U(1) symmetries which are spontaneously broken are introduced, as well as some other underlying physics. In facts, although the ALP mass might share a common origin with the QCD axion, it is possible for these particle not to be related to the dynamics of the gauge fields whatsoever. Examples include ``accidental'' axions~\cite{Choi:2006qj, Choi:2009jt, Dias:2014osa, Higaki:2014pja, Kim:2015yna, Redi:2016esr, DiLuzio:2017tjx} and axions from string theory~\cite{Svrcek:2006yi, Arvanitaki:2009fg, Acharya:2010zx, Dine:2010cr, Ringwald:2012cu, Cicoli:2012sz, Kim:2013fga, Bachlechner:2014gfa, Halverson:2017deq, Stott:2017hvl, Visinelli:2018utg} that generally arise in models with extra dimensions~\cite{ArkaniHamed:1998nn, Randall:1999vf, Binetruy:1999ut, Chung:1999xg, Caldwell:2001ja, Visinelli:2017bny}. See also Ref.~\cite{Alonso:2017avz} for the effects of wormholes to the QCD axion potential. The potential of the axion thus generated might be in tension with the recent swampland conjectures, unless some sophisticated possibilities are considered~\cite{Agrawal:2018own,Akrami:2018ylq, Marsh:2018kub, Conlon:2018eyr, Murayama:2018lie, Kinney:2018nny, Damian:2018tlf, Danielsson:2018qpa}. In all these scenarios, two energy scales emerge: the symmetry-breaking scale $\Lambda$ and the ALP decay constant $f$. Similarly to the QCD axion, the ALP field acquires a mass $m \sim \Lambda^2/f$, so that, contrarily to the QCD axion, the mass $m$ and the energy scale $f$ can be treated as independent parameters. An interesting proposed ALP is the Ultra-Light Axion (ULA), of mass $m_{\rm ULA} \approx 10^{-22}\,$eV~\cite{Baldeschi:1983, Membrado:1989bqo, Press:1990, Sin:1994, Ji:1994, Lee:1996, Guzman:2000, Sahni:2000, Peebles:2000yy, Goodman:2000, Matos:2000, Hu:2000ke}. Such a light axion, recently revised in Refs.~\cite{Hui:2016ltb, Diez-Tejedor:2017ivd}, would have a wavelength of astrophysical scale $\lambda\sim 1\,$kpc and could possibly address some controversies arising when treating small scales in the standard $\Lambda$CDM cosmology, namely the missing satellites and the cusp-core problems~(see Ref.~\cite{Weinberg:2013aya} for a review).

ALPs from global and accidental U(1) symmetries share a common cosmological history with the QCD axion and spectate inflation whenever $f > H_I / 2\pi$. One of the main results of the present paper is to show that, in the opposite regime $f < H_I / 2\pi$, the observational constraint on $H_I$ coming from the Planck mission leads to a lower bound on the ALP mass, $m \gtrsim \bar{m}_\chi$, for some limiting mass $\bar{m}_\chi$ whose value depends on the ALP susceptibility $\chi$. We find a numerical value $\bar{m}_\chi = 10\,n{\rm eV} \div 0.5{\rm \,peV}$, depending on the value of $\chi$. This means that, if the CDM is discovered to be entirely composed of an ALP of mass $m < \bar{m}_\chi$, e.g. ULAs, such particles must be already present during inflation. Instead, if an ALP is discovered with $m > \bar{m}_\chi$, both cosmological origins are possible. We also show that, when $f > H_I / 2\pi$ and the U(1) symmetry is never restored afterwards, the non-detection of axion isocurvature fluctuations by the Planck mission leads to an upper bound on the scale of inflation $H_I$, regardless of the ALP mass. Although this second result is quite straightforward to derive, it has never been stressed in the past literature.

The paper is organized as follows. In Sec.~\ref{ALPs and inflation} we review the temperature dependence of the QCD axion mass, the ALP parameter space, and we derive the lower bound on the ALP mass. In Sec.~\ref{Framing the ALP parameter space} we show results for the ALP parameter space, assuming either a cosine or a harmonic potential. In Sec.~\ref{Effects of physics beyond the Standard Model}, we discuss some exceptions to the computation used coming from the effects of some physics beyond the standard model, including the modification to the effective number of degrees of freedom, non-standard cosmologies, or entropy dilution. Conclusions are drawn in Sec.~\ref{conclusion}.

\section{ALPs and inflation} \label{ALPs and inflation}

\subsection{Reviewing the temperature dependence of the QCD axion mass}

The QCD axion mass originates from non-perturbative effects during the QCD phase transition. At zero temperature, the axion gets a mass $m_0$ from mixing with the neutral pion~\cite{Weinberg:1977ma},
\begin{equation}
	m_0 = \frac{\Lambda_a^2}{f_a} = \frac{\sqrt{z}}{1+z}\,\frac{m_\pi f_\pi}{f_a},
\end{equation}
where $z = m_u/m_d$ is the ratio of the masses of the up and down quarks, $m_\pi$ and $f_\pi$ are respectively the mass and the energy scale of the pion, and $f_a$ is the QCD axion energy scale. The energy scale $\Lambda_a$ is proportional to the QCD scale $\Lambda_{\rm QCD}$, so that the axion mass is tied to the underlying QCD theory. Using $z = 0.48(5)$, $m_\pi = 132\,$MeV, and $f_\pi = 92.3\,$MeV, the authors in Ref.~\cite{diCortona:2015ldu} obtain $\Lambda_a = 75.5\,$MeV, a value slightly smaller than what obtained in other work. For example, Ref.~\cite{Wantz:2009it} obtains $\Lambda_a = 78\,$MeV within the framework of the ``interacting instanton liquid model'', fixing the QCD scale to $\Lambda_{\rm QCD} = 400\,$MeV. Recently, more refined computations on the QCD lattice have become accessible~\cite{Borsanyi:2015cka, Borsanyi:2016ksw, Petreczky:2016vrs}.

When temperature-dependent effects become important, the QCD axion mass acquires a complicated dependence on the plasma temperature~\cite{Gross:53.43, Fox:2004kb}. Here, we model such dependence as~\cite{Turner:1986, Bae:2008ue, Wantz:2009it}
\begin{equation}
	m_a(T) =  
	\begin{cases}
	\frac{\alpha^2\Lambda_{\rm QCD}^2}{f_a}\left(\frac{\Lambda_{\rm QCD}}{T}\right)^{\chi/2}, & \text{for~} T \geq \Te,\\
	m_0,& \text{for~} T < \Te,
	\end{cases}
	\label{eq:QCDaxion_mass}
\end{equation}
where $\chi$ is the QCD axion susceptibility and $\alpha$ is a numerical factor. At present, there is no general consensus on the numerical value of the susceptibility, which depends on the particle content of the embedding theory~\cite{Davoudiasl:2006bt, Davoudiasl:2017jke}, as well as the computational technique used~\cite{Turner:1986, Wantz:2009it, Borsanyi:2015cka}. Ref.~\cite{Wantz:2009it} obtains $\chi = 6.68$ and $\alpha = (1.68\times 10^{-7})^{1/4} \approx 0.02$ while the methods in Refs.~\cite{Gross:53.43, Fox:2004kb, Beltran:2006sq, Hertzberg:2008wr} predict $\chi = 8$ and
\begin{equation}
	\alpha = \frac{\Lambda_a}{\Lambda_{\rm QCD}} C^{1/2}\(\frac{\Lambda_{\rm QCD}}{200{\rm\,MeV}}\)^{1/4} \approx 0.03\div 0.05,
\end{equation}
where $C = 0.018$, see Eq.~(4) in Ref.~\cite{Beltran:2006sq}. In addition, we have introduced the temperature scale $\Te = \Lambda_{\rm QCD} (\alpha\Lambda_{\rm QCD}/\Lambda_a)^{4/\chi}$ at which the two expressions in Eq.~\eqref{eq:QCDaxion_mass} match. This allows us to rewrite Eq.~\eqref{eq:QCDaxion_mass} as $m_a(T) = m_0\,G(T)$, with the function
\begin{equation}
	G(T) = 
	\begin{cases}
	\left(\frac{\Te}{T}\right)^{\chi/2}, & \text{for~} T \geq \Te,\\
	1 ,& \text{for~} T < \Te.
	\end{cases}
	\label{eq:axion_massG}
\end{equation}

\subsection{Observational constraints}

The QCD axion, and more generally ALPs, are suitable CDM candidates in some region of the parameter space, provided that these particles are produced non-thermally. In the following, we assume that the totality of the observed CDM budget is in the form of ALPs of mass $m$. This is equivalent to demanding that the energy density in ALPs, here $\rho_A$, is equal to the present CDM  energy density $\rho_{\rm CDM}$. We write this requirement as
\begin{equation}
	\Omega_A h^2 = \OCDM h^2 = 0.1197 \pm 0.0022
	\label{CDM}
\end{equation}
where $\Omega_A = \rho_A / \rho_{\rm crit}$ and $\OCDM = \rho_{\rm CDM} / \rho_{\rm crit}$ are, respectively, the energy densities in ALPs and in the observed CDM~\cite{Ade:2015xua} at 68\% Confidence Level (CL), both given in units of the critical density $\rho_{\rm crit} = 3H^2_0\MPl^2/8\pi$, with the Planck mass $M_{\rm Pl}=1.221 \times 10^{19}\,$GeV and where $h$ is the Hubble constant $H_0$ in units of 100\,km s$^{-1}$Mpc$^{-1}$.

Besides its mass, energy scale, and initial value of the misalignment angle, the ALP energy budget depends on the Hubble expansion rate $H_I$ at the end of inflation, which is constrained from measurements on the scalar power spectrum $\Delta^2_{\mathcal{R}}(k_0)$ and the tensor-to-scalar ratio $r_{k_0}$ at the pivotal scale $k_0$ as~\cite{Lyth:1984, Lyth:1992yy}
\begin{equation}
	H_I < \frac{\MPl}{4}\,\sqrt{\pi\,\Delta^2_{\mathcal{R}}(k_0)\,r_{k_0}} \sim 7\times 10^{13}{\rm \,GeV}.
	\label{eq:HIbound}
\end{equation}
The numerical value of the bound has been computed by using the measurements at the wave number $k_0 = 0.05\, {\rm Mpc}^{-1}$~\cite{Ade:2013zuv, Planck:2013jfk, Barkats:2013jfa, Ade:2015tva, Array:2015xqh}
\begin{eqnarray}
	\Delta^2_{\mathcal{R}}(k_0) &=& (2.215^{+0.032}_{-0.079})\times 10^{-9}, \hbox{at 68\% CL},\\ \label{eq:measure_spectrum}
	r_{k_0} &<& 0.07, \qquad\qquad\quad\qquad\,\hbox{at 95\% CL}. \label{eq:measure_r}
\end{eqnarray}

We finally comment on isocurvature perturbations. Quantum fluctuations imprint into all massless scalar field $a$ present during inflation, with variance~\cite{Lyth:1990, kolb:1994early}
\begin{equation}
	\langle|\delta a^2|\rangle = \(\frac{H_I}{2\pi}\)^2.
	\label{variance_axion}
\end{equation}
Primordial quantum fluctuations later develop into isocurvature perturbations~\cite{Kobayashi:2013nva}, which modify the number density of axions, since the gauge invariant entropy perturbation is non-zero~\cite{Axenides:1983, Linde:1985yf, Seckel:1985},
\begin{equation}
	\mathcal{S}_a = \frac{\delta\(n_a/s\)}{n_a/s} \neq 0,
\end{equation}
where $s$ is the comoving entropy and $n_a$ the axion number density. If all of the CDM is in axions, then we define~\cite{Crotty:2003rz, Beltran:2005xd, Beltran:2006sq}
\begin{equation}
	\Delta_{S, A}^2 \equiv \langle |\mathcal{S}_a|^2\rangle = \Delta^2_{\mathcal{R}}(k_0)\frac{\beta}{1-\beta},
	\label{eq:axionisocurvaturebound}
\end{equation}
where the parameter $\beta$ is constrained from Planck~\cite{Ade:2013zuv, Planck:2013jfk} at the scale $k_0 = 0.05\, {\rm Mpc}^{-1}$ as
\begin{equation}
	\beta \lesssim 0.037,\quad \hbox{at 95\% CL},
	\label{eq:constrain_beta}
\end{equation}
independently on the ALP mass. 

\subsection{Constraining the ALP mass} \label{Constraining the ALP mass}

We now consider the parameter space of ALPs produced through the vacuum realignment mechanism (VRM)~\cite{Abbott:1982af, Dine:1982ah, Preskill:1982cy}, as revised in Appendix~\ref{sec_vrm}. Although, in principle, other mechanisms in addition to the VRM like the decay of topological defects produced at the PQ phase transition through the Kibble mechanism~\cite{Kibble:1976} and the decay of parent particles into ALPs might sensibly contribute to the present abundance of cold ALPs, we do not consider them here. 

Similarly to what obtained for axions, we represent the ALP mass as $m(T) = m\,G(T)$, where $m$ is a new parameter and $G(T)$ is given in Eq.~\eqref{eq:axion_massG}. The ALP susceptibility $\chi$ might take any real non-negative value and is left here as a free parameter. An infinite susceptibility corresponds to the ALP mass abruptly jumping from zero to the value $m$ at temperature $\Te$; any finite value of $\chi$ results in a smoother transition. ALPs from string theory or arising from accidental symmetries have $\chi = 0$. The ALP energy scale $f$ is related to the ALP mass by $f = \Lambda^2/m$, where $\Lambda$ is a new energy scale specified by an underlying theory. Finally, we write $\Te = c \Lambda$, for some constant value $c$.

We review the non-thermal production of a cosmological population of  ALPs from the misalignment mechanism in the Appendix~\ref{sec_vrm}, assuming that ALPs move in the potential
\begin{equation}
	V(\theta) = f^2 m^2(T)\,\(1-\cos \theta\),
	\label{eq:ALPpotential}
\end{equation}
where $\theta = a/f$ and $a$ is the ALP field. We assume that, when the ALP field originates, the initial value of the misalignment angle is $\theta_i$. The present value of the ALP energy density obtained from the misalignment mechanism is given in Eq.~\eqref{eq:presentnumberdensity},
\begin{equation}
	\rho_A = \frac{\Lambda^4\,G(T_1)}{2}\frac{g_S(T_0)}{g_S(T_1)}\left(\frac{T_0}{T_1}\right)^3\,\langle\theta_i^2\rangle,
	\label{eq:presentnumberdensity_TD}
\end{equation}
where $\langle\theta_i^2\rangle$ is the initial value of the misalignment angle squared, averaged over our Hubble volume, while the effective number of relativistic (``$R$'') and entropy (``$S$'') degrees of freedom are defined as~\cite{kolb:1994early}
\begin{eqnarray}
	g_R(T) &=& \sum_i\(\frac{T_i}{T}\)^4\int_0^{+\infty}Q_i(x)dx,\label{eq:deg_R}\\
	g_S(T) &=& \frac{3}{4}\sum_i\(\frac{T_i}{T}\)^3\! \int_0^{+\infty}\!\!\! x^2Q_i(x)\(1 \!+\! \frac{x^2}{3(x^2 \!+\! y_i^2)}\)\!dx,\nonumber\\
	Q_i(x) &=& \frac{15g_i}{\pi^4}\frac{\sqrt{x^2+y_i^2}}{\exp\(\sqrt{x^2+y_i^2}\)+(-1)^{Q_i^f}}\label{eq:deg_S}.
\end{eqnarray}
In the expressions above, $T$ is the temperature of the plasma, and the sum runs over the $i$ species considered, each with temperature $T_i$, mass $m_i$, $y_i \equiv m_i/T_i$, and $Q_i^f = 1$ ($Q_i^f = 0$) if $i$ is a fermion (boson). Instead of computing the integrals in Eqs.~\eqref{eq:deg_R}-\eqref{eq:deg_S}, we have considered the parametrization in Refs.~\cite{Coleman:2003hs, Wantz:2009it}, where the effective number of degrees of freedom are approximated with a series of step functions, for temperatures up to $O(100{\rm\,GeV})$.

In Eq.~\eqref{eq:presentnumberdensity_TD}, we have introduced the initial value of the misalignment angle $\theta_i$, which is the ALP field in units of $f$, and angle brackets define the average over all possible values of $\theta_i$. In this scenario, $\theta_i$ takes different values within our Hubble horizon, so
\begin{equation}
	\langle \theta_i^2\rangle = \frac{1}{2\pi}\int_{-\pi}^\pi\,\theta_i^2\,F(\theta_i)\,d\theta_i,
	\label{eq:effectiveangle}
\end{equation}
where the weighting function $F(\theta_i)$ has been thoroughly discussed in the literature~\cite{Lyth:1992, Strobl:1994wk, Bae:2008ue, Visinelli:2009zm, Visinelli:2009kt, Visinelli:2014twa, Diez-Tejedor:2017ivd}. Here, we take~\cite{Diez-Tejedor:2017ivd}
\begin{equation}
	F(\theta_i) = \ln\left[\frac{e}{1-\left(\theta_i/\pi\right)^4}\right],
	\label{eq:anharmonicities}
\end{equation}
which gives $\sqrt{\langle \theta_i^2\rangle} = 2.45$.

Coherent oscillations in the ALP field begin at temperature $T_1$ given by $3H(T_1) = m$, see Eq.~\eqref{eq:masscondition} below, and the Hubble rate during radiation domination is
\begin{equation}
	H(T) = \mA(T)\frac{T^2}{3\MPl}, \qquad \mA(T) = \sqrt{\frac{4\pi^3}{5}g_*(T)}.
	\label{eq:hubble_rate_r}
\end{equation}
The temperature $T_1$ at which the coherent oscillations in the ALP field begin is
\begin{equation}
	T_1 = \Te\,
	\begin{cases}
	\left(\frac{\hat{f}}{f}\right)^{\frac{2}{4+\chi}}, & \text{for~} f \leq \hat{f},\\
	\left(\frac{\hat{f}}{f}\right)^{\frac{1}{2}},& \text{for~} f > \hat{f}.
	\end{cases}
	\label{eq:T1_ALP}
\end{equation}
where we have defined the axion energy scale
\begin{equation}
	\hat{f} \equiv \frac{\MPl}{c^2\mA(T_1)}.
\end{equation}
Inserting Eq.~\eqref{eq:T1_ALP} into Eq.~\eqref{eq:presentnumberdensity_TD}, we obtain the present ALP energy density as
\begin{equation}
	\rho_A = \hat{\rho}_A\,\langle\theta_i^2\rangle\(\frac{m}{\hat{f}}\)^{1/2}\,
	\begin{cases}
	\left(\frac{f}{\hat{f}}\right)^{\frac{16+3\chi}{2(4+\chi)}}, & \text{for~} f \leq \hat{f},\\
	\left(\frac{f}{\hat{f}}\right)^2, & \text{for~} f > \hat{f},
	\end{cases}
	\label{eq:energy_density}
\end{equation}
where we have defined
\begin{equation}
	\hat{\rho}_A = \frac{g_{*S}(T_0)}{g_{*S}(T_1)}\frac{\hat{f}}{2}\left(\frac{T_0}{c}\right)^3.
	\label{eq:energy_density_funct}
\end{equation}
If the ALP field originates after inflation, the energy density is a function of the mass $m$ and the ALP energy scale $f$ only, but it does not depend on $\theta_i$ which is averaged out. Equating $\rho_A$ in Eq.~\eqref{eq:energy_density} with the CDM energy density $\rho_{\rm CDM} = \Omega_{\rm CDM}\,\rho_{\rm crit}$ gives
\begin{equation}
	f =\hat{f} \begin{cases}
	\(\frac{\rho_{\rm CDM}}{\hat{\rho}_A\langle\theta_i^2\rangle}\)^{\frac{8+2\chi}{16+3\chi}}\(\frac{\hat{f}}{m}\)^{\frac{4+\chi}{16+3\chi}}, & \text{for~} f \leq \hat{f},\\
	\(\frac{\rho_{\rm CDM}}{\hat{\rho}_A\langle\theta_i^2\rangle}\)^{\frac{1}{2}}\(\frac{\hat{f}}{m}\)^{\frac{1}{4}}, & \text{for~} f > \hat{f}.
	\end{cases}
	\label{eq:ALPenergydensity}
\end{equation}
For any value of $m$, Eq.~\eqref{eq:ALPenergydensity} expresses the value of $f$ for which the ALP explains the observed CDM budget.

We show that lighter ALPs cannot make the totality of the CDM when produced after the end of inflation. In facts, the region where $f < H_I / 2\pi$ (which implies $f < \hat{f}$) is constrained by the bound on $H_I$ expressed in Eq.~\eqref{eq:HIbound}, which leads to the lower bound on the ALP mass,
\begin{equation}
	m \geq \bar{m}_\chi \equiv \hat{f} \[\frac{64\pi}{\Delta^2_{\mathcal{R}}(k_0)\,r_{k_0}}\(\frac{\hat{f}}{\MPl}\)^2\]^{\frac{16+3\chi}{8+2\chi}}\(\frac{\rho_{\rm CDM}}{\hat{\rho}_A\langle\theta_i^2\rangle}\)^2.
	\label{eq:ALP_mass_chi}
\end{equation}
The numerical value of $\bar{m}_\chi$ depends on the susceptibility $\chi$ and on the value of the constant $c$ in the model. Setting $c = 1$, we obtain the limiting cases $\bar{m}_0 = 10\,n{\rm eV}$ and $\bar{m}_{\infty} = 0.5{\rm \,peV}$. Axion theories where $m < \bar{m}_\chi$ must embed the axion production in the inflationary mechanism, as we discuss below. We remark that the bound in Eq.~\eqref{eq:ALP_mass_chi} only applies if the ALP field originated after the end of inflation, $f < H_I/2\pi$, and if the ALP field has originated from the breaking of a U(1) symmetry. in these scenarios, a Hubble volume contains a multitude of patches where the axion field has a different, random value. These patches are bound by topological defects which could decay and leave to an additional component of the cold ALP energy density. The inclusion of non-relativistic ALPs from the decay of topological defects would increment their number density, potentially reducing the value of $\bar{m}_\chi$ by a couple of orders of magnitude. Here, we do not consider such contribution. Notice that the result in Eq.~\eqref{eq:ALP_mass_chi} does not depend on the value of $\Lambda$.

\subsection{ALPs and inflation}

\begin{figure}[h!]
\includegraphics[width=0.5\textwidth]{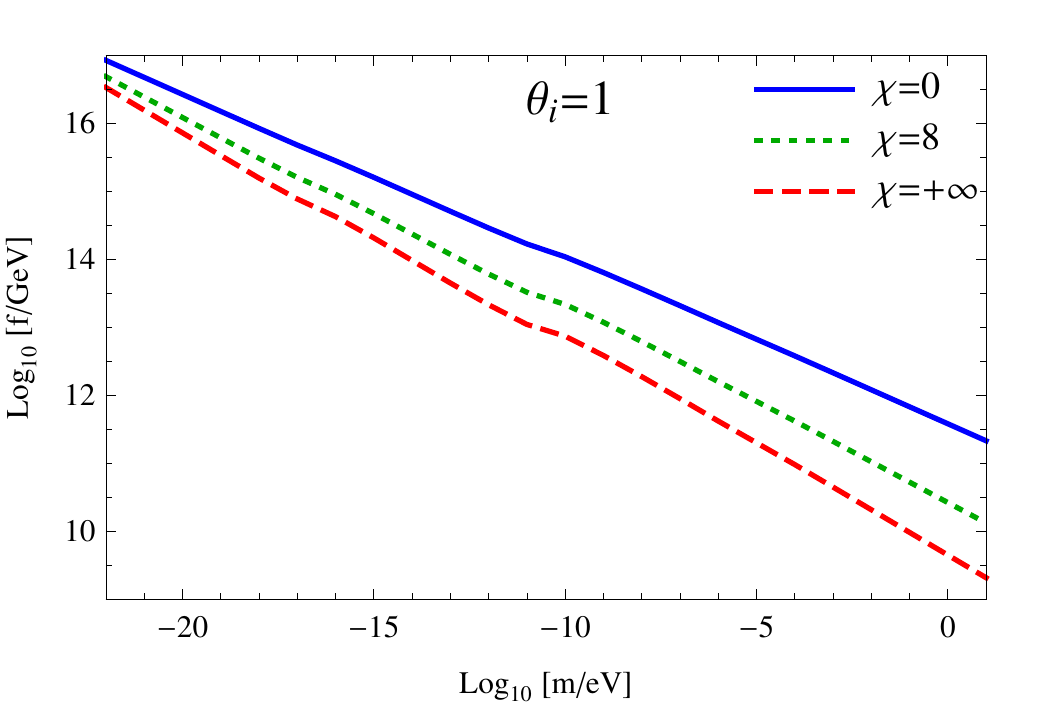}
	\caption{The energy scale $f$ as a function of the ALP mass $m$, Eq.~\eqref{eq:ALPenergydensity}. We have chosen the initial misalignment angle $\theta_i = 1$ and different values of the susceptibility: $\chi = 0$ (blue solid line), $\chi = 8$ (green dotted line), $\chi = +\infty$ (red dashed line).}
	\label{fig:axionenergyscale}
\end{figure}

ALPs of mass smaller than $\bar{m}_\chi$ can still be regarded as CDM candidates, although the related U(1) symmetry must have broken during the inflationary period, with the ALP energy scale satisfying $f > H_I/2\pi$. The cosmological properties of such ALPs would greatly differ from those described in the region $f < H_I/2\pi$, in particular no defects are present and a unique value of $\theta_i$ is singled out by the inflationary period within our Hubble volume. For example, consider the case of an ULA of mass $m_{\rm ULA}=10^{-22}\,{\rm eV}$, which is the mass scale proposed to solve some small-scale galactic problems~\cite{Baldeschi:1983, Membrado:1989bqo, Press:1990, Sin:1994, Ji:1994, Lee:1996, Guzman:2000, Sahni:2000, Peebles:2000yy, Goodman:2000, Matos:2000, Hu:2000ke} and recently has been vigorously reconsidered as a possible CDM candidate~\cite{Hui:2016ltb}. Since the mass scale $m_{\rm ULA}$ falls well within the limit excluded by Eq.~\eqref{eq:ALP_mass_chi}, ULAs must have been produced during inflation to be the CDM, with a precise relation between the initial misalignment angle and the energy scale given by Eq.~\eqref{eq:ALPenergydensity} with $\langle\theta_i^2\rangle$ replaced by $\theta_i^2\,F(\theta_i)$. The replacements accounts for the fact that the angle average $\langle \theta_i^2\rangle$ singles out a uniform value for $\theta_i$ over the entire Hubble volume. In this scenario, we expect that the initial misalignment angle is of the order of one, with smaller values of $\theta_i$ still possible albeit disfavored. In Fig.~\ref{fig:axionenergyscale}, we show the value for $f$ given in Eq.~\eqref{eq:ALPenergydensity}, as a function of the ALP mass $m$, for the value $\theta_i = 1$ and for different values of the ALP susceptibility: $\chi = 0$ (blue solid line), $\chi = 8$ (green dotted line), $\chi = +\infty$ (red dashed line). Values of $f$ of the order of the GUT scale $f \sim 10^{15}\,${\rm GeV} are expected for $m\sim 10^{-17}\div10^{-13}\,$eV, while the ULA mass $m_{\rm ULA} \sim 10^{-22}\,$eV gives $f \sim 10^{17}\,$GeV~\cite{Hui:2016ltb}. For higher values of the ALP mass, the spread among $f$ for different values of $\chi$ widens.

\section{Framing the ALP parameter space} \label{Framing the ALP parameter space}

\subsection{Cosine potential}

\begin{figure}
\includegraphics[width=0.5\textwidth]{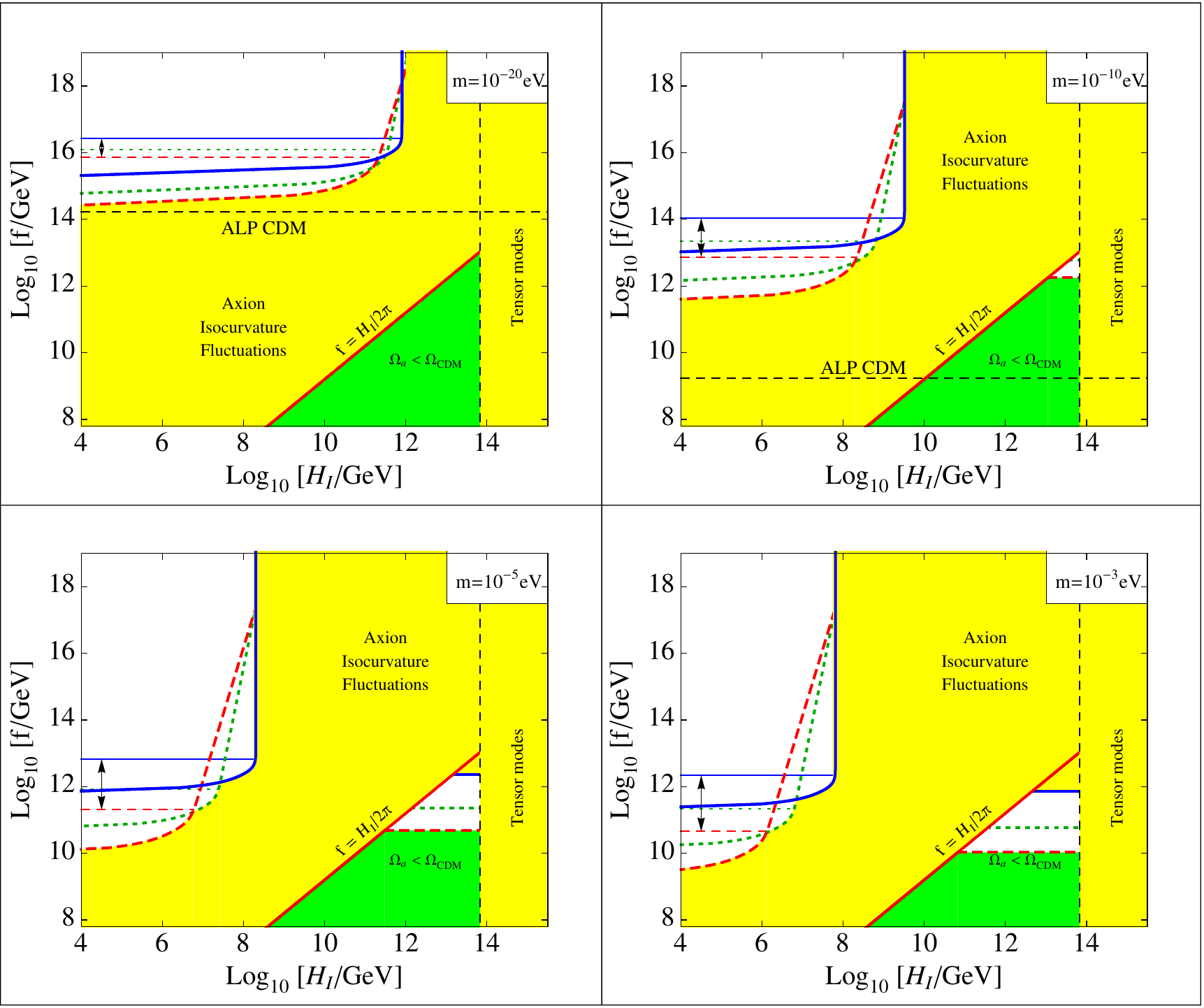}
	\caption{The APL parameter space for different values of the ALP mass. We have assumed the ALP potential in Eq.~\eqref{eq:ALPpotential}. Top left: $m=10^{-20}\,$eV. Top right: $m=10^{-10}\,$eV. Bottom left: $m=10^{-5}\,$eV. Bottom right: $m=10^{-3}\,$eV. We have shown results for different values of the susceptibility: $\chi= 0$ (solid blue), $\chi = 8$ (dotted green), $\chi = +\infty$ (dashed red). The yellow region is excluded by CDM overproduction, $\rho_A > \rho_{\rm CDM}$. The region labeled ``Axion isocurvature fluctuations'' is excluded below the curve shown. For clarity, we have shaded in yellow the region below the lowest of the three curves only. Horizontal lines show the values of $f$ for which the ALP is the CDM for each ALP mass and for each value of $\chi$. The green region is accessible, however the ALP is a subdominant CDM component, $\rho_A < \rho_{\rm CDM}$. The bound labelled ``Tensor modes'' is derived from the non-detection of primordial gravitational waves, see Eq.~\eqref{eq:HIbound}.}
	\label{fig:axionparameterspace}
\end{figure}

We apply the expression for axion isocurvature fluctuations in Eq.~\eqref{eq:axionisocurvaturebound} to the ALP scenario, to obtain~\cite{Kitajima:2014xla}
\begin{equation}
	\Delta_{S,A}^2 = \left(\frac{\partial \ln\rho_A}{\partial \theta_i}\right)^2\,\langle\delta \theta_i^2\rangle = \bigg(\frac{H_I\,\mF(\theta_i)}{\pi\,\theta_i \, f}\bigg)^2,
	\label{eq:axionisocurvaturefluctuations}
\end{equation}
where in the last step we have used Eq.~\eqref{variance_axion}, and where we defined the function
\begin{equation}
	\mF(x) = 1 + \frac{xF'(x)}{2F(x)}.
\end{equation}
Results on the various bounds on the ALP parameter space are summarized in Fig.~\ref{fig:axionparameterspace}. Since we do not consider the contribution from the decay of topological defects, the parameter space of CDM ALPs depends on six quantities, $f$, $\theta_i$, $H_I$, $m$, $c$, and $\chi$. We show how the parameter space modifies when considering different values of the ALP mass: $m=10^{-20}\,$eV (top left), $m=10^{-10}\,$eV (top right), $m=10^{-5}\,$eV (bottom left), and $m=10^{-3}\,$eV (bottom right). For each panel, the line $f=H_I/2\pi$ separates the region where the axion is present during inflation (top-left) from the region where the axion field originates after inflation (bottom-right), for a fixed value $c = 1$. This line has to be though as a qualitative bound between the two scenarios we will describe, since the exact details depend on the inflationary model, the preheating-reheating scenarios, and axion particle physics. The horizontal line labeled ``ALP CDM'' corresponds to the requirement that the primordial ALP condensate has started behaving like CDM at matter-radiation equality (See Ref.~\cite{Arias:2012az} for details),
\begin{equation}
	f \gtrsim \frac{53\,{\rm TeV}}{\pi}\,\sqrt{\frac{\rm eV}{m}}.
\end{equation}

We first discuss the scenario where $f > H_I/2\pi$. The region is bound by the non detection of axion isocurvature fluctuations, obtained from Eq.~\eqref{eq:constrain_beta} with the requirement that $\rho_A = \rho_{\rm CDM}$. We plot the bound for three different values of the susceptibility: $\chi = 0$ (blue solid line), $\chi = 8$ (green dotted line), $\chi = +\infty$ (red dashed line). For clarity, we shade in yellow the region below the minimum of the three curves although we have to bear in mind that the whole parameter space below a curve of fixed $\chi$ has been ruled out. The change in the slope corresponds to the argument of the anharmonicity function $F(\theta_i)$ approaching $\pi$. For each value of $\chi$, the horizontal lines in the allowed parameter space show the ``natural'' value of $f$ for which $\rho_A = \rho_{\rm CDM}$ and $\theta_i = 1$, as shown in Fig.~\ref{fig:axionenergyscale}. For $m=10^{-20}\,$eV, the natural value of the axion energy scale is of the order of $f \sim 10^{16}\,$GeV, corresponding to the ``ALP miracle'' discussed in Ref.~\cite{Hui:2016ltb}. For smaller values of the ALP mass, the natural value of $f$ lowers, and the spread among different values of $\chi$ widens, as shown in Fig.~\ref{fig:axionenergyscale}. The bound from isocurvature fluctuations steepens when $\theta_i$ decreases, and it is vertical for $\theta_i \ll 1$ and for $\chi = 0$, or for $f > \bar{f}$. We reformulate this constraint as an upper bound on $H_I$ for a given ALP theory, which is obtained by combining Eqs.~\eqref{eq:axionisocurvaturebound},~\eqref{eq:ALPenergydensity}, and~\eqref{eq:axionisocurvaturefluctuations} as
\begin{equation}
	H_I \leq \pi \hat{f} \(\frac{\hat{f}}{m}\)^{\frac{1}{4}}\sqrt{\frac{\rho_{\rm CDM}}{\hat{\rho}_A}\frac{\beta}{1-\beta}\Delta^2_{\mathcal{R}}(k_0)} = 10^{7}{\rm GeV}\sqrt[4]{\frac{\rm eV}{m}}.
	\label{eq:bound_HI}
\end{equation}
Isocurvature bounds have been used in the string axiverse realization discussed in Ref.~\cite{Acharya:2010zx}, neglecting the dependence on the susceptibility and the anharmonic corrections in the potential. The presence of axion isocurvatures in the CMB, whose constrain on the power spectrum leads to Eq.~\eqref{eq:bound_HI}, relies on the fact that the PQ symmetry has never been restored after the end of inflation. Caveats that allow to evade the bound from isocurvature fluctuations in Eq.~\eqref{eq:bound_HI} include the presence of more than one ALP~\cite{Kitajima:2014xla} or by identifying the inflaton with the radial component of the PQ field~\cite{Fairbairn:2014zta}. This latter technique has been embedded into the SMASH model~\cite{Ballesteros:2016xej} where, for a decay scale $f \lesssim 4\times10^{16}\,$GeV, the PQ symmetry is restored immediately after the end of inflation and isocurvature modes are absent, so that the bound in Eq.~\eqref{eq:bound_HI} does not apply.

In the second scenario $f< H_I/2\pi$, the axion is not present during inflation. In this scenario, a horizontal line gives the value of $f$ for which the ALP is the CDM for a given value of the susceptibility. ALPs with an energy scale smaller than this value are a subdominant CDM component (green region, $\rho_A < \rho_{\rm CDM}$), while values above are excluded (yellow region, $\rho_A > \rho_{\rm CDM}$). The constrain in Eq.~\eqref{eq:ALP_mass_chi} applies in this region of the parameter space, for some values of the ALP mass. For $m = 10^{-20}\,$eV, which lies below the critical value $\bar{m}_\chi$ in Eq.~\eqref{eq:ALP_mass_chi}, we always have $\rho_A < \rho_{\rm CDM}$, so the region $f< H_I/2\pi$ is shaded with green. Larger values of the ALP mass allow for $\rho_A = \rho_{\rm CDM}$ for some values of $f$ and $\chi$, avoiding the constrain in Eq.~\eqref{eq:ALP_mass_chi}.

\subsection{Harmonic potential}

\begin{figure}
\includegraphics[width=0.5\textwidth]{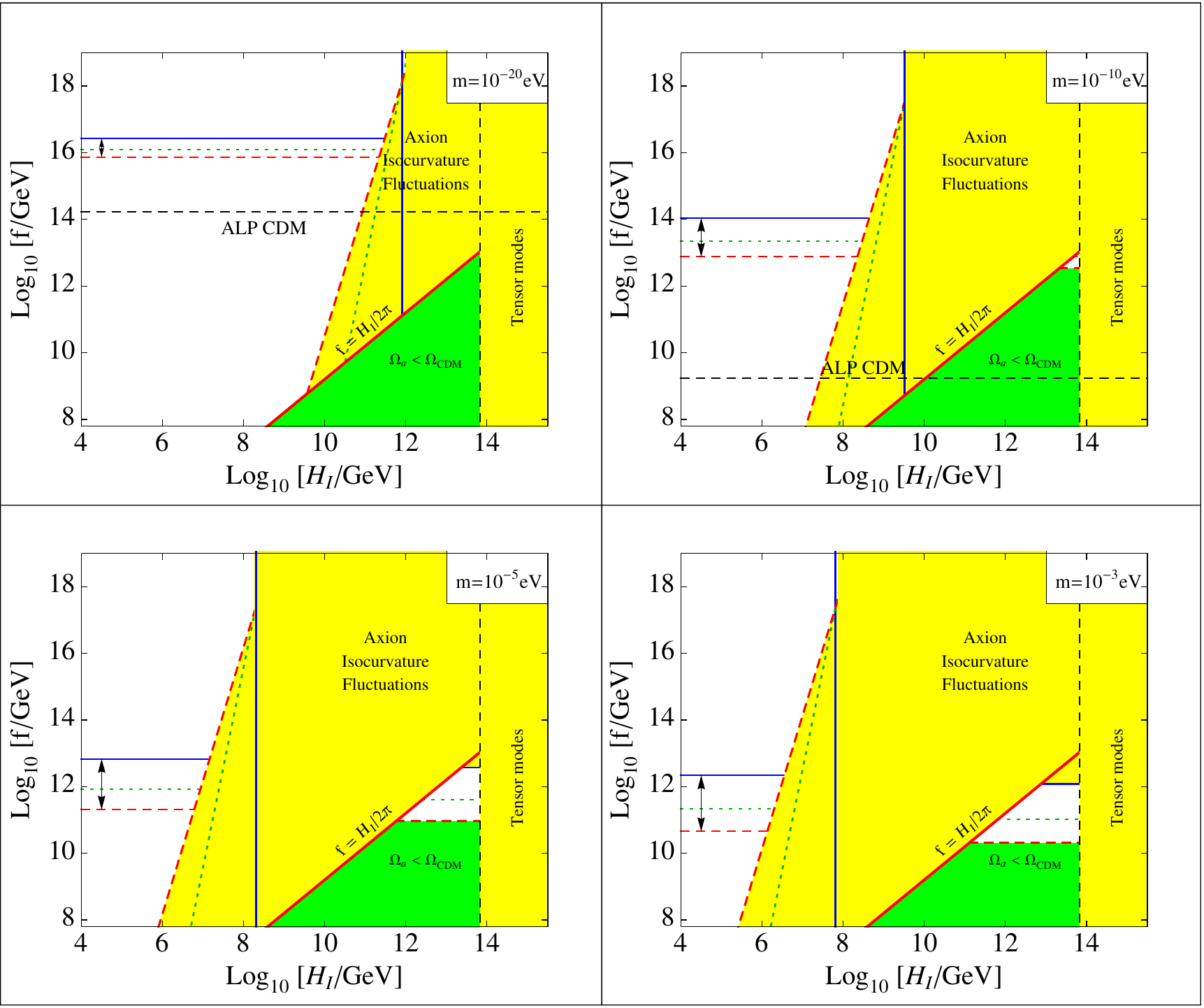}
	\caption{Same as Fig.~\ref{fig:axionparameterspace}, but for a harmonic potential in Eq.~\eqref{eq:ALPpotential_quadratic}}
	\label{fig:axionparameterspace_harmonic}
\end{figure}

In Fig.~\ref{fig:axionparameterspace}, we have shown the parameter space of ALPs moving in the cosine ALP potential in Eq.~\eqref{eq:ALPpotential}, including the non-harmonic corrections through the function $F(\theta_i)$ in Eq.~\eqref{eq:anharmonicities}. However, the ALP potential can greatly differ from what expressed in Eq.~\eqref{eq:ALPpotential}. For example, in the presence of a monodromy~\cite{Silverstein:2008sg, McAllister:2008hb, Kaloper:2008fb}, the degeneracy among the minima of the cosine potential is lifted by a quadratic potential, which might dominate the axion CDM potential~\cite{Jaeckel:2016qjp}. We repeat the computation in the previous Section for a harmonic potential, by switching off the non-harmonic corrections, setting $F(\theta_i) = 1$, considering the ALP moving in the quadratic potential
\begin{equation}
	V_H(\theta) = \frac{1}{2}f^2 m^2(T)\theta^2.
	\label{eq:ALPpotential_quadratic}
\end{equation}
Inserting Eq.~\eqref{eq:energy_density} into Eq.~\eqref{eq:axionisocurvaturefluctuations} for a harmonic potential to eliminate $\theta_i$ leads to a relation between $H_I$ and $f$,
\begin{equation}
	f = \hat{f}\,\left[\(\frac{\pi\hat{f}}{H_I}\)^2\frac{\rho_{\rm CDM}}{\hat{\rho}_A} \frac{\beta \Delta^2_{\mathcal{R}}(k_0)}{1-\beta}\right]^{\frac{8+2\chi}{8+\chi}}.
\end{equation}
We show results for the parameter space thus obtained in Fig.~\ref{fig:axionparameterspace_harmonic}. Notice that the upper left panel ($m=10^{-20}\,$eV) qualitatively reproduces the results recently obtained in Ref.~\cite{Diez-Tejedor:2017ivd} when the anharmonic corrections are neglected in the isocurvature modes. Eq.~\eqref{eq:bound_HI} describes the vertical blue line at the boundary of the region excluded by the non-observation of isocurvature fluctuations.

\section{Effects of additional physics beyond the Standard Model} \label{Effects of physics beyond the Standard Model}

Additional new physics might sensibly alter the axion parameter space presented in Fig.~\ref{fig:axionparameterspace}. Besides the QCD axion and other ALPs, examples of new physics not currently described within the framework of the Standard Model include additional particles whose presence modifies the effective number of degrees of freedom, or heavy scalar fields that might have dominated the Universe before the onset of radiation domination. We discuss some of the issues in the following. We focus on the case in which the axion mass is independent of temperature, since results can be easily generalized.

\subsection{Effective number of degrees of freedom}

The existence of particles that are still to be discovered would alter the effective number of relativistic and entropy degrees of freedom for temperatures larger than $T \gtrsim O(100){\rm\,GeV}$. For example, the maximum number of effective relativistic degrees of freedom is 106.75 in the Standard Model, and 228.75 in the Minimal Supersymmetric Standard Model~\cite{kolb:1994early}. Setting $3H(T) = m$, with $H$ given in Eq.~\eqref{eq:hubble_rate_r} and $T = 1\,$TeV, we obtain that corrections to $g_R(T)$ from physics beyond the Standard Model become important when $m \gtrsim 10^{-4}{\rm \,eV}$. We thus neglect these contributions when deriving the results in Sec.~\ref{Constraining the ALP mass}.

\subsection{Non-standard cosmological history}

The content of the Universe for temperatures larger than $\TRH \gtrsim 4{\rm\,MeV}$ is currently unknown, with lower bound being obtained from the requirement that the Big-Bang nucleosynthesis is achieved in a radiation-dominated cosmology~\cite{Kawasaki:1999na, Kawasaki:2000en, Hannestad:2004px, Ichikawa:2005vw, DeBernardis:2008zz}. However, for higher temperatures, the expansion rate of the Universe could have been dominated by some unknown form of energy, with an equation of state that differs from the one describing a relativistic fluid. A popular example is the early domination of a massive scalar field $\phi$, emerging as a by-product of the decay of the inflaton field. In the following, we refer to this modified cosmology as being $\phi$-dominated. The effect of a non-standard cosmological history might vary the present value of the axion energy density by orders of magnitude~\cite{Visinelli:2009kt}, depending on the equation of state for the fluid that dominates the expansion and the presence of an entropy dilution fact. In a nutshell, in a $\phi$-dominated Universe the ALP begins to oscillate at a temperature $T_1$ that is different from what obtained in the standard picture, because of a different relation between temperature and time in the modified cosmology. Assuming that the equation of state of the $\phi$ field in the modified cosmology is $p = w\rho$ ($w=1/3$ for radiation), for times $t$ larger than the moment $t_{\rm RH}$ at which the Universe transitions from $\phi$ domination to radiation domination, the Hubble rate is
\begin{equation}
	H = \frac{2}{3(w+1)t} = H_{\rm RH}\(\frac{T}{\TRH}\)^{\frac{3(w+1)}{2}},
\end{equation}
where the last expression is valid only if the entropy density $s=g_{S}T^3$ in a comoving volume is conserved, we have neglected the contribution from the entropy degrees of freedom, and
\begin{equation}
	H_{\rm RH} = H(t_{\rm RH}) = \mA(\TRH)\frac{\TRH^2}{3\MPl}.
\end{equation}
We consider the temperature dependence of the ALP mass as $m(T) = m(\Te/T)^{\chi/2}$, while the constant ALP mass case is obtained by setting $\chi = 0$. An early $\phi$ domination modifies the temperature at which coherent oscillations begin, Eq.~\eqref{eq:T1_ALP}, as
\begin{equation}
	T_1 = \Te \(\frac{\hat{f}_{\rm RH}}{f}\)^{\frac{2}{3(w+1)+\chi}}\,\(\frac{\TRH}{\Te}\)^{\frac{3w-1}{3(w+1)+\chi}}.
	\label{eq:t1_const_m_phi}
\end{equation}
where $\hat{f}_{\rm RH} \equiv \MPl/c^2\mA(\TRH) \approx \hat{f}$. The new value of $T_1$ modifies the present energy density, given by Eq.~\eqref{eq:presentnumberdensity_TD} when it is assumed entropy conservation from the onset of axion oscillations. The ALP energy density is
\begin{equation}
	\rho_A \!=\! \hat{\rho}_A\(\frac{m}{\hat{f}}\)^{1/2}\!\langle\theta_i^2\rangle\(\frac{c\hat{\Lambda}}{\TRH}\)^{\frac{(6\!+\!\chi)(3w\!-\!1)}{6(w\!+\!1)\!+\!2\chi}}\!\(\frac{f}{\hat{f}}\)^{\frac{2(16\!+\!3\chi)\!+\!(3w\!-\!1)(8\!+\!\chi)}{4[3(w\!+\!1)\!+\!\chi]}}\!\!\!,
\end{equation}
where $\hat{\rho}_A$ has been defined in Eq.~\eqref{eq:energy_density_funct}. Notice that, setting $w=1/3$, we obtain the energy density given in the first line in Eq.~\eqref{eq:energy_density}. The axion energy scale for which the ALP is the CDM particle reads
\begin{equation}
	f \!=\! \hat{f}\!\(\frac{\rho_{\rm CDM}}{\hat{\rho}_A\langle\theta_i^2\rangle}\sqrt{\frac{\hat{f}}{m}}\)^{\frac{4[3(w\!+\!1)\!+\!\chi]}{2(16\!+\!3\chi)\!+\!(3w\!-\!1)(8\!+\!\chi)}}\!\!\!\(\frac{\TRH}{c\hat{\Lambda}}\)^{\frac{2(6\!+\!\chi)(3w\!-\!1)}{2(16\!+\!3\chi)\!+\!(3w\!-\!1)(8\!+\!\chi)}}\!\!\!. \label{eq:f_phi}
\end{equation}
For a generic cosmological mode, the constraint in Eq.~\eqref{eq:ALP_mass_chi} for the region $f < H_I / 2\pi$ modifies as
\begin{eqnarray}
	m &\geq& \hat{f}\! \[\frac{64\pi}{\Delta^2_{\mathcal{R}}(k_0)\,r_{k_0}}\(\frac{\hat{f}}{\MPl}\)^2\]^{\frac{2(16\!+\!3\chi)\!+\!(3w\!-\!1)(8\!+\!\chi)}{2[3w(8\!+\!\chi)\!+\!\chi]}}\times \nonumber\\
	&&\times \(\frac{\rho_{\rm CDM}}{\hat{\rho}_A\langle\theta_i^2\rangle}\)^{\frac{4[3(w+1)+\chi]}{3w(8+\chi)+\chi}}\(\frac{\TRH}{c\hat{f}}\)^{\frac{2(6+\chi)(3w-1)}{3w(8+\chi)+\chi}}.
	\label{eq:mass_bound_const_phi}
\end{eqnarray}
The latter expression depends on two additional parameters $w$ and $\TRH$, and gives the result already obtained in Eq.~\eqref{eq:ALP_mass_chi} for $w = 1/3$.

For $w < 1/3$, Eq.~\eqref{eq:mass_bound_const_phi} can be restated as a lower bound on the reheating temperature, valid when assuming that the ALPs considered make up the totality of the CDM observed and that coherent oscillations in the field began after inflation, in a $\phi$-dominated cosmology. In the case of an early matter-dominated cosmology $w = 0$, the bound in Eq.~\eqref{eq:mass_bound_const_phi} can be restated as a bound on the reheating temperature as
\begin{equation}
	\TRH \!\geq\! c\hat{f}\! \(\frac{\hat{f}}{m}\)^{\frac{\chi}{2(6\!+\!\chi)}}\!\! \[\frac{64\pi}{\Delta^2_{\mathcal{R}}(k_0)\,r_{k_0}}\(\frac{\hat{f}}{\MPl}\)^2\]^{\frac{24\!+\!5\chi}{4(6\!+\!\chi)}}\!\!\!\!\(\frac{\rho_{\rm CDM}}{\hat{\rho}_A\langle\theta_i^2\rangle}\)^{\frac{6+2\chi}{6+\chi}}\!\!.
	\label{eq:mass_bound_phi_w0}
\end{equation}
If the mass is not affected by non-perturbative effects and $\chi = 0$, like for accidental ALPs, the expression above becomes independent on $m$ and yields the bound $\TRH \gtrsim 3\,$GeV, which is about three orders of magnitude more stringent than what obtained in Refs.~\cite{Kawasaki:1999na, Kawasaki:2000en, Hannestad:2004px, Ichikawa:2005vw, DeBernardis:2008zz} using BBN considerations. We nevertheless stress that the bound in Eq.~\eqref{eq:mass_bound_phi_w0} can be easily evaded, given the strong assumptions under which it has been derived.

\subsection{Dilution factor}

Some scenarios predict a violation in the conservation of the total entropy in a comoving volume, $sa^3$, due for example to the decay into lighter degrees of freedom of the $\phi$ field that dominates the Universe at that time. This is the case, for example, of a low-temperature reheating (LTR) stage~\cite{Dine:1982ah, Steinhardt:1983ia, Turner:1983, Scherrer:1985}, in which the Universe is dominated by a massive, decaying moduli field. In this situation, the relation between the scale factor and temperature changes from the simple relation $g_S^{1/3}T \sim 1/a$ to a generic relation $aT^{\delta}\sim {\rm const}$, where $\delta$ is a new constant in the model. For example, $\delta = 8/3$ in the LTR scenario~\cite{Giudice:2000ex}. A different parametrisation consists in assuming that a certain amount of entropy $\gamma$ is produced during the decaying stage~\cite{Steinhardt:1983ia, Lazarides:1987zf, Lazarides:1990xp}. See Ref.~\cite{Visinelli:2017qga} for the cosmology with a decaying kination field~\cite{Barrow:1982, Ford:1986sy, Spokoiny:1993kt, Joyce:1996cp, Matos:2000, Salati:2002md, Profumo:2003hq}. Either way, the effect of entropy dilution reduces the present energy density of axions in Eq.~\eqref{eq:energy_density} by a factor $\gamma$, and the bound on the ALP mass in Eq.~\eqref{eq:ALP_mass_chi} is lowered. In general, we obtain the ALP energy density to be diluted by a factor $\rho_A \to \rho_A / \gamma$. If $\gamma$ were to be independent on the ALP mass, we would get a reduction by $\bar{m}_\chi \to \bar{m}_\chi/\gamma^2$.

We compute the dilution factor in the LTR scenario as
\begin{eqnarray}
	\gamma &=& \frac{g_S(\TRH)a_R^3 T_R^3}{g_S(T_1)a_1^3 T_1^3} = \frac{g_S(\TRH)}{g_S(T_1)}\(\frac{T_1}{\TRH}\)^{3(\delta-1)} =\nonumber\\
	&=& \frac{g_S(\TRH)}{g_S(T_1)}\(\frac{\Te}{\TRH}\)^{3(\delta-1)}\left(\frac{\hat{f}}{f}\right)^{\frac{6(\delta-1)}{4+\chi}},
	\label{eq:dilution_gamma}
\end{eqnarray}
where in the last step we have used the expression for $T_1$ in Eq.~\eqref{eq:T1_ALP} for the case $f \leq \hat{f}$. Since we expect oscillations to begin in the $\phi$-dominated scenario, for which $T_1 > \TRH$, demanding $\delta > 1$ indeed leads to a dilution that is larger than one. For example, using $\TRH = 4\,$MeV and $m = 10^{-5}\,$eV with $\delta = 8/3$ and $\chi = 0$, we obtain $\gamma \approx 10^{20}$. This large discrepancy with respect to the standard cosmology scenario has been used in Ref.~\cite{Visinelli:2009kt} to dilute the energy density of the cosmological QCD axion, obtaining results that sensibly differ from the standard picture. Taking the expression for $\rho_A$ in Eq.~\eqref{eq:energy_density}, we rephrase the bound in Eq.~\eqref{eq:ALP_mass_chi} when the dilution in Eq.~\eqref{eq:dilution_gamma} is added as
\begin{eqnarray}
	m &\geq& \(\!\frac{g_S(T_1)}{g_S(\TRH)}\!\frac{g_R(T_1)}{g_R(T_0)}\! \frac{H_0^2}{T_0^3}\!\frac{48\Omega_{\rm CDM}}{\langle\theta_i^2\rangle\Delta^2_{\mathcal{R}}(k_0)r_{k_0}}\!\)^{\frac{2}{3\delta-2}} \times\nonumber\\
	&&\times \(\frac{\mA(T_1)}{\MPl}\)^{\frac{3(\delta-2)}{3\delta-2}} \TRH^{\frac{6(\delta-1)}{3\delta-2}} \sim 10^{-13}\,{\rm eV}.
	\label{eq:mass_bound_const_dilute}
\end{eqnarray}
We have treated separately the effects due to the modified expansion rate and dilution to obtain the bounds in Eqs.~\eqref{eq:mass_bound_const_phi} and~\eqref{eq:mass_bound_const_dilute}. A consistent derivation within a modified cosmology (say, LTR), has to consistently take into account both effects.

\section{Conclusion} \label{conclusion}

The present energy density of ALPs depends on both its mass $m$ and the energy scale $f$. In general, these parameters can be tuned so that $\rho_A = \rho_{\rm CDM}$. However, in models where the ALP field originates after inflation, we have shown in Sec.~\ref{Constraining the ALP mass} that the bound on the scale of inflation $H_I$ from the non-detection of primordial gravitational waves leads to a minimum value of the ALP mass $\bar{m}_\chi$ below which the tuning of $m$ and $f$ is no longer possible. An ALP with mass $m < \bar{m}_\chi$ can still be a CDM candidate if it spectates inflation. In this latter scenario, the scale of inflation $H_I$ is bound by the ALP mass through Eq.~\eqref{eq:bound_HI} which, although used in other work~\cite{Wilczek:2004cr, Tegmark:2005dy, Hertzberg:2008wr, Freivogel:2008qc, Mack:2009hv, Visinelli:2009zm, Acharya:2010zx}, has never been explicitly derived before. We have shown how these results affect the parameter space of the ALP for different values of the mass and of the susceptibility in Fig.~\ref{fig:axionparameterspace} (cosine potential) and Fig.~\ref{fig:axionparameterspace_harmonic} (harmonic potential). Finally, we have commented on how results are affected by the presence of additional physics beyond the standard model, focusing on the modification of the effective number of degrees of freedom, non-standard inflation and post-inflation cosmologies, and entropy dilution.

\begin{acknowledgments}
The author would like to thank the anonymous referee for the careful read and the helpful suggestions, which led to a substantial improvement of the manuscript with respect to its original version, and Javier Redondo (U. Zaragoza) for the useful discussion. The author acknowledges support by the Vetenskapsr\r{a}det (Swedish Research Council) through contract No. 638-2013-8993 and the Oskar Klein Centre for Cosmoparticle Physics.
\end{acknowledgments}

\appendix

\section{Review of the vacuum realignment mechanism} \label{sec_vrm}

\subsection{Equation of motion for the axion field}
The ALP field originates from the breaking of the PQ symmetry at a temperature of the order of $f$. The equation of motion for the angular variable of the ALP field at any time is
\begin{equation} \label{eq_motion_vrm}
\ddot{\theta} + 3H\,\dot{\theta} - \frac{\bar{\nabla}^2}{R^2}\,\theta + m^2(T)\,\sin\theta = 0,
\end{equation}
where $\theta$ is the ALP field in units of $f$, $\bar{\nabla}$ is the Laplacian operator with respect to the physical coordinates $\bar{x}$, and $R$ is the scale factor. To derive Eq.~\eqref{eq_motion_vrm}, we have considered the simplest possible ALP potential $V(\theta) = f^2m^2(T)\,\left(1-\cos\theta\right)$. The mass term in the equation of motion becomes important when the Hubble rate is comparable to the axion mass,
\begin{equation}
	H(T_1) = 3m(T_1),
	\label{eq:masscondition}
\end{equation}
whose solution gives the temperature $T_1$ when coherent oscillations begin. Setting the scale factor and the Hubble rate at temperature $T_1$ respectively as $R_1$ and $H_1$, we rescale time $t$ and scale factor $R$ as $t \to H_1t$ and $R \to R/R_1$, so that Eq.~\eqref{eq_motion_vrm} reads
\begin{equation} \label{eq_motion_vrm1}
\ddot{\theta} + 3H\,\dot{\theta} - \frac{\nabla^2}{R^2}\,\theta + 9g^2\,\sin\theta = 0,
\end{equation}
where the Laplacian operator is written with respect to the co-moving spatial coordinates $x = H_1\,R_1\,\bar{x}$ and $g=G(T)/G(T_1)$. We work in a radiation-dominated cosmology, where time and scale factor are related by $R \propto t^{1/2}$. Setting $\theta = \psi/R$, Eq.~\eqref{eq_motion_vrm1} reads
\begin{equation} \label{eq_motion_std}
\psi'' - \nabla^2\,\psi + 9g^2\,R^3\,\sin\left(\frac{\psi}{R}\right) = 0,
\end{equation}
where a prime indicates a derivation with respect to $R$. Eq.~\eqref{eq_motion_std} coincides with the results in Ref.~\cite{Kolb:1993hw}, where the conformal time $\eta$ is used as the independent variable in place of the scale factor $R$. 

Taking the Fourier transform of the axion field as
\begin{equation}
\psi(\bx) = \int e^{-i q\,\bx}\,\psi(q),
\end{equation}
we rewrite Eq.~\eqref{eq_motion_std} as
\begin{equation}\label{eq_motion_full}
\psi'' + q^2\,\psi + 9g^2\,R^3\,\sin\left(\frac{\psi}{R}\right) = 0.
\end{equation}
Eq.~\eqref{eq_motion_full} expresses the full equation of motion for the axion field in the variable $R$, conveniently written to be solved numerically.

\subsection{Approximate solutions of the equation of motion}

Analytic solutions to Eq.~\eqref{eq_motion_full} can be obtained in the limiting regime $\psi/R \ll 1$, where Eq.~\eqref{eq_motion_full} reads
\begin{equation} \label{eq_motion_small_phi}
\psi'' + \kappa^2(R)\,\psi = 0,
\end{equation}
with the wave number $\kappa^2(R) = q^2 + 9g^2R^2$. An approximate solution of Eq.~\eqref{eq_motion_small_phi}, valid in the adiabatic regime in which higher derivatives are neglected, is~\cite{Kolb:1993hw, Sikivie2008}
\begin{equation} \label{ansatz1}
\psi = \psi_0(R)\,\exp\,\left(i\,\int^R\,\kappa(R')\,dR'\right),
\end{equation}
where the amplitude $\psi_0$ is given by
\begin{equation}
|\psi_0(R)|^2\,\kappa(R) = {\rm const.}
\end{equation}
Each term appearing in $\kappa(R)$ is the leading term in a particular regime of the evolution of the axion field. We analyze these approximate behavior in depths in the following.

\begin{itemize}
\item Solution at early times, outside the horizon

At early times $t \sim R^2 \lesssim t_1$ prior to the onset of axion oscillations, the mass term in Eq.~\eqref{eq_motion_small_phi} can be neglected since $m(T) \ll m(T_1)$. Defining the physical wavelength $\lambda = R/q$, we distinguish two different regimes in this approximation, corresponding to the evolution of the modes outside the horizon ($\lambda \gtrsim t$) or inside the horizon ($\lambda \lesssim t$). In the first case $\lambda \gtrsim t$, Eq.~\eqref{eq_motion_small_phi} at early times reduces to $\psi'' = 0$, with solution ($\psi = R\phi$)
\begin{equation}\label{field_early_times}
\phi(q,t) = \phi_1(q) + \frac{\phi_2(q)}{R} = \phi_1(q) + \frac{\phi_2(q)}{t^{1/2}},
\end{equation}
the first solution being a constant in time $\phi_1(q)$, while the second solution dropping to zero. The axion field for modes larger than the horizon is ``frozen by causality''~\cite{Sikivie2008}.

\item Solution at early times, inside the horizon

Eq.~\eqref{eq_motion_small_phi} for modes that evolve inside the horizon $\lambda \lesssim t$ reduces to 
\begin{equation} \label{motion_approx_inside}
\psi'' + q^2\,\psi = 0,
\end{equation}
whose solution in a closed form, obtained through Eq.~\eqref{ansatz1} and $\phi = \psi/R$, reads
\begin{equation} \label{solution_eqmotion_q}
\phi \propto R^{-1}\,\exp\left(iq\,R\right).
\end{equation}
The dependence of the amplitude $|\phi|\sim 1/R$ in Eq.~\eqref{solution_eqmotion_q} is crucial, since it shows that the axion number density scales as cold matter,
\begin{equation}
n_A(q, t) \sim \frac{|\phi|^2}{\lambda} \sim \frac{1}{R^3}.
\end{equation}

\item Solution for the zero mode at the onset of oscillations

An approximate solution of Eq.~\eqref{eq_motion_small_phi} for the zero-momentum mode $q = 0$, valid after the onset of axion oscillations when $t \sim t_1$, is obtained by setting
\begin{equation}
\kappa(R) \approx 3gR,
\end{equation}
so that the adiabatic solution for $\psi$ in Eq.~\eqref{ansatz1} in this slowly oscillating regime gives the axion number density
\begin{equation} \label{numberdensity}
n_A^{\rm mis}(R) = \frac{1}{2}m(R)\,f^2\,\frac{\left|\psi\right|^2}{R^2} = \ns\,\left(\frac{R}{R_1}\right)^{-3},
\end{equation}
where $\ns$ is the number density of axions from the misalignment mechanism at temperature $T_1$,
\begin{equation} \label{number_density_onset}
\ns = \frac{1}{2}\,m(T_1)\,f^2\,F(\theta_i)\theta_i^2,
\end{equation}
and $F(\theta_i)$ is a function that accounts for neglecting the non-harmonic higher-order terms in the Taylor expansion of the sine function, see Eq.~\eqref{eq:anharmonicities}. Eq.~\eqref{numberdensity} shows that the axion number density of the zero modes after the onset of axion oscillations scales as cold matter, with $R^{-3}$. The present ALP energy density is found by conservation of the comoving axion number density,
\begin{equation}
	\rho_A = m\,\ns\frac{s(T_0)}{s(T_1)} = m\ns\frac{g_{*S}(T_0)}{g_{*S}(T_1)}\left(\frac{T_0}{T_1}\right)^3, 
	\label{eq:presentnumberdensity}
\end{equation}
where $s(T)$ is the entropy density and $g_{*S}(T)$ is the number of degrees of freedom at temperature $T$.

\end{itemize}

\bibliography{axBib}

\begin{thebibliography}{157}%
\makeatletter
\providecommand \@ifxundefined [1]{%
 \@ifx{#1\undefined}
}%
\providecommand \@ifnum [1]{%
 \ifnum #1\expandafter \@firstoftwo
 \else \expandafter \@secondoftwo
 \fi
}%
\providecommand \@ifx [1]{%
 \ifx #1\expandafter \@firstoftwo
 \else \expandafter \@secondoftwo
 \fi
}%
\providecommand \natexlab [1]{#1}%
\providecommand \enquote  [1]{``#1''}%
\providecommand \bibnamefont  [1]{#1}%
\providecommand \bibfnamefont [1]{#1}%
\providecommand \citenamefont [1]{#1}%
\providecommand \href@noop [0]{\@secondoftwo}%
\providecommand \href [0]{\begingroup \@sanitize@url \@href}%
\providecommand \@href[1]{\@@startlink{#1}\@@href}%
\providecommand \@@href[1]{\endgroup#1\@@endlink}%
\providecommand \@sanitize@url [0]{\catcode `\\12\catcode `\$12\catcode
  `\&12\catcode `\#12\catcode `\^12\catcode `\_12\catcode `\%12\relax}%
\providecommand \@@startlink[1]{}%
\providecommand \@@endlink[0]{}%
\providecommand \url  [0]{\begingroup\@sanitize@url \@url }%
\providecommand \@url [1]{\endgroup\@href {#1}{\urlprefix }}%
\providecommand \urlprefix  [0]{URL }%
\providecommand \Eprint [0]{\href }%
\providecommand \doibase [0]{http://dx.doi.org/}%
\providecommand \selectlanguage [0]{\@gobble}%
\providecommand \bibinfo  [0]{\@secondoftwo}%
\providecommand \bibfield  [0]{\@secondoftwo}%
\providecommand \translation [1]{[#1]}%
\providecommand \BibitemOpen [0]{}%
\providecommand \bibitemStop [0]{}%
\providecommand \bibitemNoStop [0]{.\EOS\space}%
\providecommand \EOS [0]{\spacefactor3000\relax}%
\providecommand \BibitemShut  [1]{\csname bibitem#1\endcsname}%
\let\auto@bib@innerbib\@empty
%</preamble>
\bibitem [{\citenamefont {Weinberg}(1978)}]{Weinberg:1977ma}%
  \BibitemOpen
  \bibfield  {author} {\bibinfo {author} {\bibfnamefont {S.}~\bibnamefont
  {Weinberg}},\ }\href {\doibase 10.1103/PhysRevLett.40.223} {\bibfield
  {journal} {\bibinfo  {journal} {Phys. Rev. Lett.}\ }\textbf {\bibinfo
  {volume} {40}},\ \bibinfo {pages} {223} (\bibinfo {year} {1978})}\BibitemShut
  {NoStop}%
%%CITATION = PRLTA,40,223;%%
\bibitem [{\citenamefont {Wilczek}(1978)}]{Wilczek:1977pj}%
  \BibitemOpen
  \bibfield  {author} {\bibinfo {author} {\bibfnamefont {F.}~\bibnamefont
  {Wilczek}},\ }\href {\doibase 10.1103/PhysRevLett.40.279} {\bibfield
  {journal} {\bibinfo  {journal} {Phys. Rev. Lett.}\ }\textbf {\bibinfo
  {volume} {40}},\ \bibinfo {pages} {279} (\bibinfo {year} {1978})}\BibitemShut
  {NoStop}%
%%CITATION = PRLTA,40,279;%%
\bibitem [{\citenamefont {Peccei}\ and\ \citenamefont
  {Quinn}(1977{\natexlab{a}})}]{Peccei:1977hh}%
  \BibitemOpen
  \bibfield  {author} {\bibinfo {author} {\bibfnamefont {R.~D.}\ \bibnamefont
  {Peccei}}\ and\ \bibinfo {author} {\bibfnamefont {H.~R.}\ \bibnamefont
  {Quinn}},\ }\href {\doibase 10.1103/PhysRevLett.38.1440} {\bibfield
  {journal} {\bibinfo  {journal} {Phys. Rev. Lett.}\ }\textbf {\bibinfo
  {volume} {38}},\ \bibinfo {pages} {1440} (\bibinfo {year}
  {1977}{\natexlab{a}})}\BibitemShut {NoStop}%
%%CITATION = PRLTA,38,1440;%%
\bibitem [{\citenamefont {Peccei}\ and\ \citenamefont
  {Quinn}(1977{\natexlab{b}})}]{Peccei:1977ur}%
  \BibitemOpen
  \bibfield  {author} {\bibinfo {author} {\bibfnamefont {R.~D.}\ \bibnamefont
  {Peccei}}\ and\ \bibinfo {author} {\bibfnamefont {H.~R.}\ \bibnamefont
  {Quinn}},\ }\href {\doibase 10.1103/PhysRevD.16.1791} {\bibfield  {journal}
  {\bibinfo  {journal} {Phys. Rev.}\ }\textbf {\bibinfo {volume} {D16}},\
  \bibinfo {pages} {1791} (\bibinfo {year} {1977}{\natexlab{b}})}\BibitemShut
  {NoStop}%
%%CITATION = PHRVA,D16,1791;%%
\bibitem [{\citenamefont {Raffelt}(2008)}]{Raffelt2008}%
  \BibitemOpen
  \bibfield  {author} {\bibinfo {author} {\bibfnamefont {G.~G.}\ \bibnamefont
  {Raffelt}},\ }\enquote {\bibinfo {title} {Astrophysical axion bounds},}\ in\
  \href {\doibase 10.1007/978-3-540-73518-2_3} {\emph {\bibinfo {booktitle}
  {Axions: Theory, Cosmology, and Experimental Searches}}},\ \bibinfo {editor}
  {edited by\ \bibinfo {editor} {\bibfnamefont {M.}~\bibnamefont {Kuster}},
  \bibinfo {editor} {\bibfnamefont {G.}~\bibnamefont {Raffelt}}, \ and\
  \bibinfo {editor} {\bibfnamefont {B.}~\bibnamefont {Beltr{\'a}n}}}\ (\bibinfo
   {publisher} {Springer Berlin Heidelberg},\ \bibinfo {address} {Berlin,
  Heidelberg},\ \bibinfo {year} {2008})\ pp.\ \bibinfo {pages}
  {51--71}\BibitemShut {NoStop}%
\bibitem [{\citenamefont {Kim}(1979)}]{Kim:1979if}%
  \BibitemOpen
  \bibfield  {author} {\bibinfo {author} {\bibfnamefont {J.~E.}\ \bibnamefont
  {Kim}},\ }\href {\doibase 10.1103/PhysRevLett.43.103} {\bibfield  {journal}
  {\bibinfo  {journal} {Phys. Rev. Lett.}\ }\textbf {\bibinfo {volume} {43}},\
  \bibinfo {pages} {103} (\bibinfo {year} {1979})}\BibitemShut {NoStop}%
%%CITATION = PRLTA,43,103;%%
\bibitem [{\citenamefont {Shifman}\ \emph {et~al.}(1980)\citenamefont
  {Shifman}, \citenamefont {Vainshtein},\ and\ \citenamefont
  {Zakharov}}]{Shifman1980493}%
  \BibitemOpen
  \bibfield  {author} {\bibinfo {author} {\bibfnamefont {M.}~\bibnamefont
  {Shifman}}, \bibinfo {author} {\bibfnamefont {A.}~\bibnamefont {Vainshtein}},
  \ and\ \bibinfo {author} {\bibfnamefont {V.}~\bibnamefont {Zakharov}},\
  }\href {\doibase http://dx.doi.org/10.1016/0550-3213(80)90209-6} {\bibfield
  {journal} {\bibinfo  {journal} {Nuclear Physics B}\ }\textbf {\bibinfo
  {volume} {166}},\ \bibinfo {pages} {493 } (\bibinfo {year}
  {1980})}\BibitemShut {NoStop}%
\bibitem [{\citenamefont {Dine}\ \emph {et~al.}(1981)\citenamefont {Dine},
  \citenamefont {Fischler},\ and\ \citenamefont {Srednicki}}]{Dine:1981rt}%
  \BibitemOpen
  \bibfield  {author} {\bibinfo {author} {\bibfnamefont {M.}~\bibnamefont
  {Dine}}, \bibinfo {author} {\bibfnamefont {W.}~\bibnamefont {Fischler}}, \
  and\ \bibinfo {author} {\bibfnamefont {M.}~\bibnamefont {Srednicki}},\ }\href
  {\doibase 10.1016/0370-2693(81)90590-6} {\bibfield  {journal} {\bibinfo
  {journal} {Phys. Lett.}\ }\textbf {\bibinfo {volume} {B104}},\ \bibinfo
  {pages} {199} (\bibinfo {year} {1981})}\BibitemShut {NoStop}%
%%CITATION = PHLTA,B104,199;%%
\bibitem [{\citenamefont {Zhitnitsky}(1980)}]{Zhitnitsky:1980tq}%
  \BibitemOpen
  \bibfield  {author} {\bibinfo {author} {\bibfnamefont {A.~R.}\ \bibnamefont
  {Zhitnitsky}},\ }\href@noop {} {\bibfield  {journal} {\bibinfo  {journal}
  {Sov. J. Nucl. Phys.}\ }\textbf {\bibinfo {volume} {31}},\ \bibinfo {pages}
  {260} (\bibinfo {year} {1980})}\BibitemShut {NoStop}%
%%CITATION = SJNCA,31,260;%%
\bibitem [{\citenamefont {Linde}(1988)}]{Linde:1987bx}%
  \BibitemOpen
  \bibfield  {author} {\bibinfo {author} {\bibfnamefont {A.~D.}\ \bibnamefont
  {Linde}},\ }\href {\doibase 10.1016/0370-2693(88)90597-7} {\bibfield
  {journal} {\bibinfo  {journal} {Phys. Lett.}\ }\textbf {\bibinfo {volume}
  {B201}},\ \bibinfo {pages} {437} (\bibinfo {year} {1988})}\BibitemShut
  {NoStop}%
%%CITATION = PHLTA,B201,437;%%
\bibitem [{\citenamefont {Linde}(1991)}]{Linde:1991km}%
  \BibitemOpen
  \bibfield  {author} {\bibinfo {author} {\bibfnamefont {A.~D.}\ \bibnamefont
  {Linde}},\ }\href {\doibase 10.1016/0370-2693(91)90130-I} {\bibfield
  {journal} {\bibinfo  {journal} {Phys. Lett.}\ }\textbf {\bibinfo {volume}
  {B259}},\ \bibinfo {pages} {38} (\bibinfo {year} {1991})}\BibitemShut
  {NoStop}%
%%CITATION = PHLTA,B259,38;%%
\bibitem [{\citenamefont {Turner}\ and\ \citenamefont
  {Wilczek}(1991)}]{Turner:1991}%
  \BibitemOpen
  \bibfield  {author} {\bibinfo {author} {\bibfnamefont {M.~S.}\ \bibnamefont
  {Turner}}\ and\ \bibinfo {author} {\bibfnamefont {F.}~\bibnamefont
  {Wilczek}},\ }\href {\doibase 10.1103/PhysRevLett.66.5} {\bibfield  {journal}
  {\bibinfo  {journal} {Phys. Rev. Lett.}\ }\textbf {\bibinfo {volume} {66}},\
  \bibinfo {pages} {5} (\bibinfo {year} {1991})}\BibitemShut {NoStop}%
\bibitem [{\citenamefont {Wilczek}(2004)}]{Wilczek:2004cr}%
  \BibitemOpen
  \bibfield  {author} {\bibinfo {author} {\bibfnamefont {F.}~\bibnamefont
  {Wilczek}},\ }\href@noop {} {\  (\bibinfo {year} {2004})},\ \Eprint
  {http://arxiv.org/abs/hep-ph/0408167} {arXiv:hep-ph/0408167 [hep-ph]}
  \BibitemShut {NoStop}%
%%CITATION = HEP-PH/0408167;%%
\bibitem [{\citenamefont {Tegmark}\ \emph {et~al.}(2006)\citenamefont
  {Tegmark}, \citenamefont {Aguirre}, \citenamefont {Rees},\ and\ \citenamefont
  {Wilczek}}]{Tegmark:2005dy}%
  \BibitemOpen
  \bibfield  {author} {\bibinfo {author} {\bibfnamefont {M.}~\bibnamefont
  {Tegmark}}, \bibinfo {author} {\bibfnamefont {A.}~\bibnamefont {Aguirre}},
  \bibinfo {author} {\bibfnamefont {M.}~\bibnamefont {Rees}}, \ and\ \bibinfo
  {author} {\bibfnamefont {F.}~\bibnamefont {Wilczek}},\ }\href {\doibase
  10.1103/PhysRevD.73.023505} {\bibfield  {journal} {\bibinfo  {journal} {Phys.
  Rev.}\ }\textbf {\bibinfo {volume} {D73}},\ \bibinfo {pages} {023505}
  (\bibinfo {year} {2006})},\ \Eprint {http://arxiv.org/abs/astro-ph/0511774}
  {arXiv:astro-ph/0511774 [astro-ph]} \BibitemShut {NoStop}%
%%CITATION = ASTRO-PH/0511774;%%
\bibitem [{\citenamefont {Hertzberg}\ \emph {et~al.}(2008)\citenamefont
  {Hertzberg}, \citenamefont {Tegmark},\ and\ \citenamefont
  {Wilczek}}]{Hertzberg:2008wr}%
  \BibitemOpen
  \bibfield  {author} {\bibinfo {author} {\bibfnamefont {M.~P.}\ \bibnamefont
  {Hertzberg}}, \bibinfo {author} {\bibfnamefont {M.}~\bibnamefont {Tegmark}},
  \ and\ \bibinfo {author} {\bibfnamefont {F.}~\bibnamefont {Wilczek}},\ }\href
  {\doibase 10.1103/PhysRevD.78.083507} {\bibfield  {journal} {\bibinfo
  {journal} {Phys. Rev.}\ }\textbf {\bibinfo {volume} {D78}},\ \bibinfo {pages}
  {083507} (\bibinfo {year} {2008})},\ \Eprint {http://arxiv.org/abs/0807.1726}
  {arXiv:0807.1726 [astro-ph]} \BibitemShut {NoStop}%
%%CITATION = ARXIV:0807.1726;%%
\bibitem [{\citenamefont {Freivogel}(2010)}]{Freivogel:2008qc}%
  \BibitemOpen
  \bibfield  {author} {\bibinfo {author} {\bibfnamefont {B.}~\bibnamefont
  {Freivogel}},\ }\href {\doibase 10.1088/1475-7516/2010/03/021} {\bibfield
  {journal} {\bibinfo  {journal} {JCAP}\ }\textbf {\bibinfo {volume} {1003}},\
  \bibinfo {pages} {021} (\bibinfo {year} {2010})},\ \Eprint
  {http://arxiv.org/abs/0810.0703} {arXiv:0810.0703 [hep-th]} \BibitemShut
  {NoStop}%
%%CITATION = ARXIV:0810.0703;%%
\bibitem [{\citenamefont {Mack}(2011)}]{Mack:2009hv}%
  \BibitemOpen
  \bibfield  {author} {\bibinfo {author} {\bibfnamefont {K.~J.}\ \bibnamefont
  {Mack}},\ }\href {\doibase 10.1088/1475-7516/2011/07/021} {\bibfield
  {journal} {\bibinfo  {journal} {JCAP}\ }\textbf {\bibinfo {volume} {1107}},\
  \bibinfo {pages} {021} (\bibinfo {year} {2011})},\ \Eprint
  {http://arxiv.org/abs/0911.0421} {arXiv:0911.0421 [astro-ph.CO]} \BibitemShut
  {NoStop}%
%%CITATION = ARXIV:0911.0421;%%
\bibitem [{\citenamefont {Visinelli}\ and\ \citenamefont
  {Gondolo}(2009)}]{Visinelli:2009zm}%
  \BibitemOpen
  \bibfield  {author} {\bibinfo {author} {\bibfnamefont {L.}~\bibnamefont
  {Visinelli}}\ and\ \bibinfo {author} {\bibfnamefont {P.}~\bibnamefont
  {Gondolo}},\ }\href {\doibase 10.1103/PhysRevD.80.035024} {\bibfield
  {journal} {\bibinfo  {journal} {Phys. Rev.}\ }\textbf {\bibinfo {volume}
  {D80}},\ \bibinfo {pages} {035024} (\bibinfo {year} {2009})},\ \Eprint
  {http://arxiv.org/abs/0903.4377} {arXiv:0903.4377 [astro-ph.CO]} \BibitemShut
  {NoStop}%
%%CITATION = ARXIV:0903.4377;%%
\bibitem [{\citenamefont {Acharya}\ \emph {et~al.}(2010)\citenamefont
  {Acharya}, \citenamefont {Bobkov},\ and\ \citenamefont
  {Kumar}}]{Acharya:2010zx}%
  \BibitemOpen
  \bibfield  {author} {\bibinfo {author} {\bibfnamefont {B.~S.}\ \bibnamefont
  {Acharya}}, \bibinfo {author} {\bibfnamefont {K.}~\bibnamefont {Bobkov}}, \
  and\ \bibinfo {author} {\bibfnamefont {P.}~\bibnamefont {Kumar}},\ }\href
  {\doibase 10.1007/JHEP11(2010)105} {\bibfield  {journal} {\bibinfo  {journal}
  {JHEP}\ }\textbf {\bibinfo {volume} {11}},\ \bibinfo {pages} {105} (\bibinfo
  {year} {2010})},\ \Eprint {http://arxiv.org/abs/1004.5138} {arXiv:1004.5138
  [hep-th]} \BibitemShut {NoStop}%
%%CITATION = ARXIV:1004.5138;%%
\bibitem [{\citenamefont {Hogan}\ and\ \citenamefont
  {Rees}(1988)}]{Hogan:1988mp}%
  \BibitemOpen
  \bibfield  {author} {\bibinfo {author} {\bibfnamefont {C.~J.}\ \bibnamefont
  {Hogan}}\ and\ \bibinfo {author} {\bibfnamefont {M.~J.}\ \bibnamefont
  {Rees}},\ }\href {\doibase 10.1016/0370-2693(88)91655-3} {\bibfield
  {journal} {\bibinfo  {journal} {Phys. Lett.}\ }\textbf {\bibinfo {volume}
  {B205}},\ \bibinfo {pages} {228} (\bibinfo {year} {1988})}\BibitemShut
  {NoStop}%
%%CITATION = PHLTA,B205,228;%%
\bibitem [{\citenamefont {Kolb}\ and\ \citenamefont
  {Tkachev}(1994)}]{Kolb:1993hw}%
  \BibitemOpen
  \bibfield  {author} {\bibinfo {author} {\bibfnamefont {E.~W.}\ \bibnamefont
  {Kolb}}\ and\ \bibinfo {author} {\bibfnamefont {I.~I.}\ \bibnamefont
  {Tkachev}},\ }\href {\doibase 10.1103/PhysRevD.49.5040} {\bibfield  {journal}
  {\bibinfo  {journal} {Phys. Rev.}\ }\textbf {\bibinfo {volume} {D49}},\
  \bibinfo {pages} {5040} (\bibinfo {year} {1994})},\ \Eprint
  {http://arxiv.org/abs/astro-ph/9311037} {arXiv:astro-ph/9311037 [astro-ph]}
  \BibitemShut {NoStop}%
%%CITATION = ASTRO-PH/9311037;%%
\bibitem [{\citenamefont {Kolb}\ and\ \citenamefont
  {Tkachev}(1993)}]{Kolb:1993zz}%
  \BibitemOpen
  \bibfield  {author} {\bibinfo {author} {\bibfnamefont {E.~W.}\ \bibnamefont
  {Kolb}}\ and\ \bibinfo {author} {\bibfnamefont {I.~I.}\ \bibnamefont
  {Tkachev}},\ }\href {\doibase 10.1103/PhysRevLett.71.3051} {\bibfield
  {journal} {\bibinfo  {journal} {Phys. Rev. Lett.}\ }\textbf {\bibinfo
  {volume} {71}},\ \bibinfo {pages} {3051} (\bibinfo {year} {1993})},\ \Eprint
  {http://arxiv.org/abs/hep-ph/9303313} {arXiv:hep-ph/9303313 [hep-ph]}
  \BibitemShut {NoStop}%
%%CITATION = HEP-PH/9303313;%%
\bibitem [{\citenamefont {Fairbairn}\ \emph {et~al.}(2018)\citenamefont
  {Fairbairn}, \citenamefont {Marsh}, \citenamefont {Quevillon},\ and\
  \citenamefont {Rozier}}]{Fairbairn:2017sil}%
  \BibitemOpen
  \bibfield  {author} {\bibinfo {author} {\bibfnamefont {M.}~\bibnamefont
  {Fairbairn}}, \bibinfo {author} {\bibfnamefont {D.~J.~E.}\ \bibnamefont
  {Marsh}}, \bibinfo {author} {\bibfnamefont {J.}~\bibnamefont {Quevillon}}, \
  and\ \bibinfo {author} {\bibfnamefont {S.}~\bibnamefont {Rozier}},\ }\href
  {\doibase 10.1103/PhysRevD.97.083502} {\bibfield  {journal} {\bibinfo
  {journal} {Phys. Rev.}\ }\textbf {\bibinfo {volume} {D97}},\ \bibinfo {pages}
  {083502} (\bibinfo {year} {2018})},\ \Eprint
  {http://arxiv.org/abs/1707.03310} {arXiv:1707.03310 [astro-ph.CO]}
  \BibitemShut {NoStop}%
%%CITATION = ARXIV:1707.03310;%%
\bibitem [{\citenamefont {Visinelli}\ and\ \citenamefont
  {Redondo}(2018)}]{Visinelli:2018wza}%
  \BibitemOpen
  \bibfield  {author} {\bibinfo {author} {\bibfnamefont {L.}~\bibnamefont
  {Visinelli}}\ and\ \bibinfo {author} {\bibfnamefont {J.}~\bibnamefont
  {Redondo}},\ }\href@noop {} {\  (\bibinfo {year} {2018})},\ \Eprint
  {http://arxiv.org/abs/1808.01879} {arXiv:1808.01879 [astro-ph.CO]}
  \BibitemShut {NoStop}%
%%CITATION = ARXIV:1808.01879;%%
\bibitem [{\citenamefont {Vaquero}\ \emph {et~al.}(2018)\citenamefont
  {Vaquero}, \citenamefont {Redondo},\ and\ \citenamefont
  {Stadler}}]{Vaquero:2018tib}%
  \BibitemOpen
  \bibfield  {author} {\bibinfo {author} {\bibfnamefont {A.}~\bibnamefont
  {Vaquero}}, \bibinfo {author} {\bibfnamefont {J.}~\bibnamefont {Redondo}}, \
  and\ \bibinfo {author} {\bibfnamefont {J.}~\bibnamefont {Stadler}},\
  }\href@noop {} {\  (\bibinfo {year} {2018})},\ \Eprint
  {http://arxiv.org/abs/1809.09241} {arXiv:1809.09241 [astro-ph.CO]}
  \BibitemShut {NoStop}%
%%CITATION = ARXIV:1809.09241;%%
\bibitem [{\citenamefont {Helfer}\ \emph {et~al.}(2016)\citenamefont {Helfer},
  \citenamefont {Marsh}, \citenamefont {Clough}, \citenamefont {Fairbairn},
  \citenamefont {Lim},\ and\ \citenamefont {Becerril}}]{Helfer:2016ljl}%
  \BibitemOpen
  \bibfield  {author} {\bibinfo {author} {\bibfnamefont {T.}~\bibnamefont
  {Helfer}}, \bibinfo {author} {\bibfnamefont {D.~J.~E.}\ \bibnamefont
  {Marsh}}, \bibinfo {author} {\bibfnamefont {K.}~\bibnamefont {Clough}},
  \bibinfo {author} {\bibfnamefont {M.}~\bibnamefont {Fairbairn}}, \bibinfo
  {author} {\bibfnamefont {E.~A.}\ \bibnamefont {Lim}}, \ and\ \bibinfo
  {author} {\bibfnamefont {R.}~\bibnamefont {Becerril}},\ }\href@noop {} {\
  (\bibinfo {year} {2016})},\ \Eprint {http://arxiv.org/abs/1609.04724}
  {arXiv:1609.04724 [astro-ph.CO]} \BibitemShut {NoStop}%
%%CITATION = ARXIV:1609.04724;%%
\bibitem [{\citenamefont {Braaten}\ \emph {et~al.}(2016)\citenamefont
  {Braaten}, \citenamefont {Mohapatra},\ and\ \citenamefont
  {Zhang}}]{Braaten:2015eeu}%
  \BibitemOpen
  \bibfield  {author} {\bibinfo {author} {\bibfnamefont {E.}~\bibnamefont
  {Braaten}}, \bibinfo {author} {\bibfnamefont {A.}~\bibnamefont {Mohapatra}},
  \ and\ \bibinfo {author} {\bibfnamefont {H.}~\bibnamefont {Zhang}},\ }\href
  {\doibase 10.1103/PhysRevLett.117.121801} {\bibfield  {journal} {\bibinfo
  {journal} {Phys. Rev. Lett.}\ }\textbf {\bibinfo {volume} {117}},\ \bibinfo
  {pages} {121801} (\bibinfo {year} {2016})},\ \Eprint
  {http://arxiv.org/abs/1512.00108} {arXiv:1512.00108 [hep-ph]} \BibitemShut
  {NoStop}%
%%CITATION = ARXIV:1512.00108;%%
\bibitem [{\citenamefont {Visinelli}\ \emph
  {et~al.}(2018{\natexlab{a}})\citenamefont {Visinelli}, \citenamefont {Baum},
  \citenamefont {Redondo}, \citenamefont {Freese},\ and\ \citenamefont
  {Wilczek}}]{Visinelli:2017ooc}%
  \BibitemOpen
  \bibfield  {author} {\bibinfo {author} {\bibfnamefont {L.}~\bibnamefont
  {Visinelli}}, \bibinfo {author} {\bibfnamefont {S.}~\bibnamefont {Baum}},
  \bibinfo {author} {\bibfnamefont {J.}~\bibnamefont {Redondo}}, \bibinfo
  {author} {\bibfnamefont {K.}~\bibnamefont {Freese}}, \ and\ \bibinfo {author}
  {\bibfnamefont {F.}~\bibnamefont {Wilczek}},\ }\href {\doibase
  10.1016/j.physletb.2017.12.010} {\bibfield  {journal} {\bibinfo  {journal}
  {Phys. Lett.}\ }\textbf {\bibinfo {volume} {B777}},\ \bibinfo {pages} {64}
  (\bibinfo {year} {2018}{\natexlab{a}})},\ \Eprint
  {http://arxiv.org/abs/1710.08910} {arXiv:1710.08910 [astro-ph.CO]}
  \BibitemShut {NoStop}%
%%CITATION = ARXIV:1710.08910;%%
\bibitem [{\citenamefont {Wilczek}(1987)}]{Wilczek:1987mv}%
  \BibitemOpen
  \bibfield  {author} {\bibinfo {author} {\bibfnamefont {F.}~\bibnamefont
  {Wilczek}},\ }\href {\doibase 10.1103/PhysRevLett.58.1799} {\bibfield
  {journal} {\bibinfo  {journal} {Phys. Rev. Lett.}\ }\textbf {\bibinfo
  {volume} {58}},\ \bibinfo {pages} {1799} (\bibinfo {year}
  {1987})}\BibitemShut {NoStop}%
%%CITATION = PRLTA,58,1799;%%
\bibitem [{\citenamefont {Krasnikov}(1996)}]{Krasnikov:1996bm}%
  \BibitemOpen
  \bibfield  {author} {\bibinfo {author} {\bibfnamefont {S.~V.}\ \bibnamefont
  {Krasnikov}},\ }\href {\doibase 10.1103/PhysRevLett.76.2633} {\bibfield
  {journal} {\bibinfo  {journal} {Phys. Rev. Lett.}\ }\textbf {\bibinfo
  {volume} {76}},\ \bibinfo {pages} {2633} (\bibinfo {year}
  {1996})}\BibitemShut {NoStop}%
%%CITATION = PRLTA,76,2633;%%
\bibitem [{\citenamefont {Li}\ \emph {et~al.}(2010)\citenamefont {Li},
  \citenamefont {Wang}, \citenamefont {Qi},\ and\ \citenamefont
  {Zhang}}]{Li:2009tca}%
  \BibitemOpen
  \bibfield  {author} {\bibinfo {author} {\bibfnamefont {R.}~\bibnamefont
  {Li}}, \bibinfo {author} {\bibfnamefont {J.}~\bibnamefont {Wang}}, \bibinfo
  {author} {\bibfnamefont {X.}~\bibnamefont {Qi}}, \ and\ \bibinfo {author}
  {\bibfnamefont {S.-C.}\ \bibnamefont {Zhang}},\ }\href {\doibase
  10.1038/nphys1534} {\bibfield  {journal} {\bibinfo  {journal} {Nature Phys.}\
  }\textbf {\bibinfo {volume} {6}},\ \bibinfo {pages} {284} (\bibinfo {year}
  {2010})},\ \Eprint {http://arxiv.org/abs/0908.1537} {arXiv:0908.1537
  [cond-mat.other]} \BibitemShut {NoStop}%
%%CITATION = ARXIV:0908.1537;%%
\bibitem [{\citenamefont {Visinelli}(2013)}]{visinelli:2013fia}%
  \BibitemOpen
  \bibfield  {author} {\bibinfo {author} {\bibfnamefont {L.}~\bibnamefont
  {Visinelli}},\ }\href {\doibase 10.1142/S0217732313501629} {\bibfield
  {journal} {\bibinfo  {journal} {Mod. Phys. Lett.}\ }\textbf {\bibinfo
  {volume} {A28}},\ \bibinfo {pages} {1350162} (\bibinfo {year} {2013})},\
  \Eprint {http://arxiv.org/abs/1401.0709} {arXiv:1401.0709 [physics.class-ph]}
  \BibitemShut {NoStop}%
%%CITATION = ARXIV:1401.0709;%%
\bibitem [{\citenamefont {Ter{\c c}as}\ \emph {et~al.}(2018)\citenamefont
  {Ter{\c c}as}, \citenamefont {Rodrigues},\ and\ \citenamefont {Mendon{\c
  c}a}}]{Tercas:2018gxv}%
  \BibitemOpen
  \bibfield  {author} {\bibinfo {author} {\bibfnamefont {H.}~\bibnamefont
  {Ter{\c c}as}}, \bibinfo {author} {\bibfnamefont {J.~D.}\ \bibnamefont
  {Rodrigues}}, \ and\ \bibinfo {author} {\bibfnamefont {J.~T.}\ \bibnamefont
  {Mendon{\c c}a}},\ }\href {\doibase 10.1103/PhysRevLett.120.181803}
  {\bibfield  {journal} {\bibinfo  {journal} {Phys. Rev. Lett.}\ }\textbf
  {\bibinfo {volume} {120}},\ \bibinfo {pages} {181803} (\bibinfo {year}
  {2018})},\ \Eprint {http://arxiv.org/abs/1801.06254} {arXiv:1801.06254
  [hep-ph]} \BibitemShut {NoStop}%
%%CITATION = ARXIV:1801.06254;%%
\bibitem [{\citenamefont {Visinelli}\ and\ \citenamefont {Ter{\c
  c}as}(2018)}]{Visinelli:2018zif}%
  \BibitemOpen
  \bibfield  {author} {\bibinfo {author} {\bibfnamefont {L.}~\bibnamefont
  {Visinelli}}\ and\ \bibinfo {author} {\bibfnamefont {H.}~\bibnamefont {Ter{\c
  c}as}},\ }\href@noop {} {\  (\bibinfo {year} {2018})},\ \Eprint
  {http://arxiv.org/abs/1807.06828} {arXiv:1807.06828 [hep-ph]} \BibitemShut
  {NoStop}%
%%CITATION = ARXIV:1807.06828;%%
\bibitem [{\citenamefont {Stern}(2016)}]{Stern:2016bbw}%
  \BibitemOpen
  \bibfield  {author} {\bibinfo {author} {\bibfnamefont {I.}~\bibnamefont
  {Stern}},\ }\bibfield  {booktitle} {\emph {\bibinfo {booktitle}
  {{Proceedings, 38th International Conference on High Energy Physics (ICHEP
  2016): Chicago, IL, USA, August 3-10, 2016}}},\ }\href@noop {} {\bibfield
  {journal} {\bibinfo  {journal} {PoS}\ }\textbf {\bibinfo {volume}
  {ICHEP2016}},\ \bibinfo {pages} {198} (\bibinfo {year} {2016})},\ \Eprint
  {http://arxiv.org/abs/1612.08296} {arXiv:1612.08296 [physics.ins-det]}
  \BibitemShut {NoStop}%
%%CITATION = ARXIV:1612.08296;%%
\bibitem [{\citenamefont {Raggi}\ and\ \citenamefont
  {Kozhuharov}(2014)}]{Raggi:2014zpa}%
  \BibitemOpen
  \bibfield  {author} {\bibinfo {author} {\bibfnamefont {M.}~\bibnamefont
  {Raggi}}\ and\ \bibinfo {author} {\bibfnamefont {V.}~\bibnamefont
  {Kozhuharov}},\ }\href {\doibase 10.1155/2014/959802} {\bibfield  {journal}
  {\bibinfo  {journal} {Adv. High Energy Phys.}\ }\textbf {\bibinfo {volume}
  {2014}},\ \bibinfo {pages} {959802} (\bibinfo {year} {2014})},\ \Eprint
  {http://arxiv.org/abs/1403.3041} {arXiv:1403.3041 [physics.ins-det]}
  \BibitemShut {NoStop}%
%%CITATION = ARXIV:1403.3041;%%
\bibitem [{\citenamefont {Majorovits}\ and\ \citenamefont
  {Redondo}(2017)}]{Majorovits:2016yvk}%
  \BibitemOpen
  \bibfield  {author} {\bibinfo {author} {\bibfnamefont {B.}~\bibnamefont
  {Majorovits}}\ and\ \bibinfo {author} {\bibfnamefont {J.}~\bibnamefont
  {Redondo}} (\bibinfo {collaboration} {MADMAX Working Group}),\ }in\ \href
  {\doibase 10.3204/DESY-PROC-2009-03/Majorovits_Bela} {\emph {\bibinfo
  {booktitle} {{Proceedings, 12th Patras Workshop on Axions, WIMPs and WISPs
  (PATRAS 2016): Jeju Island, South Korea, June 20-24, 2016}}}}\ (\bibinfo
  {year} {2017})\ pp.\ \bibinfo {pages} {94--97},\ \Eprint
  {http://arxiv.org/abs/1611.04549} {arXiv:1611.04549 [astro-ph.IM]}
  \BibitemShut {NoStop}%
%%CITATION = ARXIV:1611.04549;%%
\bibitem [{\citenamefont {Kahn}\ \emph {et~al.}(2016)\citenamefont {Kahn},
  \citenamefont {Safdi},\ and\ \citenamefont {Thaler}}]{Kahn:2016aff}%
  \BibitemOpen
  \bibfield  {author} {\bibinfo {author} {\bibfnamefont {Y.}~\bibnamefont
  {Kahn}}, \bibinfo {author} {\bibfnamefont {B.~R.}\ \bibnamefont {Safdi}}, \
  and\ \bibinfo {author} {\bibfnamefont {J.}~\bibnamefont {Thaler}},\ }\href
  {\doibase 10.1103/PhysRevLett.117.141801} {\bibfield  {journal} {\bibinfo
  {journal} {Phys. Rev. Lett.}\ }\textbf {\bibinfo {volume} {117}},\ \bibinfo
  {pages} {141801} (\bibinfo {year} {2016})},\ \Eprint
  {http://arxiv.org/abs/1602.01086} {arXiv:1602.01086 [hep-ph]} \BibitemShut
  {NoStop}%
%%CITATION = ARXIV:1602.01086;%%
\bibitem [{\citenamefont {Alesini}\ \emph {et~al.}(2017)\citenamefont
  {Alesini}, \citenamefont {Babusci}, \citenamefont {Di~Gioacchino},
  \citenamefont {Gatti}, \citenamefont {Lamanna},\ and\ \citenamefont
  {Ligi}}]{Alesini:2017ifp}%
  \BibitemOpen
  \bibfield  {author} {\bibinfo {author} {\bibfnamefont {D.}~\bibnamefont
  {Alesini}}, \bibinfo {author} {\bibfnamefont {D.}~\bibnamefont {Babusci}},
  \bibinfo {author} {\bibfnamefont {D.}~\bibnamefont {Di~Gioacchino}}, \bibinfo
  {author} {\bibfnamefont {C.}~\bibnamefont {Gatti}}, \bibinfo {author}
  {\bibfnamefont {G.}~\bibnamefont {Lamanna}}, \ and\ \bibinfo {author}
  {\bibfnamefont {C.}~\bibnamefont {Ligi}},\ }\href@noop {} {\  (\bibinfo
  {year} {2017})},\ \Eprint {http://arxiv.org/abs/1707.06010} {arXiv:1707.06010
  [physics.ins-det]} \BibitemShut {NoStop}%
%%CITATION = ARXIV:1707.06010;%%
\bibitem [{\citenamefont {Raffelt}(1995)}]{Raffelt:1995ym}%
  \BibitemOpen
  \bibfield  {author} {\bibinfo {author} {\bibfnamefont {G.~G.}\ \bibnamefont
  {Raffelt}},\ }in\ \href
  {https://inspirehep.net/record/392944/files/Pages_from_C95-01-22--1_159.pdf}
  {\emph {\bibinfo {booktitle} {{Dark matter in cosmology, clocks and test of
  fundamental laws. Proceedings, 30th Rencontres de Moriond, 15th Moriond
  Workshop, Villars sur Ollon, Switzerland, January 22-29, 1995}}}}\ (\bibinfo
  {year} {1995})\ pp.\ \bibinfo {pages} {159--168},\ \Eprint
  {http://arxiv.org/abs/hep-ph/9502358} {arXiv:hep-ph/9502358 [hep-ph]}
  \BibitemShut {NoStop}%
%%CITATION = HEP-PH/9502358;%%
\bibitem [{\citenamefont {Raffelt}(2007)}]{Raffelt:2006rj}%
  \BibitemOpen
  \bibfield  {author} {\bibinfo {author} {\bibfnamefont {G.~G.}\ \bibnamefont
  {Raffelt}},\ }\bibfield  {booktitle} {\emph {\bibinfo {booktitle} {{Quantum
  theories and renormalization group in gravity and cosmology: Proceedings, 2nd
  International Conference, IRGAC 2006, Barcelona, Spain, July 11-15, 2006}}},\
  }\href {\doibase 10.1088/1751-8113/40/25/S05} {\bibfield  {journal} {\bibinfo
   {journal} {J. Phys.}\ }\textbf {\bibinfo {volume} {A40}},\ \bibinfo {pages}
  {6607} (\bibinfo {year} {2007})},\ \Eprint
  {http://arxiv.org/abs/hep-ph/0611118} {arXiv:hep-ph/0611118 [hep-ph]}
  \BibitemShut {NoStop}%
%%CITATION = HEP-PH/0611118;%%
\bibitem [{\citenamefont {Sikivie}(2008{\natexlab{a}})}]{Sikivie:2006ni}%
  \BibitemOpen
  \bibfield  {author} {\bibinfo {author} {\bibfnamefont {P.}~\bibnamefont
  {Sikivie}},\ }\bibfield  {booktitle} {\emph {\bibinfo {booktitle} {{Axions:
  Theory, cosmology, and experimental searches. Proceedings, 1st Joint
  ILIAS-CERN-CAST axion training, Geneva, Switzerland, November 30-December 2,
  2005}}},\ }\href {\doibase 10.1007/978-3-540-73518-2_2} {\bibfield  {journal}
  {\bibinfo  {journal} {Lect. Notes Phys.}\ }\textbf {\bibinfo {volume}
  {741}},\ \bibinfo {pages} {19} (\bibinfo {year} {2008}{\natexlab{a}})},\
  \bibinfo {note} {[,19(2006)]},\ \Eprint
  {http://arxiv.org/abs/astro-ph/0610440} {arXiv:astro-ph/0610440 [astro-ph]}
  \BibitemShut {NoStop}%
%%CITATION = ASTRO-PH/0610440;%%
\bibitem [{\citenamefont {Kim}\ and\ \citenamefont
  {Carosi}(2010)}]{Kim:2008hd}%
  \BibitemOpen
  \bibfield  {author} {\bibinfo {author} {\bibfnamefont {J.~E.}\ \bibnamefont
  {Kim}}\ and\ \bibinfo {author} {\bibfnamefont {G.}~\bibnamefont {Carosi}},\
  }\href {\doibase 10.1103/RevModPhys.82.557} {\bibfield  {journal} {\bibinfo
  {journal} {Rev. Mod. Phys.}\ }\textbf {\bibinfo {volume} {82}},\ \bibinfo
  {pages} {557} (\bibinfo {year} {2010})},\ \Eprint
  {http://arxiv.org/abs/0807.3125} {arXiv:0807.3125 [hep-ph]} \BibitemShut
  {NoStop}%
%%CITATION = ARXIV:0807.3125;%%
\bibitem [{\citenamefont {Wantz}\ and\ \citenamefont
  {Shellard}(2010)}]{Wantz:2009it}%
  \BibitemOpen
  \bibfield  {author} {\bibinfo {author} {\bibfnamefont {O.}~\bibnamefont
  {Wantz}}\ and\ \bibinfo {author} {\bibfnamefont {E.~P.~S.}\ \bibnamefont
  {Shellard}},\ }\href {\doibase 10.1103/PhysRevD.82.123508} {\bibfield
  {journal} {\bibinfo  {journal} {Phys. Rev.}\ }\textbf {\bibinfo {volume}
  {D82}},\ \bibinfo {pages} {123508} (\bibinfo {year} {2010})},\ \Eprint
  {http://arxiv.org/abs/0910.1066} {arXiv:0910.1066 [astro-ph.CO]} \BibitemShut
  {NoStop}%
%%CITATION = ARXIV:0910.1066;%%
\bibitem [{\citenamefont {Kawasaki}\ and\ \citenamefont
  {Nakayama}(2013)}]{Kawasaki:2013ae}%
  \BibitemOpen
  \bibfield  {author} {\bibinfo {author} {\bibfnamefont {M.}~\bibnamefont
  {Kawasaki}}\ and\ \bibinfo {author} {\bibfnamefont {K.}~\bibnamefont
  {Nakayama}},\ }\href {\doibase 10.1146/annurev-nucl-102212-170536} {\bibfield
   {journal} {\bibinfo  {journal} {Ann. Rev. Nucl. Part. Sci.}\ }\textbf
  {\bibinfo {volume} {63}},\ \bibinfo {pages} {69} (\bibinfo {year} {2013})},\
  \Eprint {http://arxiv.org/abs/1301.1123} {arXiv:1301.1123 [hep-ph]}
  \BibitemShut {NoStop}%
%%CITATION = ARXIV:1301.1123;%%
\bibitem [{\citenamefont {Marsh}(2016)}]{Marsh:2015xka}%
  \BibitemOpen
  \bibfield  {author} {\bibinfo {author} {\bibfnamefont {D.~J.~E.}\
  \bibnamefont {Marsh}},\ }\href {\doibase 10.1016/j.physrep.2016.06.005}
  {\bibfield  {journal} {\bibinfo  {journal} {Phys. Rept.}\ }\textbf {\bibinfo
  {volume} {643}},\ \bibinfo {pages} {1} (\bibinfo {year} {2016})},\ \Eprint
  {http://arxiv.org/abs/1510.07633} {arXiv:1510.07633 [astro-ph.CO]}
  \BibitemShut {NoStop}%
%%CITATION = ARXIV:1510.07633;%%
\bibitem [{\citenamefont {Kim}(2017)}]{Kim:2017yqo}%
  \BibitemOpen
  \bibfield  {author} {\bibinfo {author} {\bibfnamefont {J.~E.}\ \bibnamefont
  {Kim}},\ }\bibfield  {booktitle} {\emph {\bibinfo {booktitle} {{Proceedings,
  16th Hellenic School and Workshops on Elementary Particle Physics and Gravity
  (CORFU2016): Corfu, Greece, August 31-September 23, 2016}}},\ }\href
  {\doibase 10.22323/1.292.0037} {\bibfield  {journal} {\bibinfo  {journal}
  {PoS}\ }\textbf {\bibinfo {volume} {CORFU2016}},\ \bibinfo {pages} {037}
  (\bibinfo {year} {2017})},\ \Eprint {http://arxiv.org/abs/1703.03114}
  {arXiv:1703.03114 [hep-ph]} \BibitemShut {NoStop}%
%%CITATION = ARXIV:1703.03114;%%
\bibitem [{\citenamefont {Choi}\ \emph {et~al.}(2007)\citenamefont {Choi},
  \citenamefont {Kim},\ and\ \citenamefont {Kim}}]{Choi:2006qj}%
  \BibitemOpen
  \bibfield  {author} {\bibinfo {author} {\bibfnamefont {K.-S.}\ \bibnamefont
  {Choi}}, \bibinfo {author} {\bibfnamefont {I.-W.}\ \bibnamefont {Kim}}, \
  and\ \bibinfo {author} {\bibfnamefont {J.~E.}\ \bibnamefont {Kim}},\ }\href
  {\doibase 10.1088/1126-6708/2007/03/116} {\bibfield  {journal} {\bibinfo
  {journal} {JHEP}\ }\textbf {\bibinfo {volume} {03}},\ \bibinfo {pages} {116}
  (\bibinfo {year} {2007})},\ \Eprint {http://arxiv.org/abs/hep-ph/0612107}
  {arXiv:hep-ph/0612107 [hep-ph]} \BibitemShut {NoStop}%
%%CITATION = HEP-PH/0612107;%%
\bibitem [{\citenamefont {Choi}\ \emph {et~al.}(2009)\citenamefont {Choi},
  \citenamefont {Nilles}, \citenamefont {Ramos-Sanchez},\ and\ \citenamefont
  {Vaudrevange}}]{Choi:2009jt}%
  \BibitemOpen
  \bibfield  {author} {\bibinfo {author} {\bibfnamefont {K.-S.}\ \bibnamefont
  {Choi}}, \bibinfo {author} {\bibfnamefont {H.~P.}\ \bibnamefont {Nilles}},
  \bibinfo {author} {\bibfnamefont {S.}~\bibnamefont {Ramos-Sanchez}}, \ and\
  \bibinfo {author} {\bibfnamefont {P.~K.~S.}\ \bibnamefont {Vaudrevange}},\
  }\href {\doibase 10.1016/j.physletb.2009.04.028} {\bibfield  {journal}
  {\bibinfo  {journal} {Phys. Lett.}\ }\textbf {\bibinfo {volume} {B675}},\
  \bibinfo {pages} {381} (\bibinfo {year} {2009})},\ \Eprint
  {http://arxiv.org/abs/0902.3070} {arXiv:0902.3070 [hep-th]} \BibitemShut
  {NoStop}%
%%CITATION = ARXIV:0902.3070;%%
\bibitem [{\citenamefont {Dias}\ \emph {et~al.}(2014)\citenamefont {Dias},
  \citenamefont {Machado}, \citenamefont {Nishi}, \citenamefont {Ringwald},\
  and\ \citenamefont {Vaudrevange}}]{Dias:2014osa}%
  \BibitemOpen
  \bibfield  {author} {\bibinfo {author} {\bibfnamefont {A.~G.}\ \bibnamefont
  {Dias}}, \bibinfo {author} {\bibfnamefont {A.~C.~B.}\ \bibnamefont
  {Machado}}, \bibinfo {author} {\bibfnamefont {C.~C.}\ \bibnamefont {Nishi}},
  \bibinfo {author} {\bibfnamefont {A.}~\bibnamefont {Ringwald}}, \ and\
  \bibinfo {author} {\bibfnamefont {P.}~\bibnamefont {Vaudrevange}},\ }\href
  {\doibase 10.1007/JHEP06(2014)037} {\bibfield  {journal} {\bibinfo  {journal}
  {JHEP}\ }\textbf {\bibinfo {volume} {06}},\ \bibinfo {pages} {037} (\bibinfo
  {year} {2014})},\ \Eprint {http://arxiv.org/abs/1403.5760} {arXiv:1403.5760
  [hep-ph]} \BibitemShut {NoStop}%
%%CITATION = ARXIV:1403.5760;%%
\bibitem [{\citenamefont {Higaki}\ and\ \citenamefont
  {Takahashi}(2014)}]{Higaki:2014pja}%
  \BibitemOpen
  \bibfield  {author} {\bibinfo {author} {\bibfnamefont {T.}~\bibnamefont
  {Higaki}}\ and\ \bibinfo {author} {\bibfnamefont {F.}~\bibnamefont
  {Takahashi}},\ }\href {\doibase 10.1007/JHEP07(2014)074} {\bibfield
  {journal} {\bibinfo  {journal} {JHEP}\ }\textbf {\bibinfo {volume} {07}},\
  \bibinfo {pages} {074} (\bibinfo {year} {2014})},\ \Eprint
  {http://arxiv.org/abs/1404.6923} {arXiv:1404.6923 [hep-th]} \BibitemShut
  {NoStop}%
%%CITATION = ARXIV:1404.6923;%%
\bibitem [{\citenamefont {Kim}\ and\ \citenamefont
  {Marsh}(2016)}]{Kim:2015yna}%
  \BibitemOpen
  \bibfield  {author} {\bibinfo {author} {\bibfnamefont {J.~E.}\ \bibnamefont
  {Kim}}\ and\ \bibinfo {author} {\bibfnamefont {D.~J.~E.}\ \bibnamefont
  {Marsh}},\ }\href {\doibase 10.1103/PhysRevD.93.025027} {\bibfield  {journal}
  {\bibinfo  {journal} {Phys. Rev.}\ }\textbf {\bibinfo {volume} {D93}},\
  \bibinfo {pages} {025027} (\bibinfo {year} {2016})},\ \Eprint
  {http://arxiv.org/abs/1510.01701} {arXiv:1510.01701 [hep-ph]} \BibitemShut
  {NoStop}%
%%CITATION = ARXIV:1510.01701;%%
\bibitem [{\citenamefont {Redi}\ and\ \citenamefont
  {Sato}(2016)}]{Redi:2016esr}%
  \BibitemOpen
  \bibfield  {author} {\bibinfo {author} {\bibfnamefont {M.}~\bibnamefont
  {Redi}}\ and\ \bibinfo {author} {\bibfnamefont {R.}~\bibnamefont {Sato}},\
  }\href {\doibase 10.1007/JHEP05(2016)104} {\bibfield  {journal} {\bibinfo
  {journal} {JHEP}\ }\textbf {\bibinfo {volume} {05}},\ \bibinfo {pages} {104}
  (\bibinfo {year} {2016})},\ \Eprint {http://arxiv.org/abs/1602.05427}
  {arXiv:1602.05427 [hep-ph]} \BibitemShut {NoStop}%
%%CITATION = ARXIV:1602.05427;%%
\bibitem [{\citenamefont {Di~Luzio}\ \emph {et~al.}(2017)\citenamefont
  {Di~Luzio}, \citenamefont {Nardi},\ and\ \citenamefont
  {Ubaldi}}]{DiLuzio:2017tjx}%
  \BibitemOpen
  \bibfield  {author} {\bibinfo {author} {\bibfnamefont {L.}~\bibnamefont
  {Di~Luzio}}, \bibinfo {author} {\bibfnamefont {E.}~\bibnamefont {Nardi}}, \
  and\ \bibinfo {author} {\bibfnamefont {L.}~\bibnamefont {Ubaldi}},\
  }\href@noop {} {\  (\bibinfo {year} {2017})},\ \Eprint
  {http://arxiv.org/abs/1704.01122} {arXiv:1704.01122 [hep-ph]} \BibitemShut
  {NoStop}%
%%CITATION = ARXIV:1704.01122;%%
\bibitem [{\citenamefont {Svrcek}\ and\ \citenamefont
  {Witten}(2006)}]{Svrcek:2006yi}%
  \BibitemOpen
  \bibfield  {author} {\bibinfo {author} {\bibfnamefont {P.}~\bibnamefont
  {Svrcek}}\ and\ \bibinfo {author} {\bibfnamefont {E.}~\bibnamefont
  {Witten}},\ }\href {\doibase 10.1088/1126-6708/2006/06/051} {\bibfield
  {journal} {\bibinfo  {journal} {JHEP}\ }\textbf {\bibinfo {volume} {06}},\
  \bibinfo {pages} {051} (\bibinfo {year} {2006})},\ \Eprint
  {http://arxiv.org/abs/hep-th/0605206} {arXiv:hep-th/0605206 [hep-th]}
  \BibitemShut {NoStop}%
%%CITATION = HEP-TH/0605206;%%
\bibitem [{\citenamefont {Arvanitaki}\ \emph {et~al.}(2010)\citenamefont
  {Arvanitaki}, \citenamefont {Dimopoulos}, \citenamefont {Dubovsky},
  \citenamefont {Kaloper},\ and\ \citenamefont
  {March-Russell}}]{Arvanitaki:2009fg}%
  \BibitemOpen
  \bibfield  {author} {\bibinfo {author} {\bibfnamefont {A.}~\bibnamefont
  {Arvanitaki}}, \bibinfo {author} {\bibfnamefont {S.}~\bibnamefont
  {Dimopoulos}}, \bibinfo {author} {\bibfnamefont {S.}~\bibnamefont
  {Dubovsky}}, \bibinfo {author} {\bibfnamefont {N.}~\bibnamefont {Kaloper}}, \
  and\ \bibinfo {author} {\bibfnamefont {J.}~\bibnamefont {March-Russell}},\
  }\href {\doibase 10.1103/PhysRevD.81.123530} {\bibfield  {journal} {\bibinfo
  {journal} {Phys. Rev.}\ }\textbf {\bibinfo {volume} {D81}},\ \bibinfo {pages}
  {123530} (\bibinfo {year} {2010})},\ \Eprint {http://arxiv.org/abs/0905.4720}
  {arXiv:0905.4720 [hep-th]} \BibitemShut {NoStop}%
%%CITATION = ARXIV:0905.4720;%%
\bibitem [{\citenamefont {Dine}\ \emph {et~al.}(2011)\citenamefont {Dine},
  \citenamefont {Festuccia}, \citenamefont {Kehayias},\ and\ \citenamefont
  {Wu}}]{Dine:2010cr}%
  \BibitemOpen
  \bibfield  {author} {\bibinfo {author} {\bibfnamefont {M.}~\bibnamefont
  {Dine}}, \bibinfo {author} {\bibfnamefont {G.}~\bibnamefont {Festuccia}},
  \bibinfo {author} {\bibfnamefont {J.}~\bibnamefont {Kehayias}}, \ and\
  \bibinfo {author} {\bibfnamefont {W.}~\bibnamefont {Wu}},\ }\href {\doibase
  10.1007/JHEP01(2011)012} {\bibfield  {journal} {\bibinfo  {journal} {JHEP}\
  }\textbf {\bibinfo {volume} {01}},\ \bibinfo {pages} {012} (\bibinfo {year}
  {2011})},\ \Eprint {http://arxiv.org/abs/1010.4803} {arXiv:1010.4803
  [hep-th]} \BibitemShut {NoStop}%
%%CITATION = ARXIV:1010.4803;%%
\bibitem [{\citenamefont {Ringwald}(2014)}]{Ringwald:2012cu}%
  \BibitemOpen
  \bibfield  {author} {\bibinfo {author} {\bibfnamefont {A.}~\bibnamefont
  {Ringwald}},\ }\bibfield  {booktitle} {\emph {\bibinfo {booktitle}
  {{Proceedings, 18th International Symposium on Particles, Strings and
  Cosmology (PASCOS 2012): Merida, Yucatan, Mexico, June 3-8, 2012}}},\ }\href
  {\doibase 10.1088/1742-6596/485/1/012013} {\bibfield  {journal} {\bibinfo
  {journal} {J. Phys. Conf. Ser.}\ }\textbf {\bibinfo {volume} {485}},\
  \bibinfo {pages} {012013} (\bibinfo {year} {2014})},\ \Eprint
  {http://arxiv.org/abs/1209.2299} {arXiv:1209.2299 [hep-ph]} \BibitemShut
  {NoStop}%
%%CITATION = ARXIV:1209.2299;%%
\bibitem [{\citenamefont {Cicoli}\ \emph {et~al.}(2012)\citenamefont {Cicoli},
  \citenamefont {Goodsell},\ and\ \citenamefont {Ringwald}}]{Cicoli:2012sz}%
  \BibitemOpen
  \bibfield  {author} {\bibinfo {author} {\bibfnamefont {M.}~\bibnamefont
  {Cicoli}}, \bibinfo {author} {\bibfnamefont {M.}~\bibnamefont {Goodsell}}, \
  and\ \bibinfo {author} {\bibfnamefont {A.}~\bibnamefont {Ringwald}},\ }\href
  {\doibase 10.1007/JHEP10(2012)146} {\bibfield  {journal} {\bibinfo  {journal}
  {JHEP}\ }\textbf {\bibinfo {volume} {10}},\ \bibinfo {pages} {146} (\bibinfo
  {year} {2012})},\ \Eprint {http://arxiv.org/abs/1206.0819} {arXiv:1206.0819
  [hep-th]} \BibitemShut {NoStop}%
%%CITATION = ARXIV:1206.0819;%%
\bibitem [{\citenamefont {Kim}(2013)}]{Kim:2013fga}%
  \BibitemOpen
  \bibfield  {author} {\bibinfo {author} {\bibfnamefont {J.~E.}\ \bibnamefont
  {Kim}},\ }\href {\doibase 10.1103/PhysRevLett.111.031801} {\bibfield
  {journal} {\bibinfo  {journal} {Phys. Rev. Lett.}\ }\textbf {\bibinfo
  {volume} {111}},\ \bibinfo {pages} {031801} (\bibinfo {year} {2013})},\
  \Eprint {http://arxiv.org/abs/1303.1822} {arXiv:1303.1822 [hep-ph]}
  \BibitemShut {NoStop}%
%%CITATION = ARXIV:1303.1822;%%
\bibitem [{\citenamefont {Bachlechner}\ \emph {et~al.}(2015)\citenamefont
  {Bachlechner}, \citenamefont {Long},\ and\ \citenamefont
  {McAllister}}]{Bachlechner:2014gfa}%
  \BibitemOpen
  \bibfield  {author} {\bibinfo {author} {\bibfnamefont {T.~C.}\ \bibnamefont
  {Bachlechner}}, \bibinfo {author} {\bibfnamefont {C.}~\bibnamefont {Long}}, \
  and\ \bibinfo {author} {\bibfnamefont {L.}~\bibnamefont {McAllister}},\
  }\href {\doibase 10.1007/JHEP12(2015)042} {\bibfield  {journal} {\bibinfo
  {journal} {JHEP}\ }\textbf {\bibinfo {volume} {12}},\ \bibinfo {pages} {042}
  (\bibinfo {year} {2015})},\ \Eprint {http://arxiv.org/abs/1412.1093}
  {arXiv:1412.1093 [hep-th]} \BibitemShut {NoStop}%
%%CITATION = ARXIV:1412.1093;%%
\bibitem [{\citenamefont {Halverson}\ \emph {et~al.}(2017)\citenamefont
  {Halverson}, \citenamefont {Long},\ and\ \citenamefont
  {Nath}}]{Halverson:2017deq}%
  \BibitemOpen
  \bibfield  {author} {\bibinfo {author} {\bibfnamefont {J.}~\bibnamefont
  {Halverson}}, \bibinfo {author} {\bibfnamefont {C.}~\bibnamefont {Long}}, \
  and\ \bibinfo {author} {\bibfnamefont {P.}~\bibnamefont {Nath}},\ }\href
  {\doibase 10.1103/PhysRevD.96.056025} {\bibfield  {journal} {\bibinfo
  {journal} {Phys. Rev.}\ }\textbf {\bibinfo {volume} {D96}},\ \bibinfo {pages}
  {056025} (\bibinfo {year} {2017})},\ \Eprint
  {http://arxiv.org/abs/1703.07779} {arXiv:1703.07779 [hep-ph]} \BibitemShut
  {NoStop}%
%%CITATION = ARXIV:1703.07779;%%
\bibitem [{\citenamefont {Stott}\ \emph {et~al.}(2017)\citenamefont {Stott},
  \citenamefont {Marsh}, \citenamefont {Pongkitivanichkul}, \citenamefont
  {Price},\ and\ \citenamefont {Acharya}}]{Stott:2017hvl}%
  \BibitemOpen
  \bibfield  {author} {\bibinfo {author} {\bibfnamefont {M.~J.}\ \bibnamefont
  {Stott}}, \bibinfo {author} {\bibfnamefont {D.~J.~E.}\ \bibnamefont {Marsh}},
  \bibinfo {author} {\bibfnamefont {C.}~\bibnamefont {Pongkitivanichkul}},
  \bibinfo {author} {\bibfnamefont {L.~C.}\ \bibnamefont {Price}}, \ and\
  \bibinfo {author} {\bibfnamefont {B.~S.}\ \bibnamefont {Acharya}},\
  }\href@noop {} {\  (\bibinfo {year} {2017})},\ \Eprint
  {http://arxiv.org/abs/1706.03236} {arXiv:1706.03236 [astro-ph.CO]}
  \BibitemShut {NoStop}%
%%CITATION = ARXIV:1706.03236;%%
\bibitem [{\citenamefont {Visinelli}\ and\ \citenamefont
  {Vagnozzi}(2018)}]{Visinelli:2018utg}%
  \BibitemOpen
  \bibfield  {author} {\bibinfo {author} {\bibfnamefont {L.}~\bibnamefont
  {Visinelli}}\ and\ \bibinfo {author} {\bibfnamefont {S.}~\bibnamefont
  {Vagnozzi}},\ }\href@noop {} {\  (\bibinfo {year} {2018})},\ \Eprint
  {http://arxiv.org/abs/1809.06382} {arXiv:1809.06382 [hep-ph]} \BibitemShut
  {NoStop}%
%%CITATION = ARXIV:1809.06382;%%
\bibitem [{\citenamefont {Arkani-Hamed}\ \emph {et~al.}(1999)\citenamefont
  {Arkani-Hamed}, \citenamefont {Dimopoulos},\ and\ \citenamefont
  {Dvali}}]{ArkaniHamed:1998nn}%
  \BibitemOpen
  \bibfield  {author} {\bibinfo {author} {\bibfnamefont {N.}~\bibnamefont
  {Arkani-Hamed}}, \bibinfo {author} {\bibfnamefont {S.}~\bibnamefont
  {Dimopoulos}}, \ and\ \bibinfo {author} {\bibfnamefont {G.~R.}\ \bibnamefont
  {Dvali}},\ }\href {\doibase 10.1103/PhysRevD.59.086004} {\bibfield  {journal}
  {\bibinfo  {journal} {Phys. Rev.}\ }\textbf {\bibinfo {volume} {D59}},\
  \bibinfo {pages} {086004} (\bibinfo {year} {1999})},\ \Eprint
  {http://arxiv.org/abs/hep-ph/9807344} {arXiv:hep-ph/9807344 [hep-ph]}
  \BibitemShut {NoStop}%
%%CITATION = HEP-PH/9807344;%%
\bibitem [{\citenamefont {Randall}\ and\ \citenamefont
  {Sundrum}(1999)}]{Randall:1999vf}%
  \BibitemOpen
  \bibfield  {author} {\bibinfo {author} {\bibfnamefont {L.}~\bibnamefont
  {Randall}}\ and\ \bibinfo {author} {\bibfnamefont {R.}~\bibnamefont
  {Sundrum}},\ }\href {\doibase 10.1103/PhysRevLett.83.4690} {\bibfield
  {journal} {\bibinfo  {journal} {Phys. Rev. Lett.}\ }\textbf {\bibinfo
  {volume} {83}},\ \bibinfo {pages} {4690} (\bibinfo {year} {1999})},\ \Eprint
  {http://arxiv.org/abs/hep-th/9906064} {arXiv:hep-th/9906064 [hep-th]}
  \BibitemShut {NoStop}%
%%CITATION = HEP-TH/9906064;%%
\bibitem [{\citenamefont {Binetruy}\ \emph {et~al.}(2000)\citenamefont
  {Binetruy}, \citenamefont {Deffayet},\ and\ \citenamefont
  {Langlois}}]{Binetruy:1999ut}%
  \BibitemOpen
  \bibfield  {author} {\bibinfo {author} {\bibfnamefont {P.}~\bibnamefont
  {Binetruy}}, \bibinfo {author} {\bibfnamefont {C.}~\bibnamefont {Deffayet}},
  \ and\ \bibinfo {author} {\bibfnamefont {D.}~\bibnamefont {Langlois}},\
  }\href {\doibase 10.1016/S0550-3213(99)00696-3} {\bibfield  {journal}
  {\bibinfo  {journal} {Nucl. Phys.}\ }\textbf {\bibinfo {volume} {B565}},\
  \bibinfo {pages} {269} (\bibinfo {year} {2000})},\ \Eprint
  {http://arxiv.org/abs/hep-th/9905012} {arXiv:hep-th/9905012 [hep-th]}
  \BibitemShut {NoStop}%
%%CITATION = HEP-TH/9905012;%%
\bibitem [{\citenamefont {Chung}\ and\ \citenamefont
  {Freese}(2000)}]{Chung:1999xg}%
  \BibitemOpen
  \bibfield  {author} {\bibinfo {author} {\bibfnamefont {D.~J.~H.}\
  \bibnamefont {Chung}}\ and\ \bibinfo {author} {\bibfnamefont
  {K.}~\bibnamefont {Freese}},\ }\href {\doibase 10.1103/PhysRevD.62.063513}
  {\bibfield  {journal} {\bibinfo  {journal} {Phys. Rev.}\ }\textbf {\bibinfo
  {volume} {D62}},\ \bibinfo {pages} {063513} (\bibinfo {year} {2000})},\
  \Eprint {http://arxiv.org/abs/hep-ph/9910235} {arXiv:hep-ph/9910235 [hep-ph]}
  \BibitemShut {NoStop}%
%%CITATION = HEP-PH/9910235;%%
\bibitem [{\citenamefont {Caldwell}\ and\ \citenamefont
  {Langlois}(2001)}]{Caldwell:2001ja}%
  \BibitemOpen
  \bibfield  {author} {\bibinfo {author} {\bibfnamefont {R.~R.}\ \bibnamefont
  {Caldwell}}\ and\ \bibinfo {author} {\bibfnamefont {D.}~\bibnamefont
  {Langlois}},\ }\href {\doibase 10.1016/S0370-2693(01)00631-1} {\bibfield
  {journal} {\bibinfo  {journal} {Phys. Lett.}\ }\textbf {\bibinfo {volume}
  {B511}},\ \bibinfo {pages} {129} (\bibinfo {year} {2001})},\ \Eprint
  {http://arxiv.org/abs/gr-qc/0103070} {arXiv:gr-qc/0103070 [gr-qc]}
  \BibitemShut {NoStop}%
%%CITATION = GR-QC/0103070;%%
\bibitem [{\citenamefont {Visinelli}\ \emph
  {et~al.}(2018{\natexlab{b}})\citenamefont {Visinelli}, \citenamefont
  {Bolis},\ and\ \citenamefont {Vagnozzi}}]{Visinelli:2017bny}%
  \BibitemOpen
  \bibfield  {author} {\bibinfo {author} {\bibfnamefont {L.}~\bibnamefont
  {Visinelli}}, \bibinfo {author} {\bibfnamefont {N.}~\bibnamefont {Bolis}}, \
  and\ \bibinfo {author} {\bibfnamefont {S.}~\bibnamefont {Vagnozzi}},\ }\href
  {\doibase 10.1103/PhysRevD.97.064039} {\bibfield  {journal} {\bibinfo
  {journal} {Phys. Rev.}\ }\textbf {\bibinfo {volume} {D97}},\ \bibinfo {pages}
  {064039} (\bibinfo {year} {2018}{\natexlab{b}})},\ \Eprint
  {http://arxiv.org/abs/1711.06628} {arXiv:1711.06628 [gr-qc]} \BibitemShut
  {NoStop}%
%%CITATION = ARXIV:1711.06628;%%
\bibitem [{\citenamefont {Alonso}\ and\ \citenamefont
  {Urbano}(2017)}]{Alonso:2017avz}%
  \BibitemOpen
  \bibfield  {author} {\bibinfo {author} {\bibfnamefont {R.}~\bibnamefont
  {Alonso}}\ and\ \bibinfo {author} {\bibfnamefont {A.}~\bibnamefont
  {Urbano}},\ }\href@noop {} {\  (\bibinfo {year} {2017})},\ \Eprint
  {http://arxiv.org/abs/1706.07415} {arXiv:1706.07415 [hep-ph]} \BibitemShut
  {NoStop}%
%%CITATION = ARXIV:1706.07415;%%
\bibitem [{\citenamefont {Agrawal}\ \emph {et~al.}(2018)\citenamefont
  {Agrawal}, \citenamefont {Obied}, \citenamefont {Steinhardt},\ and\
  \citenamefont {Vafa}}]{Agrawal:2018own}%
  \BibitemOpen
  \bibfield  {author} {\bibinfo {author} {\bibfnamefont {P.}~\bibnamefont
  {Agrawal}}, \bibinfo {author} {\bibfnamefont {G.}~\bibnamefont {Obied}},
  \bibinfo {author} {\bibfnamefont {P.~J.}\ \bibnamefont {Steinhardt}}, \ and\
  \bibinfo {author} {\bibfnamefont {C.}~\bibnamefont {Vafa}},\ }\href {\doibase
  10.1016/j.physletb.2018.07.040} {\bibfield  {journal} {\bibinfo  {journal}
  {Phys. Lett.}\ }\textbf {\bibinfo {volume} {B784}},\ \bibinfo {pages} {271}
  (\bibinfo {year} {2018})},\ \Eprint {http://arxiv.org/abs/1806.09718}
  {arXiv:1806.09718 [hep-th]} \BibitemShut {NoStop}%
%%CITATION = ARXIV:1806.09718;%%
\bibitem [{\citenamefont {Akrami}\ \emph {et~al.}(2018)\citenamefont {Akrami},
  \citenamefont {Kallosh}, \citenamefont {Linde},\ and\ \citenamefont
  {Vardanyan}}]{Akrami:2018ylq}%
  \BibitemOpen
  \bibfield  {author} {\bibinfo {author} {\bibfnamefont {Y.}~\bibnamefont
  {Akrami}}, \bibinfo {author} {\bibfnamefont {R.}~\bibnamefont {Kallosh}},
  \bibinfo {author} {\bibfnamefont {A.}~\bibnamefont {Linde}}, \ and\ \bibinfo
  {author} {\bibfnamefont {V.}~\bibnamefont {Vardanyan}},\ }\href@noop {} {\
  (\bibinfo {year} {2018})},\ \Eprint {http://arxiv.org/abs/1808.09440}
  {arXiv:1808.09440 [hep-th]} \BibitemShut {NoStop}%
%%CITATION = ARXIV:1808.09440;%%
\bibitem [{\citenamefont {Marsh}(2018)}]{Marsh:2018kub}%
  \BibitemOpen
  \bibfield  {author} {\bibinfo {author} {\bibfnamefont {M.~C.~D.}\
  \bibnamefont {Marsh}},\ }\href@noop {} {\  (\bibinfo {year} {2018})},\
  \Eprint {http://arxiv.org/abs/1809.00726} {arXiv:1809.00726 [hep-th]}
  \BibitemShut {NoStop}%
%%CITATION = ARXIV:1809.00726;%%
\bibitem [{\citenamefont {Conlon}(2018)}]{Conlon:2018eyr}%
  \BibitemOpen
  \bibfield  {author} {\bibinfo {author} {\bibfnamefont {J.~P.}\ \bibnamefont
  {Conlon}},\ }\href@noop {} {\  (\bibinfo {year} {2018})},\ \Eprint
  {http://arxiv.org/abs/1808.05040} {arXiv:1808.05040 [hep-th]} \BibitemShut
  {NoStop}%
%%CITATION = ARXIV:1808.05040;%%
\bibitem [{\citenamefont {Murayama}\ \emph {et~al.}(2018)\citenamefont
  {Murayama}, \citenamefont {Yamazaki},\ and\ \citenamefont
  {Yanagida}}]{Murayama:2018lie}%
  \BibitemOpen
  \bibfield  {author} {\bibinfo {author} {\bibfnamefont {H.}~\bibnamefont
  {Murayama}}, \bibinfo {author} {\bibfnamefont {M.}~\bibnamefont {Yamazaki}},
  \ and\ \bibinfo {author} {\bibfnamefont {T.~T.}\ \bibnamefont {Yanagida}},\
  }\href@noop {} {\  (\bibinfo {year} {2018})},\ \Eprint
  {http://arxiv.org/abs/1809.00478} {arXiv:1809.00478 [hep-th]} \BibitemShut
  {NoStop}%
%%CITATION = ARXIV:1809.00478;%%
\bibitem [{\citenamefont {Kinney}\ \emph {et~al.}(2018)\citenamefont {Kinney},
  \citenamefont {Vagnozzi},\ and\ \citenamefont {Visinelli}}]{Kinney:2018nny}%
  \BibitemOpen
  \bibfield  {author} {\bibinfo {author} {\bibfnamefont {W.~H.}\ \bibnamefont
  {Kinney}}, \bibinfo {author} {\bibfnamefont {S.}~\bibnamefont {Vagnozzi}}, \
  and\ \bibinfo {author} {\bibfnamefont {L.}~\bibnamefont {Visinelli}},\
  }\href@noop {} {\  (\bibinfo {year} {2018})},\ \Eprint
  {http://arxiv.org/abs/1808.06424} {arXiv:1808.06424 [astro-ph.CO]}
  \BibitemShut {NoStop}%
%%CITATION = ARXIV:1808.06424;%%
\bibitem [{\citenamefont {Loaiza-Brito}\ and\ \citenamefont
  {Loaiza-Brito}(2018)}]{Damian:2018tlf}%
  \BibitemOpen
  \bibfield  {author} {\bibinfo {author} {\bibfnamefont {O.}~\bibnamefont
  {Loaiza-Brito}}\ and\ \bibinfo {author} {\bibfnamefont {O.}~\bibnamefont
  {Loaiza-Brito}},\ }\href@noop {} {\  (\bibinfo {year} {2018})},\ \Eprint
  {http://arxiv.org/abs/1808.03397} {arXiv:1808.03397 [hep-th]} \BibitemShut
  {NoStop}%
%%CITATION = ARXIV:1808.03397;%%
\bibitem [{\citenamefont {Danielsson}(2018)}]{Danielsson:2018qpa}%
  \BibitemOpen
  \bibfield  {author} {\bibinfo {author} {\bibfnamefont {U.}~\bibnamefont
  {Danielsson}},\ }\href@noop {} {\  (\bibinfo {year} {2018})},\ \Eprint
  {http://arxiv.org/abs/1809.04512} {arXiv:1809.04512 [hep-th]} \BibitemShut
  {NoStop}%
%%CITATION = ARXIV:1809.04512;%%
\bibitem [{\citenamefont {Baldeschi}\ \emph {et~al.}(1983)\citenamefont
  {Baldeschi}, \citenamefont {Gelmini},\ and\ \citenamefont
  {Ruffini}}]{Baldeschi:1983}%
  \BibitemOpen
  \bibfield  {author} {\bibinfo {author} {\bibfnamefont {M.}~\bibnamefont
  {Baldeschi}}, \bibinfo {author} {\bibfnamefont {G.}~\bibnamefont {Gelmini}},
  \ and\ \bibinfo {author} {\bibfnamefont {R.}~\bibnamefont {Ruffini}},\ }\href
  {\doibase http://dx.doi.org/10.1016/0370-2693(83)90688-3} {\bibfield
  {journal} {\bibinfo  {journal} {Physics Letters B}\ }\textbf {\bibinfo
  {volume} {122}},\ \bibinfo {pages} {221 } (\bibinfo {year}
  {1983})}\BibitemShut {NoStop}%
\bibitem [{\citenamefont {Membrado}\ \emph {et~al.}(1989)\citenamefont
  {Membrado}, \citenamefont {Pacheco},\ and\ \citenamefont
  {Sa{\~n}udo}}]{Membrado:1989bqo}%
  \BibitemOpen
  \bibfield  {author} {\bibinfo {author} {\bibfnamefont {M.}~\bibnamefont
  {Membrado}}, \bibinfo {author} {\bibfnamefont {A.~F.}\ \bibnamefont
  {Pacheco}}, \ and\ \bibinfo {author} {\bibfnamefont {J.}~\bibnamefont
  {Sa{\~n}udo}},\ }\href {\doibase 10.1103/PhysRevA.39.4207} {\bibfield
  {journal} {\bibinfo  {journal} {Phys. Rev.}\ }\textbf {\bibinfo {volume}
  {A39}},\ \bibinfo {pages} {4207} (\bibinfo {year} {1989})}\BibitemShut
  {NoStop}%
%%CITATION = PHRVA,A39,4207;%%
\bibitem [{\citenamefont {Press}\ \emph {et~al.}(1990)\citenamefont {Press},
  \citenamefont {Ryden},\ and\ \citenamefont {Spergel}}]{Press:1990}%
  \BibitemOpen
  \bibfield  {author} {\bibinfo {author} {\bibfnamefont {W.~H.}\ \bibnamefont
  {Press}}, \bibinfo {author} {\bibfnamefont {B.~S.}\ \bibnamefont {Ryden}}, \
  and\ \bibinfo {author} {\bibfnamefont {D.~N.}\ \bibnamefont {Spergel}},\
  }\href {\doibase 10.1103/PhysRevLett.64.1084} {\bibfield  {journal} {\bibinfo
   {journal} {Phys. Rev. Lett.}\ }\textbf {\bibinfo {volume} {64}},\ \bibinfo
  {pages} {1084} (\bibinfo {year} {1990})}\BibitemShut {NoStop}%
\bibitem [{\citenamefont {Sin}(1994)}]{Sin:1994}%
  \BibitemOpen
  \bibfield  {author} {\bibinfo {author} {\bibfnamefont {S.-J.}\ \bibnamefont
  {Sin}},\ }\href {\doibase 10.1103/PhysRevD.50.3650} {\bibfield  {journal}
  {\bibinfo  {journal} {Phys. Rev. D}\ }\textbf {\bibinfo {volume} {50}},\
  \bibinfo {pages} {3650} (\bibinfo {year} {1994})}\BibitemShut {NoStop}%
\bibitem [{\citenamefont {Ji}\ and\ \citenamefont {Sin}(1994)}]{Ji:1994}%
  \BibitemOpen
  \bibfield  {author} {\bibinfo {author} {\bibfnamefont {S.~U.}\ \bibnamefont
  {Ji}}\ and\ \bibinfo {author} {\bibfnamefont {S.~J.}\ \bibnamefont {Sin}},\
  }\href {\doibase 10.1103/PhysRevD.50.3655} {\bibfield  {journal} {\bibinfo
  {journal} {Phys. Rev. D}\ }\textbf {\bibinfo {volume} {50}},\ \bibinfo
  {pages} {3655} (\bibinfo {year} {1994})}\BibitemShut {NoStop}%
\bibitem [{\citenamefont {Lee}\ and\ \citenamefont {Koh}(1996)}]{Lee:1996}%
  \BibitemOpen
  \bibfield  {author} {\bibinfo {author} {\bibfnamefont {J.-w.}\ \bibnamefont
  {Lee}}\ and\ \bibinfo {author} {\bibfnamefont {I.~G.}\ \bibnamefont {Koh}},\
  }\href {\doibase 10.1103/PhysRevD.53.2236} {\bibfield  {journal} {\bibinfo
  {journal} {Phys. Rev. D}\ }\textbf {\bibinfo {volume} {53}},\ \bibinfo
  {pages} {2236} (\bibinfo {year} {1996})}\BibitemShut {NoStop}%
\bibitem [{\citenamefont {Guzm{\'a}n}\ and\ \citenamefont
  {Matos}(2000)}]{Guzman:2000}%
  \BibitemOpen
  \bibfield  {author} {\bibinfo {author} {\bibfnamefont {F.~S.}\ \bibnamefont
  {Guzm{\'a}n}}\ and\ \bibinfo {author} {\bibfnamefont {T.}~\bibnamefont
  {Matos}},\ }\href {http://stacks.iop.org/0264-9381/17/i=1/a=102} {\bibfield
  {journal} {\bibinfo  {journal} {Classical and Quantum Gravity}\ }\textbf
  {\bibinfo {volume} {17}},\ \bibinfo {pages} {L9} (\bibinfo {year}
  {2000})}\BibitemShut {NoStop}%
\bibitem [{\citenamefont {Sahni}\ and\ \citenamefont
  {Wang}(2000)}]{Sahni:2000}%
  \BibitemOpen
  \bibfield  {author} {\bibinfo {author} {\bibfnamefont {V.}~\bibnamefont
  {Sahni}}\ and\ \bibinfo {author} {\bibfnamefont {L.}~\bibnamefont {Wang}},\
  }\href {\doibase 10.1103/PhysRevD.62.103517} {\bibfield  {journal} {\bibinfo
  {journal} {Phys. Rev. D}\ }\textbf {\bibinfo {volume} {62}},\ \bibinfo
  {pages} {103517} (\bibinfo {year} {2000})}\BibitemShut {NoStop}%
\bibitem [{\citenamefont {Peebles}(2000)}]{Peebles:2000yy}%
  \BibitemOpen
  \bibfield  {author} {\bibinfo {author} {\bibfnamefont {P.~J.~E.}\
  \bibnamefont {Peebles}},\ }\href {\doibase 10.1086/312677} {\bibfield
  {journal} {\bibinfo  {journal} {Astrophys. J.}\ }\textbf {\bibinfo {volume}
  {534}},\ \bibinfo {pages} {L127} (\bibinfo {year} {2000})},\ \Eprint
  {http://arxiv.org/abs/astro-ph/0002495} {arXiv:astro-ph/0002495 [astro-ph]}
  \BibitemShut {NoStop}%
%%CITATION = ASTRO-PH/0002495;%%
\bibitem [{\citenamefont {Goodman}(2000)}]{Goodman:2000}%
  \BibitemOpen
  \bibfield  {author} {\bibinfo {author} {\bibfnamefont {J.}~\bibnamefont
  {Goodman}},\ }\href {\doibase
  http://dx.doi.org/10.1016/S1384-1076(00)00015-4} {\bibfield  {journal}
  {\bibinfo  {journal} {New Astronomy}\ }\textbf {\bibinfo {volume} {5}},\
  \bibinfo {pages} {103 } (\bibinfo {year} {2000})}\BibitemShut {NoStop}%
\bibitem [{\citenamefont {Matos}\ and\ \citenamefont
  {Ure{\~n}a-L{\'o}pez}(2000)}]{Matos:2000}%
  \BibitemOpen
  \bibfield  {author} {\bibinfo {author} {\bibfnamefont {T.}~\bibnamefont
  {Matos}}\ and\ \bibinfo {author} {\bibfnamefont {L.~A.}\ \bibnamefont
  {Ure{\~n}a-L{\'o}pez}},\ }\href
  {http://stacks.iop.org/0264-9381/17/i=13/a=101} {\bibfield  {journal}
  {\bibinfo  {journal} {Classical and Quantum Gravity}\ }\textbf {\bibinfo
  {volume} {17}},\ \bibinfo {pages} {L75} (\bibinfo {year} {2000})}\BibitemShut
  {NoStop}%
\bibitem [{\citenamefont {Hu}\ \emph {et~al.}(2000)\citenamefont {Hu},
  \citenamefont {Barkana},\ and\ \citenamefont {Gruzinov}}]{Hu:2000ke}%
  \BibitemOpen
  \bibfield  {author} {\bibinfo {author} {\bibfnamefont {W.}~\bibnamefont
  {Hu}}, \bibinfo {author} {\bibfnamefont {R.}~\bibnamefont {Barkana}}, \ and\
  \bibinfo {author} {\bibfnamefont {A.}~\bibnamefont {Gruzinov}},\ }\href
  {\doibase 10.1103/PhysRevLett.85.1158} {\bibfield  {journal} {\bibinfo
  {journal} {Phys. Rev. Lett.}\ }\textbf {\bibinfo {volume} {85}},\ \bibinfo
  {pages} {1158} (\bibinfo {year} {2000})},\ \Eprint
  {http://arxiv.org/abs/astro-ph/0003365} {arXiv:astro-ph/0003365 [astro-ph]}
  \BibitemShut {NoStop}%
%%CITATION = ASTRO-PH/0003365;%%
\bibitem [{\citenamefont {Hui}\ \emph {et~al.}(2016)\citenamefont {Hui},
  \citenamefont {Ostriker}, \citenamefont {Tremaine},\ and\ \citenamefont
  {Witten}}]{Hui:2016ltb}%
  \BibitemOpen
  \bibfield  {author} {\bibinfo {author} {\bibfnamefont {L.}~\bibnamefont
  {Hui}}, \bibinfo {author} {\bibfnamefont {J.~P.}\ \bibnamefont {Ostriker}},
  \bibinfo {author} {\bibfnamefont {S.}~\bibnamefont {Tremaine}}, \ and\
  \bibinfo {author} {\bibfnamefont {E.}~\bibnamefont {Witten}},\ }\href@noop {}
  {\  (\bibinfo {year} {2016})},\ \Eprint {http://arxiv.org/abs/1610.08297}
  {arXiv:1610.08297 [astro-ph.CO]} \BibitemShut {NoStop}%
%%CITATION = ARXIV:1610.08297;%%
\bibitem [{\citenamefont {Diez-Tejedor}\ and\ \citenamefont
  {Marsh}(2017)}]{Diez-Tejedor:2017ivd}%
  \BibitemOpen
  \bibfield  {author} {\bibinfo {author} {\bibfnamefont {A.}~\bibnamefont
  {Diez-Tejedor}}\ and\ \bibinfo {author} {\bibfnamefont {D.~J.~E.}\
  \bibnamefont {Marsh}},\ }\href@noop {} {\  (\bibinfo {year} {2017})},\
  \Eprint {http://arxiv.org/abs/1702.02116} {arXiv:1702.02116 [hep-ph]}
  \BibitemShut {NoStop}%
%%CITATION = ARXIV:1702.02116;%%
\bibitem [{\citenamefont {Weinberg}\ \emph {et~al.}(2014)\citenamefont
  {Weinberg}, \citenamefont {Bullock}, \citenamefont {Governato}, \citenamefont
  {Kuzio~de Naray},\ and\ \citenamefont {Peter}}]{Weinberg:2013aya}%
  \BibitemOpen
  \bibfield  {author} {\bibinfo {author} {\bibfnamefont {D.~H.}\ \bibnamefont
  {Weinberg}}, \bibinfo {author} {\bibfnamefont {J.~S.}\ \bibnamefont
  {Bullock}}, \bibinfo {author} {\bibfnamefont {F.}~\bibnamefont {Governato}},
  \bibinfo {author} {\bibfnamefont {R.}~\bibnamefont {Kuzio~de Naray}}, \ and\
  \bibinfo {author} {\bibfnamefont {A.~H.~G.}\ \bibnamefont {Peter}},\
  }\bibfield  {booktitle} {\emph {\bibinfo {booktitle} {{Sackler Colloquium:
  Dark Matter Universe: On the Threshhold of Discovery Irvine, USA, October
  18-20, 2012}}},\ }\href {\doibase 10.1073/pnas.1308716112} {\bibfield
  {journal} {\bibinfo  {journal} {Proc. Nat. Acad. Sci.}\ }\textbf {\bibinfo
  {volume} {112}},\ \bibinfo {pages} {12249} (\bibinfo {year} {2014})},\
  \Eprint {http://arxiv.org/abs/1306.0913} {arXiv:1306.0913 [astro-ph.CO]}
  \BibitemShut {NoStop}%
%%CITATION = ARXIV:1306.0913;%%
\bibitem [{\citenamefont {Grilli~di Cortona}\ \emph {et~al.}(2016)\citenamefont
  {Grilli~di Cortona}, \citenamefont {Hardy}, \citenamefont {Pardo~Vega},\ and\
  \citenamefont {Villadoro}}]{diCortona:2015ldu}%
  \BibitemOpen
  \bibfield  {author} {\bibinfo {author} {\bibfnamefont {G.}~\bibnamefont
  {Grilli~di Cortona}}, \bibinfo {author} {\bibfnamefont {E.}~\bibnamefont
  {Hardy}}, \bibinfo {author} {\bibfnamefont {J.}~\bibnamefont {Pardo~Vega}}, \
  and\ \bibinfo {author} {\bibfnamefont {G.}~\bibnamefont {Villadoro}},\ }\href
  {\doibase 10.1007/JHEP01(2016)034} {\bibfield  {journal} {\bibinfo  {journal}
  {JHEP}\ }\textbf {\bibinfo {volume} {01}},\ \bibinfo {pages} {034} (\bibinfo
  {year} {2016})},\ \Eprint {http://arxiv.org/abs/1511.02867} {arXiv:1511.02867
  [hep-ph]} \BibitemShut {NoStop}%
%%CITATION = ARXIV:1511.02867;%%
\bibitem [{\citenamefont {Borsanyi}\ \emph
  {et~al.}(2016{\natexlab{a}})\citenamefont {Borsanyi}, \citenamefont
  {Dierigl}, \citenamefont {Fodor}, \citenamefont {Katz}, \citenamefont
  {Mages}, \citenamefont {Nogradi}, \citenamefont {Redondo}, \citenamefont
  {Ringwald},\ and\ \citenamefont {Szabo}}]{Borsanyi:2015cka}%
  \BibitemOpen
  \bibfield  {author} {\bibinfo {author} {\bibfnamefont {S.}~\bibnamefont
  {Borsanyi}}, \bibinfo {author} {\bibfnamefont {M.}~\bibnamefont {Dierigl}},
  \bibinfo {author} {\bibfnamefont {Z.}~\bibnamefont {Fodor}}, \bibinfo
  {author} {\bibfnamefont {S.~D.}\ \bibnamefont {Katz}}, \bibinfo {author}
  {\bibfnamefont {S.~W.}\ \bibnamefont {Mages}}, \bibinfo {author}
  {\bibfnamefont {D.}~\bibnamefont {Nogradi}}, \bibinfo {author} {\bibfnamefont
  {J.}~\bibnamefont {Redondo}}, \bibinfo {author} {\bibfnamefont
  {A.}~\bibnamefont {Ringwald}}, \ and\ \bibinfo {author} {\bibfnamefont
  {K.~K.}\ \bibnamefont {Szabo}},\ }\href {\doibase
  10.1016/j.physletb.2015.11.020} {\bibfield  {journal} {\bibinfo  {journal}
  {Phys. Lett.}\ }\textbf {\bibinfo {volume} {B752}},\ \bibinfo {pages} {175}
  (\bibinfo {year} {2016}{\natexlab{a}})},\ \Eprint
  {http://arxiv.org/abs/1508.06917} {arXiv:1508.06917 [hep-lat]} \BibitemShut
  {NoStop}%
%%CITATION = ARXIV:1508.06917;%%
\bibitem [{\citenamefont {Borsanyi}\ \emph
  {et~al.}(2016{\natexlab{b}})\citenamefont {Borsanyi} \emph
  {et~al.}}]{Borsanyi:2016ksw}%
  \BibitemOpen
  \bibfield  {author} {\bibinfo {author} {\bibfnamefont {S.}~\bibnamefont
  {Borsanyi}} \emph {et~al.},\ }\href {\doibase 10.1038/nature20115} {\bibfield
   {journal} {\bibinfo  {journal} {Nature}\ }\textbf {\bibinfo {volume}
  {539}},\ \bibinfo {pages} {69} (\bibinfo {year} {2016}{\natexlab{b}})},\
  \Eprint {http://arxiv.org/abs/1606.07494} {arXiv:1606.07494 [hep-lat]}
  \BibitemShut {NoStop}%
%%CITATION = ARXIV:1606.07494;%%
\bibitem [{\citenamefont {Petreczky}\ \emph {et~al.}(2016)\citenamefont
  {Petreczky}, \citenamefont {Schadler},\ and\ \citenamefont
  {Sharma}}]{Petreczky:2016vrs}%
  \BibitemOpen
  \bibfield  {author} {\bibinfo {author} {\bibfnamefont {P.}~\bibnamefont
  {Petreczky}}, \bibinfo {author} {\bibfnamefont {H.-P.}\ \bibnamefont
  {Schadler}}, \ and\ \bibinfo {author} {\bibfnamefont {S.}~\bibnamefont
  {Sharma}},\ }\href {\doibase 10.1016/j.physletb.2016.09.063} {\bibfield
  {journal} {\bibinfo  {journal} {Phys. Lett.}\ }\textbf {\bibinfo {volume}
  {B762}},\ \bibinfo {pages} {498} (\bibinfo {year} {2016})},\ \Eprint
  {http://arxiv.org/abs/1606.03145} {arXiv:1606.03145 [hep-lat]} \BibitemShut
  {NoStop}%
%%CITATION = ARXIV:1606.03145;%%
\bibitem [{\citenamefont {Gross}\ \emph {et~al.}(1981)\citenamefont {Gross},
  \citenamefont {Pisarski},\ and\ \citenamefont {Yaffe}}]{Gross:53.43}%
  \BibitemOpen
  \bibfield  {author} {\bibinfo {author} {\bibfnamefont {D.~J.}\ \bibnamefont
  {Gross}}, \bibinfo {author} {\bibfnamefont {R.~D.}\ \bibnamefont {Pisarski}},
  \ and\ \bibinfo {author} {\bibfnamefont {L.~G.}\ \bibnamefont {Yaffe}},\
  }\href {\doibase 10.1103/RevModPhys.53.43} {\bibfield  {journal} {\bibinfo
  {journal} {Rev. Mod. Phys.}\ }\textbf {\bibinfo {volume} {53}},\ \bibinfo
  {pages} {43} (\bibinfo {year} {1981})}\BibitemShut {NoStop}%
\bibitem [{\citenamefont {Fox}\ \emph {et~al.}(2004)\citenamefont {Fox},
  \citenamefont {Pierce},\ and\ \citenamefont {Thomas}}]{Fox:2004kb}%
  \BibitemOpen
  \bibfield  {author} {\bibinfo {author} {\bibfnamefont {P.}~\bibnamefont
  {Fox}}, \bibinfo {author} {\bibfnamefont {A.}~\bibnamefont {Pierce}}, \ and\
  \bibinfo {author} {\bibfnamefont {S.~D.}\ \bibnamefont {Thomas}},\
  }\href@noop {} {\  (\bibinfo {year} {2004})},\ \Eprint
  {http://arxiv.org/abs/hep-th/0409059} {arXiv:hep-th/0409059 [hep-th]}
  \BibitemShut {NoStop}%
%%CITATION = HEP-TH/0409059;%%
\bibitem [{\citenamefont {Turner}(1986)}]{Turner:1986}%
  \BibitemOpen
  \bibfield  {author} {\bibinfo {author} {\bibfnamefont {M.~S.}\ \bibnamefont
  {Turner}},\ }\href {\doibase 10.1103/PhysRevD.33.889} {\bibfield  {journal}
  {\bibinfo  {journal} {Phys. Rev. D}\ }\textbf {\bibinfo {volume} {33}},\
  \bibinfo {pages} {889} (\bibinfo {year} {1986})}\BibitemShut {NoStop}%
\bibitem [{\citenamefont {Bae}\ \emph {et~al.}(2008)\citenamefont {Bae},
  \citenamefont {Huh},\ and\ \citenamefont {Kim}}]{Bae:2008ue}%
  \BibitemOpen
  \bibfield  {author} {\bibinfo {author} {\bibfnamefont {K.~J.}\ \bibnamefont
  {Bae}}, \bibinfo {author} {\bibfnamefont {J.-H.}\ \bibnamefont {Huh}}, \ and\
  \bibinfo {author} {\bibfnamefont {J.~E.}\ \bibnamefont {Kim}},\ }\href
  {\doibase 10.1088/1475-7516/2008/09/005} {\bibfield  {journal} {\bibinfo
  {journal} {JCAP}\ }\textbf {\bibinfo {volume} {0809}},\ \bibinfo {pages}
  {005} (\bibinfo {year} {2008})},\ \Eprint {http://arxiv.org/abs/0806.0497}
  {arXiv:0806.0497 [hep-ph]} \BibitemShut {NoStop}%
%%CITATION = ARXIV:0806.0497;%%
\bibitem [{\citenamefont {Davoudiasl}(2007)}]{Davoudiasl:2006bt}%
  \BibitemOpen
  \bibfield  {author} {\bibinfo {author} {\bibfnamefont {H.}~\bibnamefont
  {Davoudiasl}},\ }\href {\doibase 10.1016/j.physletb.2007.01.023} {\bibfield
  {journal} {\bibinfo  {journal} {Phys. Lett.}\ }\textbf {\bibinfo {volume}
  {B646}},\ \bibinfo {pages} {172} (\bibinfo {year} {2007})},\ \Eprint
  {http://arxiv.org/abs/hep-ph/0607111} {arXiv:hep-ph/0607111 [hep-ph]}
  \BibitemShut {NoStop}%
%%CITATION = HEP-PH/0607111;%%
\bibitem [{\citenamefont {Davoudiasl}\ and\ \citenamefont
  {Murphy}(2017)}]{Davoudiasl:2017jke}%
  \BibitemOpen
  \bibfield  {author} {\bibinfo {author} {\bibfnamefont {H.}~\bibnamefont
  {Davoudiasl}}\ and\ \bibinfo {author} {\bibfnamefont {C.~W.}\ \bibnamefont
  {Murphy}},\ }\href {\doibase 10.1103/PhysRevLett.118.141801} {\bibfield
  {journal} {\bibinfo  {journal} {Phys. Rev. Lett.}\ }\textbf {\bibinfo
  {volume} {118}},\ \bibinfo {pages} {141801} (\bibinfo {year} {2017})},\
  \Eprint {http://arxiv.org/abs/1701.01136} {arXiv:1701.01136 [hep-ph]}
  \BibitemShut {NoStop}%
%%CITATION = ARXIV:1701.01136;%%
\bibitem [{\citenamefont {Beltran}\ \emph {et~al.}(2007)\citenamefont
  {Beltran}, \citenamefont {Garcia-Bellido},\ and\ \citenamefont
  {Lesgourgues}}]{Beltran:2006sq}%
  \BibitemOpen
  \bibfield  {author} {\bibinfo {author} {\bibfnamefont {M.}~\bibnamefont
  {Beltran}}, \bibinfo {author} {\bibfnamefont {J.}~\bibnamefont
  {Garcia-Bellido}}, \ and\ \bibinfo {author} {\bibfnamefont {J.}~\bibnamefont
  {Lesgourgues}},\ }\href {\doibase 10.1103/PhysRevD.75.103507} {\bibfield
  {journal} {\bibinfo  {journal} {Phys. Rev.}\ }\textbf {\bibinfo {volume}
  {D75}},\ \bibinfo {pages} {103507} (\bibinfo {year} {2007})},\ \Eprint
  {http://arxiv.org/abs/hep-ph/0606107} {arXiv:hep-ph/0606107 [hep-ph]}
  \BibitemShut {NoStop}%
%%CITATION = HEP-PH/0606107;%%
\bibitem [{\citenamefont {Ade}\ \emph {et~al.}(2016{\natexlab{a}})\citenamefont
  {Ade} \emph {et~al.}}]{Ade:2015xua}%
  \BibitemOpen
  \bibfield  {author} {\bibinfo {author} {\bibfnamefont {P.~A.~R.}\
  \bibnamefont {Ade}} \emph {et~al.} (\bibinfo {collaboration} {Planck}),\
  }\href {\doibase 10.1051/0004-6361/201525830} {\bibfield  {journal} {\bibinfo
   {journal} {Astron. Astrophys.}\ }\textbf {\bibinfo {volume} {594}},\
  \bibinfo {pages} {A13} (\bibinfo {year} {2016}{\natexlab{a}})},\ \Eprint
  {http://arxiv.org/abs/1502.01589} {arXiv:1502.01589 [astro-ph.CO]}
  \BibitemShut {NoStop}%
%%CITATION = ARXIV:1502.01589;%%
\bibitem [{\citenamefont {Lyth}(1984)}]{Lyth:1984}%
  \BibitemOpen
  \bibfield  {author} {\bibinfo {author} {\bibfnamefont {D.}~\bibnamefont
  {Lyth}},\ }\href {\doibase http://dx.doi.org/10.1016/0370-2693(84)91391-1}
  {\bibfield  {journal} {\bibinfo  {journal} {Physics Letters B}\ }\textbf
  {\bibinfo {volume} {147}},\ \bibinfo {pages} {403 } (\bibinfo {year}
  {1984})}\BibitemShut {NoStop}%
\bibitem [{\citenamefont {Lyth}\ and\ \citenamefont
  {Stewart}(1992)}]{Lyth:1992yy}%
  \BibitemOpen
  \bibfield  {author} {\bibinfo {author} {\bibfnamefont {D.~H.}\ \bibnamefont
  {Lyth}}\ and\ \bibinfo {author} {\bibfnamefont {E.~D.}\ \bibnamefont
  {Stewart}},\ }\href {\doibase http://dx.doi.org/10.1016/0370-2693(92)90006-P}
  {\bibfield  {journal} {\bibinfo  {journal} {Physics Letters B}\ }\textbf
  {\bibinfo {volume} {283}},\ \bibinfo {pages} {189 } (\bibinfo {year}
  {1992})}\BibitemShut {NoStop}%
\bibitem [{\citenamefont {Ade}\ \emph {et~al.}(2014{\natexlab{a}})\citenamefont
  {Ade} \emph {et~al.}}]{Ade:2013zuv}%
  \BibitemOpen
  \bibfield  {author} {\bibinfo {author} {\bibfnamefont {P.~A.~R.}\
  \bibnamefont {Ade}} \emph {et~al.} (\bibinfo {collaboration} {Planck}),\
  }\href {\doibase 10.1051/0004-6361/201321591} {\bibfield  {journal} {\bibinfo
   {journal} {Astron. Astrophys.}\ }\textbf {\bibinfo {volume} {571}},\
  \bibinfo {pages} {A16} (\bibinfo {year} {2014}{\natexlab{a}})},\ \Eprint
  {http://arxiv.org/abs/1303.5076} {arXiv:1303.5076 [astro-ph.CO]} \BibitemShut
  {NoStop}%
%%CITATION = ARXIV:1303.5076;%%
\bibitem [{\citenamefont {Ade}\ \emph {et~al.}(2014{\natexlab{b}})\citenamefont
  {Ade} \emph {et~al.}}]{Planck:2013jfk}%
  \BibitemOpen
  \bibfield  {author} {\bibinfo {author} {\bibfnamefont {P.~A.~R.}\
  \bibnamefont {Ade}} \emph {et~al.} (\bibinfo {collaboration} {Planck}),\
  }\href {\doibase 10.1051/0004-6361/201321569} {\bibfield  {journal} {\bibinfo
   {journal} {Astron. Astrophys.}\ }\textbf {\bibinfo {volume} {571}},\
  \bibinfo {pages} {A22} (\bibinfo {year} {2014}{\natexlab{b}})},\ \Eprint
  {http://arxiv.org/abs/1303.5082} {arXiv:1303.5082 [astro-ph.CO]} \BibitemShut
  {NoStop}%
%%CITATION = ARXIV:1303.5082;%%
\bibitem [{\citenamefont {Barkats}\ \emph {et~al.}(2014)\citenamefont {Barkats}
  \emph {et~al.}}]{Barkats:2013jfa}%
  \BibitemOpen
  \bibfield  {author} {\bibinfo {author} {\bibfnamefont {D.}~\bibnamefont
  {Barkats}} \emph {et~al.} (\bibinfo {collaboration} {BICEP1}),\ }\href
  {\doibase 10.1088/0004-637X/783/2/67} {\bibfield  {journal} {\bibinfo
  {journal} {Astrophys. J.}\ }\textbf {\bibinfo {volume} {783}},\ \bibinfo
  {pages} {67} (\bibinfo {year} {2014})},\ \Eprint
  {http://arxiv.org/abs/1310.1422} {arXiv:1310.1422 [astro-ph.CO]} \BibitemShut
  {NoStop}%
%%CITATION = ARXIV:1310.1422;%%
\bibitem [{\citenamefont {Ade}\ \emph {et~al.}(2015)\citenamefont {Ade} \emph
  {et~al.}}]{Ade:2015tva}%
  \BibitemOpen
  \bibfield  {author} {\bibinfo {author} {\bibfnamefont {P.~A.~R.}\
  \bibnamefont {Ade}} \emph {et~al.} (\bibinfo {collaboration} {BICEP2,
  Planck}),\ }\href {\doibase 10.1103/PhysRevLett.114.101301} {\bibfield
  {journal} {\bibinfo  {journal} {Phys. Rev. Lett.}\ }\textbf {\bibinfo
  {volume} {114}},\ \bibinfo {pages} {101301} (\bibinfo {year} {2015})},\
  \Eprint {http://arxiv.org/abs/1502.00612} {arXiv:1502.00612 [astro-ph.CO]}
  \BibitemShut {NoStop}%
%%CITATION = ARXIV:1502.00612;%%
\bibitem [{\citenamefont {Ade}\ \emph {et~al.}(2016{\natexlab{b}})\citenamefont
  {Ade} \emph {et~al.}}]{Array:2015xqh}%
  \BibitemOpen
  \bibfield  {author} {\bibinfo {author} {\bibfnamefont {P.~A.~R.}\
  \bibnamefont {Ade}} \emph {et~al.} (\bibinfo {collaboration} {BICEP2, Keck
  Array}),\ }\href {\doibase 10.1103/PhysRevLett.116.031302} {\bibfield
  {journal} {\bibinfo  {journal} {Phys. Rev. Lett.}\ }\textbf {\bibinfo
  {volume} {116}},\ \bibinfo {pages} {031302} (\bibinfo {year}
  {2016}{\natexlab{b}})},\ \Eprint {http://arxiv.org/abs/1510.09217}
  {arXiv:1510.09217 [astro-ph.CO]} \BibitemShut {NoStop}%
%%CITATION = ARXIV:1510.09217;%%
\bibitem [{\citenamefont {Lyth}(1990)}]{Lyth:1990}%
  \BibitemOpen
  \bibfield  {author} {\bibinfo {author} {\bibfnamefont {D.~H.}\ \bibnamefont
  {Lyth}},\ }\href {\doibase http://dx.doi.org/10.1016/0370-2693(90)90374-F}
  {\bibfield  {journal} {\bibinfo  {journal} {Physics Letters B}\ }\textbf
  {\bibinfo {volume} {236}},\ \bibinfo {pages} {408 } (\bibinfo {year}
  {1990})}\BibitemShut {NoStop}%
\bibitem [{\citenamefont {Kolb}\ and\ \citenamefont
  {Turner}(1994)}]{kolb:1994early}%
  \BibitemOpen
  \bibfield  {author} {\bibinfo {author} {\bibfnamefont {E.}~\bibnamefont
  {Kolb}}\ and\ \bibinfo {author} {\bibfnamefont {M.}~\bibnamefont {Turner}},\
  }\href {https://books.google.se/books?id=E58\_BAAAQBAJ} {\emph {\bibinfo
  {title} {The Early Universe}}},\ Frontiers in physics\ (\bibinfo  {publisher}
  {Avalon Publishing},\ \bibinfo {year} {1994})\BibitemShut {NoStop}%
\bibitem [{\citenamefont {Kobayashi}\ \emph {et~al.}(2013)\citenamefont
  {Kobayashi}, \citenamefont {Kurematsu},\ and\ \citenamefont
  {Takahashi}}]{Kobayashi:2013nva}%
  \BibitemOpen
  \bibfield  {author} {\bibinfo {author} {\bibfnamefont {T.}~\bibnamefont
  {Kobayashi}}, \bibinfo {author} {\bibfnamefont {R.}~\bibnamefont
  {Kurematsu}}, \ and\ \bibinfo {author} {\bibfnamefont {F.}~\bibnamefont
  {Takahashi}},\ }\href {\doibase 10.1088/1475-7516/2013/09/032} {\bibfield
  {journal} {\bibinfo  {journal} {JCAP}\ }\textbf {\bibinfo {volume} {1309}},\
  \bibinfo {pages} {032} (\bibinfo {year} {2013})},\ \Eprint
  {http://arxiv.org/abs/1304.0922} {arXiv:1304.0922 [hep-ph]} \BibitemShut
  {NoStop}%
%%CITATION = ARXIV:1304.0922;%%
\bibitem [{\citenamefont {Axenides}\ \emph {et~al.}(1983)\citenamefont
  {Axenides}, \citenamefont {Brandenberger},\ and\ \citenamefont
  {Turner}}]{Axenides:1983}%
  \BibitemOpen
  \bibfield  {author} {\bibinfo {author} {\bibfnamefont {M.}~\bibnamefont
  {Axenides}}, \bibinfo {author} {\bibfnamefont {R.}~\bibnamefont
  {Brandenberger}}, \ and\ \bibinfo {author} {\bibfnamefont {M.}~\bibnamefont
  {Turner}},\ }\href {\doibase http://dx.doi.org/10.1016/0370-2693(83)90586-5}
  {\bibfield  {journal} {\bibinfo  {journal} {Physics Letters B}\ }\textbf
  {\bibinfo {volume} {126}},\ \bibinfo {pages} {178 } (\bibinfo {year}
  {1983})}\BibitemShut {NoStop}%
\bibitem [{\citenamefont {Linde}(1985)}]{Linde:1985yf}%
  \BibitemOpen
  \bibfield  {author} {\bibinfo {author} {\bibfnamefont {A.~D.}\ \bibnamefont
  {Linde}},\ }\href {\doibase 10.1016/0370-2693(85)90436-8} {\bibfield
  {journal} {\bibinfo  {journal} {Phys. Lett.}\ }\textbf {\bibinfo {volume}
  {B158}},\ \bibinfo {pages} {375} (\bibinfo {year} {1985})}\BibitemShut
  {NoStop}%
%%CITATION = PHLTA,B158,375;%%
\bibitem [{\citenamefont {Seckel}\ and\ \citenamefont
  {Turner}(1985)}]{Seckel:1985}%
  \BibitemOpen
  \bibfield  {author} {\bibinfo {author} {\bibfnamefont {D.}~\bibnamefont
  {Seckel}}\ and\ \bibinfo {author} {\bibfnamefont {M.~S.}\ \bibnamefont
  {Turner}},\ }\href {\doibase 10.1103/PhysRevD.32.3178} {\bibfield  {journal}
  {\bibinfo  {journal} {Phys. Rev. D}\ }\textbf {\bibinfo {volume} {32}},\
  \bibinfo {pages} {3178} (\bibinfo {year} {1985})}\BibitemShut {NoStop}%
\bibitem [{\citenamefont {Crotty}\ \emph {et~al.}(2003)\citenamefont {Crotty},
  \citenamefont {Garcia-Bellido}, \citenamefont {Lesgourgues},\ and\
  \citenamefont {Riazuelo}}]{Crotty:2003rz}%
  \BibitemOpen
  \bibfield  {author} {\bibinfo {author} {\bibfnamefont {P.}~\bibnamefont
  {Crotty}}, \bibinfo {author} {\bibfnamefont {J.}~\bibnamefont
  {Garcia-Bellido}}, \bibinfo {author} {\bibfnamefont {J.}~\bibnamefont
  {Lesgourgues}}, \ and\ \bibinfo {author} {\bibfnamefont {A.}~\bibnamefont
  {Riazuelo}},\ }\href {\doibase 10.1103/PhysRevLett.91.171301} {\bibfield
  {journal} {\bibinfo  {journal} {Phys. Rev. Lett.}\ }\textbf {\bibinfo
  {volume} {91}},\ \bibinfo {pages} {171301} (\bibinfo {year} {2003})},\
  \Eprint {http://arxiv.org/abs/astro-ph/0306286} {arXiv:astro-ph/0306286
  [astro-ph]} \BibitemShut {NoStop}%
%%CITATION = ASTRO-PH/0306286;%%
\bibitem [{\citenamefont {Beltran}\ \emph {et~al.}(2005)\citenamefont
  {Beltran}, \citenamefont {Garcia-Bellido}, \citenamefont {Lesgourgues},
  \citenamefont {Liddle},\ and\ \citenamefont {Slosar}}]{Beltran:2005xd}%
  \BibitemOpen
  \bibfield  {author} {\bibinfo {author} {\bibfnamefont {M.}~\bibnamefont
  {Beltran}}, \bibinfo {author} {\bibfnamefont {J.}~\bibnamefont
  {Garcia-Bellido}}, \bibinfo {author} {\bibfnamefont {J.}~\bibnamefont
  {Lesgourgues}}, \bibinfo {author} {\bibfnamefont {A.~R.}\ \bibnamefont
  {Liddle}}, \ and\ \bibinfo {author} {\bibfnamefont {A.}~\bibnamefont
  {Slosar}},\ }\href {\doibase 10.1103/PhysRevD.71.063532} {\bibfield
  {journal} {\bibinfo  {journal} {Phys. Rev.}\ }\textbf {\bibinfo {volume}
  {D71}},\ \bibinfo {pages} {063532} (\bibinfo {year} {2005})},\ \Eprint
  {http://arxiv.org/abs/astro-ph/0501477} {arXiv:astro-ph/0501477 [astro-ph]}
  \BibitemShut {NoStop}%
%%CITATION = ASTRO-PH/0501477;%%
\bibitem [{\citenamefont {Abbott}\ and\ \citenamefont
  {Sikivie}(1983)}]{Abbott:1982af}%
  \BibitemOpen
  \bibfield  {author} {\bibinfo {author} {\bibfnamefont {L.~F.}\ \bibnamefont
  {Abbott}}\ and\ \bibinfo {author} {\bibfnamefont {P.}~\bibnamefont
  {Sikivie}},\ }\href {\doibase 10.1016/0370-2693(83)90638-X} {\bibfield
  {journal} {\bibinfo  {journal} {Phys. Lett.}\ }\textbf {\bibinfo {volume}
  {B120}},\ \bibinfo {pages} {133} (\bibinfo {year} {1983})}\BibitemShut
  {NoStop}%
%%CITATION = PHLTA,B120,133;%%
\bibitem [{\citenamefont {Dine}\ and\ \citenamefont
  {Fischler}(1983)}]{Dine:1982ah}%
  \BibitemOpen
  \bibfield  {author} {\bibinfo {author} {\bibfnamefont {M.}~\bibnamefont
  {Dine}}\ and\ \bibinfo {author} {\bibfnamefont {W.}~\bibnamefont
  {Fischler}},\ }\href {\doibase 10.1016/0370-2693(83)90639-1} {\bibfield
  {journal} {\bibinfo  {journal} {Phys. Lett.}\ }\textbf {\bibinfo {volume}
  {B120}},\ \bibinfo {pages} {137} (\bibinfo {year} {1983})}\BibitemShut
  {NoStop}%
%%CITATION = PHLTA,B120,137;%%
\bibitem [{\citenamefont {Preskill}\ \emph {et~al.}(1983)\citenamefont
  {Preskill}, \citenamefont {Wise},\ and\ \citenamefont
  {Wilczek}}]{Preskill:1982cy}%
  \BibitemOpen
  \bibfield  {author} {\bibinfo {author} {\bibfnamefont {J.}~\bibnamefont
  {Preskill}}, \bibinfo {author} {\bibfnamefont {M.~B.}\ \bibnamefont {Wise}},
  \ and\ \bibinfo {author} {\bibfnamefont {F.}~\bibnamefont {Wilczek}},\ }\href
  {\doibase 10.1016/0370-2693(83)90637-8} {\bibfield  {journal} {\bibinfo
  {journal} {Phys. Lett.}\ }\textbf {\bibinfo {volume} {B120}},\ \bibinfo
  {pages} {127} (\bibinfo {year} {1983})}\BibitemShut {NoStop}%
%%CITATION = PHLTA,B120,127;%%
\bibitem [{\citenamefont {{Kibble}}(1976)}]{Kibble:1976}%
  \BibitemOpen
  \bibfield  {author} {\bibinfo {author} {\bibfnamefont {T.~W.~B.}\
  \bibnamefont {{Kibble}}},\ }\href {\doibase 10.1088/0305-4470/9/8/029}
  {\bibfield  {journal} {\bibinfo  {journal} {Journal of Physics A Mathematical
  General}\ }\textbf {\bibinfo {volume} {9}},\ \bibinfo {pages} {1387}
  (\bibinfo {year} {1976})}\BibitemShut {NoStop}%
\bibitem [{\citenamefont {Coleman}\ and\ \citenamefont
  {Roos}(2003)}]{Coleman:2003hs}%
  \BibitemOpen
  \bibfield  {author} {\bibinfo {author} {\bibfnamefont {T.~S.}\ \bibnamefont
  {Coleman}}\ and\ \bibinfo {author} {\bibfnamefont {M.}~\bibnamefont {Roos}},\
  }\href {\doibase 10.1103/PhysRevD.68.027702} {\bibfield  {journal} {\bibinfo
  {journal} {Phys. Rev.}\ }\textbf {\bibinfo {volume} {D68}},\ \bibinfo {pages}
  {027702} (\bibinfo {year} {2003})},\ \Eprint
  {http://arxiv.org/abs/astro-ph/0304281} {arXiv:astro-ph/0304281 [astro-ph]}
  \BibitemShut {NoStop}%
%%CITATION = ASTRO-PH/0304281;%%
\bibitem [{\citenamefont {Lyth}(1992)}]{Lyth:1992}%
  \BibitemOpen
  \bibfield  {author} {\bibinfo {author} {\bibfnamefont {D.~H.}\ \bibnamefont
  {Lyth}},\ }\href {\doibase 10.1103/PhysRevD.45.3394} {\bibfield  {journal}
  {\bibinfo  {journal} {Phys. Rev. D}\ }\textbf {\bibinfo {volume} {45}},\
  \bibinfo {pages} {3394} (\bibinfo {year} {1992})}\BibitemShut {NoStop}%
\bibitem [{\citenamefont {Strobl}\ and\ \citenamefont
  {Weiler}(1994)}]{Strobl:1994wk}%
  \BibitemOpen
  \bibfield  {author} {\bibinfo {author} {\bibfnamefont {K.}~\bibnamefont
  {Strobl}}\ and\ \bibinfo {author} {\bibfnamefont {T.~J.}\ \bibnamefont
  {Weiler}},\ }\href {\doibase 10.1103/PhysRevD.50.7690} {\bibfield  {journal}
  {\bibinfo  {journal} {Phys. Rev.}\ }\textbf {\bibinfo {volume} {D50}},\
  \bibinfo {pages} {7690} (\bibinfo {year} {1994})},\ \Eprint
  {http://arxiv.org/abs/astro-ph/9405028} {arXiv:astro-ph/9405028 [astro-ph]}
  \BibitemShut {NoStop}%
%%CITATION = ASTRO-PH/9405028;%%
\bibitem [{\citenamefont {Visinelli}\ and\ \citenamefont
  {Gondolo}(2010)}]{Visinelli:2009kt}%
  \BibitemOpen
  \bibfield  {author} {\bibinfo {author} {\bibfnamefont {L.}~\bibnamefont
  {Visinelli}}\ and\ \bibinfo {author} {\bibfnamefont {P.}~\bibnamefont
  {Gondolo}},\ }\href {\doibase 10.1103/PhysRevD.81.063508} {\bibfield
  {journal} {\bibinfo  {journal} {Phys. Rev.}\ }\textbf {\bibinfo {volume}
  {D81}},\ \bibinfo {pages} {063508} (\bibinfo {year} {2010})},\ \Eprint
  {http://arxiv.org/abs/0912.0015} {arXiv:0912.0015 [astro-ph.CO]} \BibitemShut
  {NoStop}%
%%CITATION = ARXIV:0912.0015;%%
\bibitem [{\citenamefont {Gondolo}\ and\ \citenamefont
  {Visinelli}(2014)}]{Visinelli:2014twa}%
  \BibitemOpen
  \bibfield  {author} {\bibinfo {author} {\bibfnamefont {P.}~\bibnamefont
  {Gondolo}}\ and\ \bibinfo {author} {\bibfnamefont {L.}~\bibnamefont
  {Visinelli}},\ }\href {\doibase 10.1103/PhysRevLett.113.011802} {\bibfield
  {journal} {\bibinfo  {journal} {Phys. Rev. Lett.}\ }\textbf {\bibinfo
  {volume} {113}},\ \bibinfo {pages} {011802} (\bibinfo {year} {2014})},\
  \Eprint {http://arxiv.org/abs/1403.4594} {arXiv:1403.4594 [hep-ph]}
  \BibitemShut {NoStop}%
%%CITATION = ARXIV:1403.4594;%%
\bibitem [{\citenamefont {Kitajima}\ and\ \citenamefont
  {Takahashi}(2015)}]{Kitajima:2014xla}%
  \BibitemOpen
  \bibfield  {author} {\bibinfo {author} {\bibfnamefont {N.}~\bibnamefont
  {Kitajima}}\ and\ \bibinfo {author} {\bibfnamefont {F.}~\bibnamefont
  {Takahashi}},\ }\href {\doibase 10.1088/1475-7516/2015/01/032} {\bibfield
  {journal} {\bibinfo  {journal} {JCAP}\ }\textbf {\bibinfo {volume} {1501}},\
  \bibinfo {pages} {032} (\bibinfo {year} {2015})},\ \Eprint
  {http://arxiv.org/abs/1411.2011} {arXiv:1411.2011 [hep-ph]} \BibitemShut
  {NoStop}%
%%CITATION = ARXIV:1411.2011;%%
\bibitem [{\citenamefont {Arias}\ \emph {et~al.}(2012)\citenamefont {Arias},
  \citenamefont {Cadamuro}, \citenamefont {Goodsell}, \citenamefont {Jaeckel},
  \citenamefont {Redondo},\ and\ \citenamefont {Ringwald}}]{Arias:2012az}%
  \BibitemOpen
  \bibfield  {author} {\bibinfo {author} {\bibfnamefont {P.}~\bibnamefont
  {Arias}}, \bibinfo {author} {\bibfnamefont {D.}~\bibnamefont {Cadamuro}},
  \bibinfo {author} {\bibfnamefont {M.}~\bibnamefont {Goodsell}}, \bibinfo
  {author} {\bibfnamefont {J.}~\bibnamefont {Jaeckel}}, \bibinfo {author}
  {\bibfnamefont {J.}~\bibnamefont {Redondo}}, \ and\ \bibinfo {author}
  {\bibfnamefont {A.}~\bibnamefont {Ringwald}},\ }\href {\doibase
  10.1088/1475-7516/2012/06/013} {\bibfield  {journal} {\bibinfo  {journal}
  {JCAP}\ }\textbf {\bibinfo {volume} {1206}},\ \bibinfo {pages} {013}
  (\bibinfo {year} {2012})},\ \Eprint {http://arxiv.org/abs/1201.5902}
  {arXiv:1201.5902 [hep-ph]} \BibitemShut {NoStop}%
%%CITATION = ARXIV:1201.5902;%%
\bibitem [{\citenamefont {Fairbairn}\ \emph {et~al.}(2015)\citenamefont
  {Fairbairn}, \citenamefont {Hogan},\ and\ \citenamefont
  {Marsh}}]{Fairbairn:2014zta}%
  \BibitemOpen
  \bibfield  {author} {\bibinfo {author} {\bibfnamefont {M.}~\bibnamefont
  {Fairbairn}}, \bibinfo {author} {\bibfnamefont {R.}~\bibnamefont {Hogan}}, \
  and\ \bibinfo {author} {\bibfnamefont {D.~J.~E.}\ \bibnamefont {Marsh}},\
  }\href {\doibase 10.1103/PhysRevD.91.023509} {\bibfield  {journal} {\bibinfo
  {journal} {Phys. Rev.}\ }\textbf {\bibinfo {volume} {D91}},\ \bibinfo {pages}
  {023509} (\bibinfo {year} {2015})},\ \Eprint {http://arxiv.org/abs/1410.1752}
  {arXiv:1410.1752 [hep-ph]} \BibitemShut {NoStop}%
%%CITATION = ARXIV:1410.1752;%%
\bibitem [{\citenamefont {Ballesteros}\ \emph {et~al.}(2017)\citenamefont
  {Ballesteros}, \citenamefont {Redondo}, \citenamefont {Ringwald},\ and\
  \citenamefont {Tamarit}}]{Ballesteros:2016xej}%
  \BibitemOpen
  \bibfield  {author} {\bibinfo {author} {\bibfnamefont {G.}~\bibnamefont
  {Ballesteros}}, \bibinfo {author} {\bibfnamefont {J.}~\bibnamefont
  {Redondo}}, \bibinfo {author} {\bibfnamefont {A.}~\bibnamefont {Ringwald}}, \
  and\ \bibinfo {author} {\bibfnamefont {C.}~\bibnamefont {Tamarit}},\ }\href
  {\doibase 10.1088/1475-7516/2017/08/001} {\bibfield  {journal} {\bibinfo
  {journal} {JCAP}\ }\textbf {\bibinfo {volume} {1708}},\ \bibinfo {pages}
  {001} (\bibinfo {year} {2017})},\ \Eprint {http://arxiv.org/abs/1610.01639}
  {arXiv:1610.01639 [hep-ph]} \BibitemShut {NoStop}%
%%CITATION = ARXIV:1610.01639;%%
\bibitem [{\citenamefont {Silverstein}\ and\ \citenamefont
  {Westphal}(2008)}]{Silverstein:2008sg}%
  \BibitemOpen
  \bibfield  {author} {\bibinfo {author} {\bibfnamefont {E.}~\bibnamefont
  {Silverstein}}\ and\ \bibinfo {author} {\bibfnamefont {A.}~\bibnamefont
  {Westphal}},\ }\href {\doibase 10.1103/PhysRevD.78.106003} {\bibfield
  {journal} {\bibinfo  {journal} {Phys. Rev.}\ }\textbf {\bibinfo {volume}
  {D78}},\ \bibinfo {pages} {106003} (\bibinfo {year} {2008})},\ \Eprint
  {http://arxiv.org/abs/0803.3085} {arXiv:0803.3085 [hep-th]} \BibitemShut
  {NoStop}%
%%CITATION = ARXIV:0803.3085;%%
\bibitem [{\citenamefont {McAllister}\ \emph {et~al.}(2010)\citenamefont
  {McAllister}, \citenamefont {Silverstein},\ and\ \citenamefont
  {Westphal}}]{McAllister:2008hb}%
  \BibitemOpen
  \bibfield  {author} {\bibinfo {author} {\bibfnamefont {L.}~\bibnamefont
  {McAllister}}, \bibinfo {author} {\bibfnamefont {E.}~\bibnamefont
  {Silverstein}}, \ and\ \bibinfo {author} {\bibfnamefont {A.}~\bibnamefont
  {Westphal}},\ }\href {\doibase 10.1103/PhysRevD.82.046003} {\bibfield
  {journal} {\bibinfo  {journal} {Phys. Rev.}\ }\textbf {\bibinfo {volume}
  {D82}},\ \bibinfo {pages} {046003} (\bibinfo {year} {2010})},\ \Eprint
  {http://arxiv.org/abs/0808.0706} {arXiv:0808.0706 [hep-th]} \BibitemShut
  {NoStop}%
%%CITATION = ARXIV:0808.0706;%%
\bibitem [{\citenamefont {Kaloper}\ and\ \citenamefont
  {Sorbo}(2009)}]{Kaloper:2008fb}%
  \BibitemOpen
  \bibfield  {author} {\bibinfo {author} {\bibfnamefont {N.}~\bibnamefont
  {Kaloper}}\ and\ \bibinfo {author} {\bibfnamefont {L.}~\bibnamefont
  {Sorbo}},\ }\href {\doibase 10.1103/PhysRevLett.102.121301} {\bibfield
  {journal} {\bibinfo  {journal} {Phys. Rev. Lett.}\ }\textbf {\bibinfo
  {volume} {102}},\ \bibinfo {pages} {121301} (\bibinfo {year} {2009})},\
  \Eprint {http://arxiv.org/abs/0811.1989} {arXiv:0811.1989 [hep-th]}
  \BibitemShut {NoStop}%
%%CITATION = ARXIV:0811.1989;%%
\bibitem [{\citenamefont {Jaeckel}\ \emph {et~al.}(2017)\citenamefont
  {Jaeckel}, \citenamefont {Mehta},\ and\ \citenamefont
  {Witkowski}}]{Jaeckel:2016qjp}%
  \BibitemOpen
  \bibfield  {author} {\bibinfo {author} {\bibfnamefont {J.}~\bibnamefont
  {Jaeckel}}, \bibinfo {author} {\bibfnamefont {V.~M.}\ \bibnamefont {Mehta}},
  \ and\ \bibinfo {author} {\bibfnamefont {L.~T.}\ \bibnamefont {Witkowski}},\
  }\href {\doibase 10.1088/1475-7516/2017/01/036} {\bibfield  {journal}
  {\bibinfo  {journal} {JCAP}\ }\textbf {\bibinfo {volume} {1701}},\ \bibinfo
  {pages} {036} (\bibinfo {year} {2017})},\ \Eprint
  {http://arxiv.org/abs/1605.01367} {arXiv:1605.01367 [hep-ph]} \BibitemShut
  {NoStop}%
%%CITATION = ARXIV:1605.01367;%%
\bibitem [{\citenamefont {Kawasaki}\ \emph {et~al.}(1999)\citenamefont
  {Kawasaki}, \citenamefont {Kohri},\ and\ \citenamefont
  {Sugiyama}}]{Kawasaki:1999na}%
  \BibitemOpen
  \bibfield  {author} {\bibinfo {author} {\bibfnamefont {M.}~\bibnamefont
  {Kawasaki}}, \bibinfo {author} {\bibfnamefont {K.}~\bibnamefont {Kohri}}, \
  and\ \bibinfo {author} {\bibfnamefont {N.}~\bibnamefont {Sugiyama}},\ }\href
  {\doibase 10.1103/PhysRevLett.82.4168} {\bibfield  {journal} {\bibinfo
  {journal} {Phys. Rev. Lett.}\ }\textbf {\bibinfo {volume} {82}},\ \bibinfo
  {pages} {4168} (\bibinfo {year} {1999})},\ \Eprint
  {http://arxiv.org/abs/astro-ph/9811437} {arXiv:astro-ph/9811437} \BibitemShut
  {NoStop}%
%%CITATION = ASTRO-PH/9811437;%%
\bibitem [{\citenamefont {Kawasaki}\ \emph {et~al.}(2000)\citenamefont
  {Kawasaki}, \citenamefont {Kohri},\ and\ \citenamefont
  {Sugiyama}}]{Kawasaki:2000en}%
  \BibitemOpen
  \bibfield  {author} {\bibinfo {author} {\bibfnamefont {M.}~\bibnamefont
  {Kawasaki}}, \bibinfo {author} {\bibfnamefont {K.}~\bibnamefont {Kohri}}, \
  and\ \bibinfo {author} {\bibfnamefont {N.}~\bibnamefont {Sugiyama}},\ }\href
  {\doibase 10.1103/PhysRevD.62.023506} {\bibfield  {journal} {\bibinfo
  {journal} {Phys. Rev.}\ }\textbf {\bibinfo {volume} {D62}},\ \bibinfo {pages}
  {023506} (\bibinfo {year} {2000})},\ \Eprint
  {http://arxiv.org/abs/astro-ph/0002127} {arXiv:astro-ph/0002127} \BibitemShut
  {NoStop}%
%%CITATION = ASTRO-PH/0002127;%%
\bibitem [{\citenamefont {Hannestad}(2004)}]{Hannestad:2004px}%
  \BibitemOpen
  \bibfield  {author} {\bibinfo {author} {\bibfnamefont {S.}~\bibnamefont
  {Hannestad}},\ }\href {\doibase 10.1103/PhysRevD.70.043506} {\bibfield
  {journal} {\bibinfo  {journal} {Phys. Rev.}\ }\textbf {\bibinfo {volume}
  {D70}},\ \bibinfo {pages} {043506} (\bibinfo {year} {2004})},\ \Eprint
  {http://arxiv.org/abs/astro-ph/0403291} {arXiv:astro-ph/0403291} \BibitemShut
  {NoStop}%
%%CITATION = ASTRO-PH/0403291;%%
\bibitem [{\citenamefont {Ichikawa}\ \emph {et~al.}(2005)\citenamefont
  {Ichikawa}, \citenamefont {Kawasaki},\ and\ \citenamefont
  {Takahashi}}]{Ichikawa:2005vw}%
  \BibitemOpen
  \bibfield  {author} {\bibinfo {author} {\bibfnamefont {K.}~\bibnamefont
  {Ichikawa}}, \bibinfo {author} {\bibfnamefont {M.}~\bibnamefont {Kawasaki}},
  \ and\ \bibinfo {author} {\bibfnamefont {F.}~\bibnamefont {Takahashi}},\
  }\href {\doibase 10.1103/PhysRevD.72.043522} {\bibfield  {journal} {\bibinfo
  {journal} {Phys. Rev.}\ }\textbf {\bibinfo {volume} {D72}},\ \bibinfo {pages}
  {043522} (\bibinfo {year} {2005})},\ \Eprint
  {http://arxiv.org/abs/astro-ph/0505395} {arXiv:astro-ph/0505395} \BibitemShut
  {NoStop}%
%%CITATION = ASTRO-PH/0505395;%%
\bibitem [{\citenamefont {Bernardis}\ \emph {et~al.}(2008)\citenamefont
  {Bernardis}, \citenamefont {Pagano},\ and\ \citenamefont
  {Melchiorri}}]{DeBernardis:2008zz}%
  \BibitemOpen
  \bibfield  {author} {\bibinfo {author} {\bibfnamefont {F.~D.}\ \bibnamefont
  {Bernardis}}, \bibinfo {author} {\bibfnamefont {L.}~\bibnamefont {Pagano}}, \
  and\ \bibinfo {author} {\bibfnamefont {A.}~\bibnamefont {Melchiorri}},\
  }\href {\doibase 10.1016/j.astropartphys.2008.09.005} {\bibfield  {journal}
  {\bibinfo  {journal} {Astropart. Phys.}\ }\textbf {\bibinfo {volume} {30}},\
  \bibinfo {pages} {192} (\bibinfo {year} {2008})}\BibitemShut {NoStop}%
%%CITATION = APHYE,30,192;%%
\bibitem [{\citenamefont {Steinhardt}\ and\ \citenamefont
  {Turner}(1983)}]{Steinhardt:1983ia}%
  \BibitemOpen
  \bibfield  {author} {\bibinfo {author} {\bibfnamefont {P.~J.}\ \bibnamefont
  {Steinhardt}}\ and\ \bibinfo {author} {\bibfnamefont {M.~S.}\ \bibnamefont
  {Turner}},\ }\href {\doibase 10.1016/0370-2693(83)90727-X} {\bibfield
  {journal} {\bibinfo  {journal} {Phys. Lett.}\ }\textbf {\bibinfo {volume}
  {B129}},\ \bibinfo {pages} {51} (\bibinfo {year} {1983})}\BibitemShut
  {NoStop}%
%%CITATION = PHLTA,B129,51;%%
\bibitem [{\citenamefont {Turner}(1983)}]{Turner:1983}%
  \BibitemOpen
  \bibfield  {author} {\bibinfo {author} {\bibfnamefont {M.~S.}\ \bibnamefont
  {Turner}},\ }\href {\doibase 10.1103/PhysRevD.28.1243} {\bibfield  {journal}
  {\bibinfo  {journal} {Phys. Rev. D}\ }\textbf {\bibinfo {volume} {28}},\
  \bibinfo {pages} {1243} (\bibinfo {year} {1983})}\BibitemShut {NoStop}%
\bibitem [{\citenamefont {Scherrer}\ and\ \citenamefont
  {Turner}(1985)}]{Scherrer:1985}%
  \BibitemOpen
  \bibfield  {author} {\bibinfo {author} {\bibfnamefont {R.~J.}\ \bibnamefont
  {Scherrer}}\ and\ \bibinfo {author} {\bibfnamefont {M.~S.}\ \bibnamefont
  {Turner}},\ }\href {\doibase 10.1103/PhysRevD.31.681} {\bibfield  {journal}
  {\bibinfo  {journal} {Phys. Rev. D}\ }\textbf {\bibinfo {volume} {31}},\
  \bibinfo {pages} {681} (\bibinfo {year} {1985})}\BibitemShut {NoStop}%
\bibitem [{\citenamefont {Giudice}\ \emph {et~al.}(2001)\citenamefont
  {Giudice}, \citenamefont {Kolb},\ and\ \citenamefont
  {Riotto}}]{Giudice:2000ex}%
  \BibitemOpen
  \bibfield  {author} {\bibinfo {author} {\bibfnamefont {G.~F.}\ \bibnamefont
  {Giudice}}, \bibinfo {author} {\bibfnamefont {E.~W.}\ \bibnamefont {Kolb}}, \
  and\ \bibinfo {author} {\bibfnamefont {A.}~\bibnamefont {Riotto}},\ }\href
  {\doibase 10.1103/PhysRevD.64.023508} {\bibfield  {journal} {\bibinfo
  {journal} {Phys. Rev.}\ }\textbf {\bibinfo {volume} {D64}},\ \bibinfo {pages}
  {023508} (\bibinfo {year} {2001})},\ \Eprint
  {http://arxiv.org/abs/hep-ph/0005123} {arXiv:hep-ph/0005123 [hep-ph]}
  \BibitemShut {NoStop}%
%%CITATION = HEP-PH/0005123;%%
\bibitem [{\citenamefont {Lazarides}\ \emph {et~al.}(1987)\citenamefont
  {Lazarides}, \citenamefont {Panagiotakopoulos},\ and\ \citenamefont
  {Shafi}}]{Lazarides:1987zf}%
  \BibitemOpen
  \bibfield  {author} {\bibinfo {author} {\bibfnamefont {G.}~\bibnamefont
  {Lazarides}}, \bibinfo {author} {\bibfnamefont {C.}~\bibnamefont
  {Panagiotakopoulos}}, \ and\ \bibinfo {author} {\bibfnamefont
  {Q.}~\bibnamefont {Shafi}},\ }\href {\doibase 10.1016/0370-2693(87)90115-8}
  {\bibfield  {journal} {\bibinfo  {journal} {Phys. Lett.}\ }\textbf {\bibinfo
  {volume} {B192}},\ \bibinfo {pages} {323} (\bibinfo {year}
  {1987})}\BibitemShut {NoStop}%
%%CITATION = PHLTA,B192,323;%%
\bibitem [{\citenamefont {Lazarides}\ \emph {et~al.}(1990)\citenamefont
  {Lazarides}, \citenamefont {Schaefer}, \citenamefont {Seckel},\ and\
  \citenamefont {Shafi}}]{Lazarides:1990xp}%
  \BibitemOpen
  \bibfield  {author} {\bibinfo {author} {\bibfnamefont {G.}~\bibnamefont
  {Lazarides}}, \bibinfo {author} {\bibfnamefont {R.~K.}\ \bibnamefont
  {Schaefer}}, \bibinfo {author} {\bibfnamefont {D.}~\bibnamefont {Seckel}}, \
  and\ \bibinfo {author} {\bibfnamefont {Q.}~\bibnamefont {Shafi}},\ }\href
  {\doibase 10.1016/0550-3213(90)90244-8} {\bibfield  {journal} {\bibinfo
  {journal} {Nucl. Phys.}\ }\textbf {\bibinfo {volume} {B346}},\ \bibinfo
  {pages} {193} (\bibinfo {year} {1990})}\BibitemShut {NoStop}%
%%CITATION = NUPHA,B346,193;%%
\bibitem [{\citenamefont {Visinelli}(2018)}]{Visinelli:2017qga}%
  \BibitemOpen
  \bibfield  {author} {\bibinfo {author} {\bibfnamefont {L.}~\bibnamefont
  {Visinelli}},\ }\href {\doibase 10.3390/sym10110546} {\bibfield  {journal}
  {\bibinfo  {journal} {Symmetry}\ }\textbf {\bibinfo {volume} {10}},\ \bibinfo
  {pages} {546} (\bibinfo {year} {2018})},\ \Eprint
  {http://arxiv.org/abs/1710.11006} {arXiv:1710.11006 [astro-ph.CO]}
  \BibitemShut {NoStop}%
%%CITATION = ARXIV:1710.11006;%%
\bibitem [{\citenamefont {Barrow}(1982)}]{Barrow:1982}%
  \BibitemOpen
  \bibfield  {author} {\bibinfo {author} {\bibfnamefont {J.~D.}\ \bibnamefont
  {Barrow}},\ }\href@noop {} {\bibfield  {journal} {\bibinfo  {journal} {Nucl.
  Phys.}\ }\textbf {\bibinfo {volume} {B208}},\ \bibinfo {pages} {501}
  (\bibinfo {year} {1982})}\BibitemShut {NoStop}%
\bibitem [{\citenamefont {Ford}(1987)}]{Ford:1986sy}%
  \BibitemOpen
  \bibfield  {author} {\bibinfo {author} {\bibfnamefont {L.~H.}\ \bibnamefont
  {Ford}},\ }\href {\doibase 10.1103/PhysRevD.35.2955} {\bibfield  {journal}
  {\bibinfo  {journal} {Phys. Rev.}\ }\textbf {\bibinfo {volume} {D35}},\
  \bibinfo {pages} {2955} (\bibinfo {year} {1987})}\BibitemShut {NoStop}%
%%CITATION = PHRVA,D35,2955;%%
\bibitem [{\citenamefont {Spokoiny}(1993)}]{Spokoiny:1993kt}%
  \BibitemOpen
  \bibfield  {author} {\bibinfo {author} {\bibfnamefont {B.}~\bibnamefont
  {Spokoiny}},\ }\href {\doibase 10.1016/0370-2693(93)90155-B} {\bibfield
  {journal} {\bibinfo  {journal} {Phys. Lett.}\ }\textbf {\bibinfo {volume}
  {B315}},\ \bibinfo {pages} {40} (\bibinfo {year} {1993})},\ \Eprint
  {http://arxiv.org/abs/gr-qc/9306008} {arXiv:gr-qc/9306008 [gr-qc]}
  \BibitemShut {NoStop}%
%%CITATION = GR-QC/9306008;%%
\bibitem [{\citenamefont {Joyce}(1997)}]{Joyce:1996cp}%
  \BibitemOpen
  \bibfield  {author} {\bibinfo {author} {\bibfnamefont {M.}~\bibnamefont
  {Joyce}},\ }\href {\doibase 10.1103/PhysRevD.55.1875} {\bibfield  {journal}
  {\bibinfo  {journal} {Phys. Rev.}\ }\textbf {\bibinfo {volume} {D55}},\
  \bibinfo {pages} {1875} (\bibinfo {year} {1997})},\ \Eprint
  {http://arxiv.org/abs/hep-ph/9606223} {arXiv:hep-ph/9606223 [hep-ph]}
  \BibitemShut {NoStop}%
%%CITATION = HEP-PH/9606223;%%
\bibitem [{\citenamefont {Salati}(2003)}]{Salati:2002md}%
  \BibitemOpen
  \bibfield  {author} {\bibinfo {author} {\bibfnamefont {P.}~\bibnamefont
  {Salati}},\ }\href {\doibase 10.1016/j.physletb.2003.07.073} {\bibfield
  {journal} {\bibinfo  {journal} {Phys. Lett.}\ }\textbf {\bibinfo {volume}
  {B571}},\ \bibinfo {pages} {121} (\bibinfo {year} {2003})},\ \Eprint
  {http://arxiv.org/abs/astro-ph/0207396} {arXiv:astro-ph/0207396 [astro-ph]}
  \BibitemShut {NoStop}%
%%CITATION = ASTRO-PH/0207396;%%
\bibitem [{\citenamefont {Profumo}\ and\ \citenamefont
  {Ullio}(2003)}]{Profumo:2003hq}%
  \BibitemOpen
  \bibfield  {author} {\bibinfo {author} {\bibfnamefont {S.}~\bibnamefont
  {Profumo}}\ and\ \bibinfo {author} {\bibfnamefont {P.}~\bibnamefont
  {Ullio}},\ }\href {\doibase 10.1088/1475-7516/2003/11/006} {\bibfield
  {journal} {\bibinfo  {journal} {JCAP}\ }\textbf {\bibinfo {volume} {0311}},\
  \bibinfo {pages} {006} (\bibinfo {year} {2003})},\ \Eprint
  {http://arxiv.org/abs/hep-ph/0309220} {arXiv:hep-ph/0309220 [hep-ph]}
  \BibitemShut {NoStop}%
%%CITATION = HEP-PH/0309220;%%
\bibitem [{\citenamefont {Sikivie}(2008{\natexlab{b}})}]{Sikivie2008}%
  \BibitemOpen
  \bibfield  {author} {\bibinfo {author} {\bibfnamefont {P.}~\bibnamefont
  {Sikivie}},\ }\enquote {\bibinfo {title} {Axion cosmology},}\ in\ \href
  {\doibase 10.1007/978-3-540-73518-2_2} {\emph {\bibinfo {booktitle} {Axions:
  Theory, Cosmology, and Experimental Searches}}},\ \bibinfo {editor} {edited
  by\ \bibinfo {editor} {\bibfnamefont {M.}~\bibnamefont {Kuster}}, \bibinfo
  {editor} {\bibfnamefont {G.}~\bibnamefont {Raffelt}}, \ and\ \bibinfo
  {editor} {\bibfnamefont {B.}~\bibnamefont {Beltr{\'a}n}}}\ (\bibinfo
  {publisher} {Springer Berlin Heidelberg},\ \bibinfo {address} {Berlin,
  Heidelberg},\ \bibinfo {year} {2008})\ pp.\ \bibinfo {pages}
  {19--50}\BibitemShut {NoStop}%
\end{thebibliography}%

\end{document}